\pdfoutput=1
\documentclass[11pt]{article}

\usepackage[utf8]{inputenc}
\usepackage[english]{babel}
\usepackage{graphicx,amssymb,amsmath,amsfonts}
\usepackage{hyperref}
\usepackage[margin=1in]{geometry}
\usepackage[noadjust]{cite}
\usepackage{xcolor}
\usepackage{graphicx,amssymb,amsmath,amsfonts}

\usepackage{enumerate}
\usepackage{rotating}
\numberwithin{equation}{section}

\usepackage[utf8]{inputenc}

\date{October 15, 2026}

\def\be{\begin{equation}}
\def\ee{\end{equation}}

\newcommand{\avg}[1]{\langle #1 \rangle}

\newmuskip\pFqmuskip

\newcommand*\pFq[6][8]{%
  \begingroup 
  \pFqmuskip=#1mu\relax
  \mathchardef\normalcomma=\mathcode`,
  \mathcode`\,=\string"8000
  \begingroup\lccode`\~=`\,
  \lowercase{\endgroup\let~}\pFqcomma
  {}_{#2}F_{#3}{\left(\genfrac..{0pt}{}{#4}{#5}\Big | #6\right)}%
  \endgroup
}
\newcommand{\pFqcomma}{{\normalcomma}\mskip\pFqmuskip}

\newcommand{\pmat}{\begin{pmatrix}}
\newcommand{\fpmat}{\end{pmatrix}}
\newcommand{\eq}{\begin{equation}}
\newcommand{\feq}{\end{equation}}
\newcommand{\cas}{\begin{cases}}
\newcommand{\fcas}{\end{cases}}

\newcommand{\eqarray}{\begin{eqnarray}}
\newcommand{\feqarray}{\end{eqnarray}}

\def\be{\begin{equation}}
\def\ee{\end{equation}}

\begin{document}
\begin{titlepage}
\title{\textbf{Statistical Characterization of Erratic and Chaotic Processes through Their Extrema 
\\ I. Thermodynamics}\vspace{5mm}}

\author{ 
\textbf{Jacob Sonnenschein} \\ \href{mailto:cobi@tauex.tau.ac.il}{cobi@tauex.tau.ac.il}  
}

\date{
 \emph{The Raymond and Beverly Sackler School of Physics and Astronomy},\\
\emph{Tel Aviv University, Ramat Aviv 69978, Tel Aviv, Israel}
\today}
	
\maketitle

\begin{abstract}
 We propose a general statistical framework for characterizing erratic functions through the statistics of their extrema, with the aim of identifying statistical signatures associated with chaotic processes.
   It is based on analyzing the extrema points of   erratic functions  associated with these processes. One is instructed to determine the probability distribution functions ( PDF)s of the locations of the extrema  points, the spacings between the locations, in particular  the  nearest neighbors,  and the ratios between adjacent spacings.  We then fit these PDFs with PDFs known to relate to chaotic behavior like of random matrix theory. 

 The extrema considered establish a cloud of points in general in $D$ dimensions for  erratic function of several variables. For certain systems one finds also  extrema curves. 
 
 We further determine the ``thermodynamic" description of the erratic functions by computing the Shannon or differential entropy and by equating the PDFs to the Gibbs one.  With this map we determine the ``energy function" and for certain cases also the ``temperature".

We analyze three benchmark spacing PDFs: the Bernoulli shift map, the Poisson  and the Gaussian beta ensemble and several others. 
 We want to emphasize that the map to thermodynamics can be applied to variety of chaotic processes including non-physics ones. 

 We  distinguish marginal spacing distributions from joint distributions of consecutive spacings. The latter probe correlations and statistical organization that are invisible in one-spacing PDFs and may provide stronger signatures for distinguishing structured chaotic-like behavior from uncorrelated irregularity.
\end{abstract}

\end{titlepage}

\flushbottom

\tableofcontents

\clearpage
\section{Introduction} 
Chaotic processes are very common. One encounters them in classical physics, in quantum mechanical systems\cite{BohigasGiannoniSchmit1984}  \cite{BerryTabor1977}, in quantum field theory  and string theory. In fact they show up also in many non-physics circumstances like in  meteorology, biology, economics population dynamics and more.

In physics there are measures that characterize  classical chaos like
Lyapunov exponent and Kolmogorov Sinai entropy\cite{Walters1982} and those related to quantum systems like the map of the energy eigenvalues  to those of random matrix theory, spectral form factor and  spectral rigidity.  

There are however,certain properties that can bridge between classical and quantum chaos like  the Lyapunov exponent related to the OTOC growth rate,  classical fidelity and Loschmidt echo, mutual information,  recurrence and more.  

In this paper we propose a general statistical framework for characterizing erratic functions through the statistics of their extrema, with the aim of identifying statistical signatures associated with chaotic processes.
 The main idea is  to allocate  in  any given suspected chaotic behavior    erratic functions and analyze them. We argue that the analysis of these erratic functions\footnote{We do not introduce a definition of an erratic function. The analysis presented here can be applied to any function. Only for erratic ones it will turn out to be relevant for chaos.} 
provides useful information that together with other measures may enable  us to determine whether the process admits  chaotic behavior  and to classify the latter. 
Differently stating we suggest a general statistical framework for characterizing erratic functions
 associated with chaos.

We consider here real erratic function of real variables. In a similar way one can consider complex erratic functions of complex variables.

The erratic functions can depend on a single variable or several ones. We refer to these cases as one-dimensional and multi-dimensional respectively. 

For one dimensional erratic functions the analysis of a given process is done using  the following steps:
\begin{itemize}
\item
(i) Identifying erratic functions that are part of the description of the process.
\item
(ii) Determining ensemble of sets of extrema points of the each erratic function.
\item
(iii) Assigning  probability density functions (PDF)s to  properties of the  extrema points in particular  their locations, spacings between the  locations, ratios of adjacent spacing   etc. Obviously these are one-dimensional PDFs. 
\item 
(iv) Examining the fits of the  PDFs that were found,  to known PDFs that  are related to chaotic processes, for instance those that  are related to RMT. 
\item
(v) Computing the Shannon or differential  entropy of   the various PDFs.
\item
(vi) Mapping the PDFs to  Gibbs thermodynamical PDFs and determining the dimensionless ``energy function" $\beta E$.
\item 
(vii) For certain cases by using families of PDFs one can also determine the temperature.
\item 
(viii) Introducing  joint probability distributions  of spacings, rather then the PDF for a single spacing. This relates to the  statistical interactions among the spacings.
\item
(ix) Comparing the thermodynamics properties in particular the entropy and the energy between different processes and in particular known   chaotic ones.

Steps (i) up to (vi) were applied to decays and scattering  of high excited string states,  and phase shifts of the leaky torus \cite{Bianchi:2022mhs},\cite{Bianchi:2023uby} and \cite{Bianchi:2024fsi}. 
in\cite{Ageev:2025yiq} \cite{Ageev:2026jvx} these steps were using to analyze correlation functions for quenched QFTs.

Higher-dimensional erratic functions admit much richer structure. The latter include  special points but also curves, surfaces etc. 
The special points include extrema points, namely, maxima and minima but also saddle points. 
The analysis of the isolated special  points in higher dimensions follows similar steps as those for the one-dimensional case.  An obvious difference is that the locations and spacings are now vectors. At this point there are two options: (i) consider PDFs which are functions of vectors and (ii) dimensional reduce  the vectors  to one dimension. In (\ref{Rtodv}) several methods of dimensional reduction are described. These methods were applied in \cite{Bianchi:2025kna} to the cases of quantum pinball scattering and a toy model of the electric potential of a set of charges  randomly distributed on a plane. 

 On top of the isolated points, there are also higher dimensional special structures.  For example in two dimensions,    there may exist curves on which a derivative with respect to one  of the variables vanish but not with respect to the other. 
 In higher dimensions this will be generalized to sub-manifolds were derivatives with respect to certain variables vanish. 

Following the analysis performed in \cite{Bianchi:2026jui} for the string scattering of two high excited and two tachyons, we propose to assign to each curve the area below this curve. We then compute the differences between the areas of adjacent curves and assign to them a PDF.  From there on the procedure is the same as described above. 

Whereas for the one dimensional erratic functions one can invoke a map between the locations of the extrema and the eigenvalues of random matrices and similarly for the spacings and the ratios of spacings, in the multi-dimensional case the story is very different.
For the latter case the locations and spacings are vectors which have no analog in random matrix theory\footnote{It was conjectured\cite{Bianchi:2025kna}  that they could be mapped to eigenvalues of random tensors.}

It is well known that the major measure of chaotic behavior is the map of the energy eigenvalues and those of random matrix theory RMT. It may seem that this measure has nothing to do with the measure that uses the extrema points of erratic functions. It turns out that in fact the former can be viewed as a special case of the latter. The erratic function of a given quantum system can be either (i) the resolvant of the density of states (ii) the scattering phase shifts.  

We  distinguish marginal spacing distributions from joint distributions of consecutive spacings. The latter probe correlations and statistical organization that are invisible in one-spacing PDFs and may provide stronger signatures for distinguishing structured chaotic-like behavior from uncorrelated irregularity.

We briefly review  applications of the universal approach on several examples. The latter include systems studied in the past like: the classical and quantum  pinball scattering\cite{Bianchi:2025kna}, the phase shifts of a ``leaky torus", electric potential of randomly distributed charges\cite{Bianchi:2025kna},  quenched massless field theory\cite{Ageev:2025yiq}, and  decays of  $HES\rightarrow T+T$\cite{Bianchi:2022mhs} and scattering of $HES+T\rightarrow HES+ T$ \cite{Bianchi:2026jui} in bosonic string theory. We also bring examples of system for which  the use of the method have not been implemented in the past, in particular the Bernoulli shift map and the spectra that can be mapped to RMT. 

An important tool in analyzing the extrema points both for one and higher dimensional erratic functions is the spectral form factor (SFF).  The latter was intensively discussed in \cite{Bianchi:2024fsi}. We will not elaborate more about it in this paper. 

To get further insight on the erratic functions we propose a prescription to assign to each PDF a thermodynamical description.  This is done following these steps:

(I) We compute the  Shannon entropy for discrete PDFs and differential entropy for the continuous cases. 

(II) We equate the PDF with the Gibbs thermodynamic probability distribution  which enables us to determine the $\beta E$ where $E$ energy function and the internal energy. 

(III) To determine the temperature separately  we will need to embed each PDF in a  ``family" of PDFs characterized by certain parameters. 

(IV) The last step is to write down the first law of thermodynamics. 


   We then  determine and describe   the  thermodynamics of known distributions. 
   This will add tools  to relate a PDF that follows numerically from a given erratic function with known PDF. 
   We divide the  known PDFs  to the following classes:
\item 
Locations of the points-
  These include
 (i)  Poisson (ii)  Gaussian  (iii) flat (iv) Dirac delta function (iii) Bernoulli map (iv) Wigner semi-circle PDF.  
\item 
Spacings-
Including: (i) Poisson (ii) GOE (iii)$G\beta E$ (iv) Log normal (v) Logistic

\item 
Ratios of spacings-
(i) Poisson (ii) $G\beta E$

\item 
``Heights"-
(i) $G\beta E$ (ii) Porter Thomas distribution
\item 
Higher-dimensional PDFs-
(i) Multivaiate Gaussian (ii)  $G\beta E$. 

In addition to the thermodynamics of each PDF one can also use tools from information theory like conditional entropy, mutual information relative entropy etc. These will be addressed in a sequel paper.





The paper is organized as follows:
In the next section (\ref{Feftcp})
We  first present the properties of the erratic function for which we  will assign PDFs. We separate the presentation to (i) the one dimensional cases and the multi-dimensional ones. The former include the locations of the extrema, the spacings between locations, the ratios between adjacent locations, the ``eigenvectors" and the integrated spacings. The subsection of  multi-dimensional cases includes special points, higher dimensional spacings between extrema points, reduction to one dimensional variables and special curves. We then discuss the determination of the PDFs associated with the various special points. These include one-dimensional discrete PDfs, the passage from discrete to continuous ones, the unfolding process,  the multi-dimensional discrete and continuous PDFs and finally the procedure of fitting PDFs.
In section (\ref{EfatcP}) we present several examples of erratic functions associated with chaotic processes and their corresponding PDFs. We start with the Bernoulli map. Then we proceed to the classical pinball scattering and the quantum one. Next we show how to present  the energy  eigenvalues that are mapped to RMT to erratic functions in terms of resolvant and scattering cross sections. The phase shifts of the leaky torus and the zeros of the Riemann zeta function come next. We describe a toy model of the electric potential of randomly located charges.
In the context of QFTs we describe correlation function in quenched massless scalar field theory. The last examples are decay and scattering processes of highly excited  bosonic string theory states.
In the following section (\ref{FPtTg} ) we state the generalities of mapping the PDFs to thermodynamics. The first property computed is the Shannon entropy the discrete and differential ones. We then  determine the energy function and when possible suggest a definition of temperature. We show that the thermodynamics of the erratic functions follows the first law. We discuss the question of whether there is a thermodynamics map for any PDF and determine the necessary condition to have a valid map.  Next is section (\ref{TosBs}) where we present   benchmark systems of the  thermodynamics of spacings.
We present the analysis of (I) The Bernoulli shift map (II) The Poisson PDF and (iii) The cases of $G\beta E$.
Section (\ref{Tfjdoaeoas} ) is devoted to the thermodynamics for joint distribution of all extrema or all spacings. The meaning of $H$ and the temperature are discussed. Constraints are shown to have natural meaning and the dynamics of extrema and the one hidden in spatial sequence  is spelled out. The underlying function ${\cal A}(x)$ and the origin of the effective Hamiltonian are discussed. We present a concrete hierarchy.

We then summarize in (\ref{Sotd}) the thermodynamics descriptions  in particular comparison between locations, spacings and spacings ratios. We wight the added values of these descriptions. What do S and T tell us about chaoticity.
The last section about future directions (\ref{Cafd})  suggests looking on how does changing temperature affect the erratic functions.  Potential phase transitions and cross overs. The information theory perspective of  erratic functions and chaotic behavior. 
how does changing the temperature affect the PDFs and brief discussion about cross over and

Appendix (\ref{ToPotlote}) describes the 
thermodynamics of PDFs of the locations of the extrema . This includes 
the Poisson probability mass function, the Gaussian distribution, the discrete and continuous uniform PDF, the Dirac delta function  and the Wigner semi-circle surmise.
The thermodynamics of PDFs of spacings is discussed next(\ref{ToPos}). The first is  the zeros of the Riemann zeta function. We end this section with the cases of the Log normal and logistic PDFs.
Section (\ref{Toros}) deals with the ratios of spacings both for the Poisson PDF and the $G\beta E$ one.
In the following section (\ref{ToPoe})
we describe the  thermodynamics of the extrema heights. In particular the  GOE and $G\beta E$ and of the Porter Thomas distribution. 
In section (\ref{TfhdP} we present the thermodynamics associated with the multivariate Gaussian and the higher dimensional $G\beta E$
\section{From erratic functions to continuous probability functions}\label{Feftcp}
Chaotic processes often admit erratic functions\footnote{ We do not introduce here a definition of an ``erratic function". The discussion applies to any function, but only for ``rapidly changing functions" it is non-trivial}. We divide the analysis of the erratic functions into two parts: (i) functions of a single  variable referred to as one-dimensional cases and to (ii) those that depend on several variables referred to as multi-dimensional cases.
We consider here real functions of real variables, but a similar type of analysis can apply also to complex functions that depend on complex variables .

There are various ways to analyze and classify erratic functions.   Our analysis is based on identifying  various types of mathematical objects  and determine  their probability distribution function (PDF). 

In the one-dimensional case these objects are: (i) Locations of the extrema points of the functions. (ii) The spacings between the extrema. These could be the spacing between  nearest neighbor (NN) locations or more generally $k$ spacings. (iii) The ratios of adjacent spacings or the ratios of $k$ spacings. (iv) The values of the functions at each extremum point  which we refer to as `` heights" (v) The spacings between the ``heights" both NN and $k$ spacings. (vi) The integrated spacings which are the integrals of the function between the extrema ( again NN and $k$ spacings. (vii) The rations of the integrated spacings. 

In the multi-dimensional case the kee player are: (i) Locations of the extrema points maxima, minima and saddle points. (ii) Spacing vectors between the latter. (iii) Projecting the spacing vectors to certain directions (iv)  All pair spacings namely the absolute value of the vectors between all pair of points (v) The nearest neighbors, namely, the set of all the minimal distances from each point. (vi) The spacings and ratios along a greedy-path.

It is well known from graph theory that there are many more types of measures that one can apply on a cloud of points\cite{Barabasi2016NetworkScience}\cite{WassermanFaust1994SocialNetworkAnalysis}\cite{Newman2018Networks}\cite{Boccaletti2006ComplexNetworks}\cite{Newman2003StructureFunction}\cite{Costa2007Characterization}.

In the following subsections we present the key players of the analysis of erratic functions. We  check the  probability distribution of each type of the players a associate to it a probability distribution function (PDF
A basic reference to the discussion  in this section is \cite{DaleyVereJones2003}
\subsection{One dimensional case}
For the one-dimensional case ${\cal A}(x)$  
one can define  the following sets of discrete values:
\subsubsection{ The locations of the extrema}
The locations of the extrema points where
\be\label{locex} 
 x_n  \qquad {\cal F}_x(x=x_n)= \frac {d Log[{\cal A}]}{dx}(x=x_n)=0 \qquad x_1<x_2<x_3...<x_N
\ee
Assuming ${\cal A}(x)$ is non-negative and ${\cal A}(x_n) \neq 0$.\footnote{We use here the definition used in \cite{Bianchi:2022mhs}. Obviously one can instead use$\frac {d {\cal A}}{dx}(x=x_n)=0$ } One can further separate the extrema points into $x_n^-$ and maxima  $x_n^+$in the form of   
\be
{\cal F}^-_x(x=x^-_n)=0 \qquad  
\frac{d{\cal F}^-_x(x=x^-_n)}{dx} >0
\ee
\be
{\cal F}^+_x(x=x^+_n)=0 \qquad \frac{d{\cal F}^-_x(x=x^-_n)}{dx} <0.
\ee
The spacings also can be defined separately for the minima and maxima points.

For non-differential functions  we will use the notion of peaks given by 
\[
x_{n-1}<x_n>x_{n+1}.
\]
( See for instance in
\subsubsection{The spacings}
Next we define  the spacings between adjacent extrema points
\be
\delta_n = x_{n+1} -x_n
\ee

In a similar manner we define the $k$ spacings and the ratios of them as follows
\be
\delta_{k,n} = x_{n+k} -x_n
\ee

\subsubsection{The ratios of spacings}
For ordered set of spacings $\delta_n$ one defines 
 the ratios between adjacent spacings $r_n$ and $\tilde r_n$ as follows

\be
r_n\equiv \frac{\delta_{n+1}}{\delta_n} \qquad \tilde r_n\equiv Min\{r_n,\frac{1}{r_n}\}
\ee
and similarly also to the $\delta_{k,n}$

\be
r_{k,n}\equiv \frac{\delta_{k,n+k}}{\delta_{k,n}}=\frac{x_{n+2k}-x_{n+k}}{x_{n+k}-x_n} \qquad \tilde r_{k,n}\equiv Min\{r_{k,n},\frac{1}{r_{k,n}}\}
\ee
\subsubsection{The extrema heights}
The function at extrema points
In RMT there are two basic distribution function: (i) the pdf of the eigenvalues of matrices and (ii) the pdf of the eigenvectors. 
In the context of erratic functions, we associated with the eigenvalues the locations of the extrema points. We would consider an analog of the eigenvectors in the form of the value of the erratic function at the extrema point.  If we consider the points of maximum, we take the set

\be
{\cal A}_n= {\cal A}(x^+_n)\qquad n=1,2...N
\ee
One can also consider the differences between the values of adjacent peaks
\be\label{difev}
s^{{\cal A}}_n= {\cal A}_{n+1} - {\cal A}_{n}
\ee
and similar to the $k$ spacing also 

\be
s^{{\cal A}}_{n,k}= {\cal A}_{n+k} - {\cal A}_{n}
\ee

\subsubsection{The integrated spacings}
In adition to the analysis of the extrema points of the original function ${\cal A}(x)$, and the spacing between them, one can also define the ``integrated spacing" as follows
\be
I_n= \int_{x_{n}}^{x_{n+1}} {\cal A}(x) dx 
\ee
where $x_n$ and $x_{n+1}$ are two consecutive extrema points of ${\cal A}(x)$ and similarly
\be
I_{k,n}= \int_{x_{n}}^{x_{n+k}} {\cal A}(x) dx 
\ee
  Once again one can define also the ratios of the adjacent  and $k$ spacings 
\be
r_{I_n}=\frac{I_{n+1}}{I_n}
\ee
and
\be
r_{I_{k,n}}=\frac{I_{n+k}}{I_n}
\ee

\subsubsection{The set of all the variables}

We can  now define the set of all the variables
\be
Z_n =( \delta_n, \delta_{k,n},r_n,r_{k,n}, s^{{\cal A}}_n, s^{{\cal A}}_{k,n}, r^{{\cal A}}_n,r^{{\cal A}}_{k,n},I_n,I_{k,n},r^I_n,r^I_{k,n}, \kappa_n,\sigma_n )\label{variables}
\ee
where $\kappa_n={\cal A}_n"$ and $\sigma_n$ Maximum/minimum label.
\subsection{Multi-dimensional case}
The generalization  of the erratic function from the one-dimensional case  ${\cal A}(x)$  to the multi-dimensional cases  ${\cal A}(\vec x)$ seem straightforward, but in fact as will see shortly the latter cases admit much richer set of structures. 
For simplicity we take assume that $\vec x = (x,y)$
\subsubsection{Special points}
\item
We start with the  set of {\bf extrema points} $\vec x_n=(x_n,y_n)$.
They obey 
\be 
  {\cal F}_x(x=x_n,y=y_n)= \frac {\partial Log[{\cal A}(x,y)]}{\partial  x}(x=x_n,y=y_n)=0 \ee
  \be  {\cal F}_y(x=x_n,y=y_n)= \frac {\partial Log[{\cal A}(x,y)]}{\partial y}(x=x_n,y=y_n)=0 
\ee
\item
We then define the Hessian 
\be
{\cal D}= {\cal F}_{xx} {\cal F}_{yy} -\left({\cal F}_{xy} \right)^2
\ee
\be
{\cal F}_{xx}= \partial_x{\cal F}_x \qquad {\cal F}_{yy}= \partial_y{\cal F}_y \qquad {\cal F}_{xy}= \partial_x{\cal F}_y
\ee
The extrema  points could be  {\bf maxima points} ${\vec x}^+_n= (x^+_n,y^+_n) $ if 
\be 
 {\cal F}_{xx}(x=x^+_n,y=y^+_n) < 0 \qquad   {\cal D}(x=x^+_n,y=y^+_n)   >0 
\ee
where 

\item
These points could be  {\bf minima points} $\vec x^-_n= (x^-_n,y^-_n) $ if

\be 
 {\cal F}_{xx}(x=x^-_n,y=y^-_n) > 0 \qquad   {\cal D}(x=x^-_n,y=y^-_n)   >0 
\ee
 
\item
In addition there are {\bf saddle points}
$\vec x^s_n= (x^s_n,y^s_n) $ if
\be{\cal D}(x=x^s_n,y=y^s_n)   <0 
\ee

\item 
The special points in two dimensions can obviously be generalized to higher dimensions.

Let \(f: U \subset \mathbb{R}^n \to \mathbb{R}\) be a \(C^2\) function, and let \(\vec x_m \in U\) be a critical point of \(f\), i.e.
\be
\nabla f(\vec x_m) = 0.
\ee
Then \(\vec x^s\) is  a {\bf saddle point} if the Hessian matrix
\be
H_f(\vec x^s) = \left[\frac{\partial^2 f}{\partial x_i \partial x_j}(\vec x^s)\right]_{i,j=1}^n
\ee
is \emph{indefinite}, meaning there exist vectors \(u,v \in \mathbb{R}^n\) such that
\be
u^T H_f(\vec x^s) u > 0
\quad \text{and} \quad
v^T H_f(\vec x^s) v < 0.
\ee
```
 \subsubsection{Higher-dimensional spacings between the extrema points}\label{tdsoe}
In two or higher dimensions, the special points $\vec x^\pm_n$  are not  ordered,  and so one cannot define consequtive spacing. However, one can define the ``all pairs" spacings as follows   
\be
\vec\delta^\pm_{m.n}=\vec x^\pm_{m}-\vec x^\pm_n
\ee

\subsubsection{Reductions to one dimensional variables}\label{Rtodv}
Instead of the vector spacing  $\vec \delta^\pm_{m,n}$ we can consider the distance between the locations of the extrema, namely,
\be
\delta^\pm_{mn} \equiv |{\vec x ^\pm}_m-{\vec x^\pm_n}|\,, \qquad m < n     
\ee
There are types of distances  that one can define.   We will examine the following procedures
\paragraph{I. Spacings of projections on an axis:}

Given $\vec x_n = (x_n,y_n)$\footnote{Where $\vec x$ stands for $\vec x^\pm$}, after we have chosen the $x$ and $y$ axes, we can consider the distribution of spacings of the projections on the two axes. If one orders the eigenvalues such that $x_1 \leq x_2 \leq ...$, then one can analyze the distribution of the (positive) variable
\begin{equation} \delta^{(1)}_n = x_{n+1}-x_n  \: .\end{equation}
Similarly one can consider the distribution of $\delta^{(2)}_n=y_{n+1}-y_n$ after ordering the eigenvalues by their $y$-value.

This method has the disadvantage that, if one does not choose the correct axes, one can fail to see the level repulsion characteristic of chaotic spectra, as we will show in the following.

\paragraph{II. Spacings for all possible pairs:}

We consider  now the distributions associated with the distances between  all the  $N(N-1)/2$ spacings defined by:
\begin{equation}
\delta_{mn} \equiv |{\vec x}_m-{\vec x_n}|\,, \qquad m < n     
\end{equation} 
This has the advantage that we do not need to define an ordering of the eigenvalues before using it. However, since they are not ordered, we cannot define the spacing ratios, and therefore we will need to unfold the spectra to uncover the universal behavior.

\paragraph{III. Spacings of nearest neighbors (NN):}

This would be the distribution of
\begin{equation}
\delta^{(\text{NN})}_{n} \equiv \min_{m\neq n}\left(|{\vec x}_n-{\vec x_m}|\right)    
\end{equation} 
Namely for each point we take the distance between it and its nearest neighbor. This means looking at only a small subset of all the spacings $\delta_{mn}$. Note that each point has exactly one nearest neighbor by definition, but a given point can be the nearest neighbor of more than one of the other points.


\paragraph{IV. Consecutive spacings along a greedy path:}\label{greedy}

We can use the following algorithm to define an ordering and the notion of successive spacings:
\begin{enumerate}
    \item Begin by choosing an initial point, ${\vec x}_1$.
    \item The next point, ${\vec x}_2$, is the nearest point to ${\vec x}_1$ in the sense that $|{\vec x}_1-\vec{x}_2|$ is minimized.
    \item Repeat: ${\vec x}_{k+1}$ is defined as the point nearest to ${\vec x }_k$, excluding points that have previously been chosen.
\end{enumerate}
This algorithm defines a path that starts from a chosen initial point, and visits all other points once, in a determined order. There are $N$ possible paths, one per each initial point. There is some degree of overlap between these paths, but there can be considerable differences between one path and the other. There are several possible ways to address this issue: one can always choose a special reference point such as the point nearest to the origin, the point with the smallest value of $x$, the point nearest to the center of the distribution, and so forth. Conversely, one can choose an initial point at random, or perform an average over all possible paths.

The spacings defined as
\begin{equation} \delta^{Greedy}_n = |{\vec x}_{n+1}-{\vec x}_n| \end{equation}
will be analogous to successive level spacings in 1D. This algorithm also allows us to define the usual spacing ratios as \begin{equation} r_n = \delta_{n+1}/\delta_n \end{equation}
One disadvantage of this definition is that it is difficult to perform any analytic computations to predict the distribution of $\delta_n$ and $r_n$.

\subsubsection{The set of all the variables in higher dimension}
We can collect the whole set of variables associated with the extrema points of higher dimensional erratic functions as follows:

\be
Z_n =( \vec x_n, \vec\delta_{m,n} , \delta_{m,n}, \delta_n^{(1)},\dots\delta_n^{(NN)},\dots  \delta^{Greedy}_n,\dots , ,s^{{\cal A}}_n, s^{{\cal A}}_{k,n}, r^{{\cal A}}_n,r^{{\cal A}}_{k,n},I_n,I_{k,n},r^I_n,r^I_{k,n}, \kappa_n,\sigma_n )\label{variablesmd}
\ee
where $\dots$ stands for all the variables that can be extracted from the one-dimensional variable to the left,  $\kappa_n=\nabla^2{\cal A}_n$ and $\sigma_n$ Maximum/minimum or saddle point  label.

\subsubsection{Special curves}
We define  curves associated with   an erratic function  of two variables ${\cal A}(x,y)$ as  configurations  
 $y^{(x)}(x)$ and $y^{(y)}(x)$ that solves the equations correspondingly
    \begin{equation}
        {\cal F}_x\big(x,y^{(x)}(x)\big) = 0 \qquad {\cal F}_y\big(x,y^{(y)(x)}\big) = 0 
    \end{equation}
    
     where 
    \be
    {\cal F}_i\big(x,y\big)= \partial _iLog[{\cal A}(x,y)]
    \ee 
    for $i=x,y$.
     
    The  curves of type $y^{(i)}(x)$  obey   
\begin{equation}
        y^{(i)\prime}(x) = -\frac{\partial_x {\cal F}_i}{\partial_y {\cal F}_i}
    \end{equation}

We will argue that the curves associated with chaotic processes are non-intersecting. This can be phrased as a ``repulsion " between curves. To quantify this property we need to  measure the distances between adjacent curves We propose to do it by defining the areas under each curve $y^{(i)}(x)$:\footnote{To simplify the notations we denote $y^{(i)}(x)$ by $y(x)$}

    \begin{equation}
        A_k = \int_{-1}^{1} dx \,(y_k(x) + 1)
    \end{equation}
    We will be interested in the areas between adjacent curves, so we define the spacings of these ``area eigenvalues'', analyzing the distribution of the spacings and their ratios:
    \begin{equation}\label{areaspac}
        \delta_k = A_{k+1}-A_k\,,\qquad r_k = \frac{\delta_{k+1}}{\delta_k}
    \end{equation}

\subsection{From special points to  probability distribution functions}
Next we associate to each type of special points, whether one-dimensional or higher-dimensional  ones,  a probability distribution function PDF. 
\subsubsection{One dimensional  discrete PDFs}
Let us denote in any    one dimensional case  by $z_n\ n=1...N$ the set   discrete variables that can be either
\be
x_n, \delta_n, r_n \qquad  or \
\delta_{n,k}, r_{k,n} \qquad 
or\  I_n, r_{I_n} 
\ee
and in fact any discrete variable.

For each erratic function that we analyze, for each $z-n$  we determine numerically  the probability  distribution function (PDF) $p_n(z_n)$ such that 

\be 
Z :=\{z_n,p_n(z_n)\}\qquad  n\in\{1,...N\} \qquad \sum_{n=1}^N p_n=1
\ee

\subsubsection{From Discrete to continuous variables }

For the case that the number of 
 discrete variables $z_n$    is very large  $N>>1$, we  can transfer the discrete points into a continuous variable 
\be
z_n = \sqrt{N}{\cal Z}(z) \qquad z=\frac{n}{N} \qquad
\frac{1}{N} \sum_n  \rightarrow \int \rho(z) dz \qquad \rho=\frac{d {\cal Z}} {dz}
\ee 
where $\rho(z)$  denotes the density of  the $z$ points.
We emphasize that when the number of points $N$ is large the transfer to a  continuous distribution function  can  be done for  all the variables discussed above $x_n$,
  $\delta_n$, $r_n$ , $\delta_{kn}$, $r_{kn}$, $I_n$, $r_{I_n}$.

For the continuous variable $z$, the probability  distribution $p(z)$ is defined by 
\be 
Z :=\{z,p(z)\}\qquad z\in\{0, \infty\} \qquad \int_0^\infty p(z) dz =1
\ee
The boundaries of the domain of $z$ can be altered for different cases.
 
\subsubsection{ Unfolding variables and their PDF}\label{unfolding_pdf}
Given a set of points $z_n$ and a pdf $\rho(z)$ One defines the unfolded set of points $ {z_n^{uf}}$ as follows
\be
 z_n^{uf}= \bar N(z_n) = \int_0^{z_n} \bar \rho(t) dt
\ee
where \(\overline\rho\) is the smooth mean density of extrema.
In a similar manner in the continuous case the unfolded variable $ z^{uf}(z)$ associated with $z$ is given by the CDF ( comulative distribution function): 
\be
 z^{uf}= \int_0^{z} \rho(z') dz'
\ee
 The unfolded gaps are then
\[
s_n=z^{uf}_{n+1}-z^{uf}_n.
\]
It is a universal property of the unfolding transform, which is the probability integral transform, that it  always yield a {\bf flat PDF}  for the unfolded variable. 
\be
\rho^{uf}(z^{uf}) = 1 \quad \text{for } z^{uf} \in [0, 1]
\ee 
This follows from the fact that 
\be
\rho^{uf}(z^{uf}) = \rho (z) \left| \frac{dz}{d z^{uf}} \right| \qquad and\qquad   \left| \frac{d z}{d z^{uf}} \right| = \frac{1}{\rho(z)}
\ee

\subsubsection{Multi-dimensional PDFs}

In higher dimensions the basic players are  $\vec x_n$, 
 $\vec \delta_{mn}$ , $\delta_{mn}$, $\delta_n^{(NN)}$  and $\delta_n, r_n$.
We can denote the vector ones as $\vec z_n$.
One can now associate to each value $\vec z_n$ a probability  distribution function (PDF) $p_n(z_n)$ as follows

\be 
Z :=\{\vec z_n,p_n(\vec z_n)\}\qquad  n\in\{1,...N\} \qquad \sum_{n=1}^N p_n=1
\ee along a path.

For a collection of $N$ points
$\vec z_n\in\mathbb{R}^d$, the empirical
probability measure is defined by
\begin{equation}
\rho_N(\vec z)
=
\frac{1}{N}\sum_{n=1}^{N}
\delta^{(d)}(\vec z-\vec z_n).
\label{eq:empirical_density}
\end{equation}

If, in the limit $N\to\infty$, the empirical
measure converges to a probability measure
admitting a density $\rho(\vec z)$, then
\begin{equation}
\lim_{N\to\infty}
\frac{1}{N}\sum_{n=1}^{N}f(\vec z_n)
=
\int_{\mathbb{R}^d}
f(\vec z)\rho(\vec z)\,d^d z
\label{eq:continuous_density}
\end{equation}
for suitable test functions $f$.

The probability density satisfies
\begin{equation}
\rho(\vec z)\geq 0,
\qquad
\int_{\mathbb{R}^d}
\rho(\vec z)\,d^d z=1.
\end{equation}


We emphasize that when the number of points $N$ is large the transfer to a  continuous distribution function  can  be done for  all the variables discussed above $\vec x_n$, 
 $\vec \delta_{mn}$ , $\delta_{mn}$, $\delta_n^{(NN)}$  and $\delta_n, r_n$.

 One can associate probability distribution  to the continuous variable $\vec z$ in the same way it was done for the one-dimensional case
\be 
Z :=\{\vec z,p(\vec z)\}\qquad x\in\{0, \infty\} \qquad \int_0^\infty p(\vec z) d \vec z =1
\ee
The boundaries of the domain of $\vec z$ can be altered for different cases.

\subsubsection{Fitting the PDFs}
We have started our journey by pointing out several types of spacial points both discrete and continuous that associate  with a given erratic function. The analysis of the erratic function means determining ( usually numerically) the PDFs that corresponds to the various types of special points. The next task is comparing each of these PDFs to known PDFs and determining which one fits the best to the  examined PDF  that associate with the special points.   

In the next section  we describe several examples  of erratic functions and their corresponding   fitted PDFs.

\section{ Erratic functions and their corresponding PDFs- examples}\label{EfatcP}
In the previous section  we have described various types mathematical objects that can be used to analyze erratic functions that associate with chaotic processes. To each of these mathematical object one assigns a probability distribution function (PDF).    It included both one dimensional erratic functions as well as multi-dimensional ones. 
Here in this section we give examples of some erratic functions that occur   in classical and quantum  physics and also in non physics chaotic processes.  The examples include  one and multi-dimensional cases.
The detailed analysis of these examples is not done here and we refer the reader to the relevant references.

The landmark characterization of quantum chaos is  the Wigner Dyson\cite{Dyson:1962b} map of the spectra of a quantum  system to that of RMT.
In this section we will present this map as a special case  of certain erratic functions related to these quantum systems.

The examples that are briefly reviewed  here are: (i) The Bernoulli shift map, (ii) The classical pinball scattering,(iii) The quantum pinball scattering, (iv) The quantum spectra viewed via description in terms of a resolvant and scattering phase shifts. (v) The phase shifts of the leaky torus and the zeros of the Riemann zeta function (vi) The electric potential of randomly distributed charges. (vii) Correlators in a quenched QFT. (viii) The decay of a high excited string state into two tachyons.
(ix) A similar decay into three tachyons and scattering that involve high excited states. 

Is worth mentioning that one should  distinguish more explicitly between three kinds of examples: systems whose chaotic dynamics is independently established, systems displaying statistical signatures associated with chaos, and exploratory applications for which such a connection remains conjectural. The Bernoulli shift map clearly falls into the first category.

\subsection{ The Bernoulli shift map}\label{TBsm}
 A well known mathematical erratic function is 
 the Bernoulli shift map\cite{Walters1982ErgodicTheory}\cite{LindMarcus1995}
  on the unit interval [0, 1) is defined as:
  \be
T (x) = 2x (mod 1)
\ee

Equivalently,
\be
T(x)=
\begin{cases}
2x, & 0\leq x<\frac12,\\
2x-1, & \frac12\leq x<1.
\end{cases}
\ee
For almost every initial condition \(x_0\), the invariant probability
density is uniform,
\be
\rho(x)=1,
\qquad 0\leq x<1.
\ee

In binary notation,
\be
x_0=0.b_1b_2b_3\cdots,
\qquad b_i\in\{0,1\},
\ee
and iteration of the Bernoulli map simply shifts the binary sequence:
\be
x_n=0.b_{n+1}b_{n+2}b_{n+3}\cdots.
\ee

For a typical orbit, the binary digits behave as independent Bernoulli
random variables,
\be
P(b_i=0)=P(b_i=1)=\frac12.
\ee

\item Definition of a peak

We call the point \(x_n\) a local peak if
\be
x_{n-1}<x_n>x_{n+1}.
\ee

Let
\be
n_1<n_2<n_3<\cdots
\ee
denote the iteration numbers at which peaks occur.

\item Numerical determination of the peaks

For numerical purposes it is preferable to generate the Bernoulli
orbit symbolically rather than repeatedly evaluate
\be
x_{n+1}=2x_n\pmod 1
\ee
in floating-point arithmetic.

The reason is that ordinary finite-precision arithmetic eventually
turns the binary expansion into zero.

Instead, generate a long sequence
\be
b_1,b_2,\ldots,b_N,
\qquad
b_i=0,1
\ee
with equal probabilities and construct
\be
x_n
=
\sum_{k=1}^{M}
\frac{b_{n+k}}{2^k},
\ee
where \(M\) is sufficiently large, for example \(M=50\)--\(60\).

A peak is recorded whenever
\be
x_{n-1}<x_n>x_{n+1}.
\ee

In a numerical calculation with approximately
$N=2\times10^6$
iterations, approximately $5.0\times10^5$
peaks are obtained.

Thus the density of peaks is approximately
\be
\rho_{\rm peak}\simeq0.25.
\ee

\item Spacing distribution

Let
\be
S_j=n_{j+1}-n_j.
\ee
Notice that \(S_j\) is integer-valued.
The smallest possible separation is
\be
S=2.
\ee

\item The probability mass function-

Thus the probability mass function of the peak spacings is
\be
p_s=
\frac{s-1}{2^s},
\qquad s\geq2.
\ee

Write the binary expansion as
\[
x_n=0.b_{n+1}b_{n+2}\ldots.
\]

For the doubling map, a peak occurs precisely when the consecutive symbols corresponding to \(x_{n-1}\) and \(x_n\) form the pattern \(01\).
The peak-spacing problem is therefore a waiting-time problem for successive occurrences of the word \(01\) in an independent fair binary sequence.
This yields the probability generating function
\[
G(z)=\sum_{s=2}^{\infty}p_s z^s
=\frac{z^2}{(2-z)^2},
\]

and hence
\[
p_s=\frac{s-1}{2^s}.
\]
The mean spacing is
\be
\langle S\rangle
=
\sum_{s=2}^{\infty}
s\,\frac{s-1}{2^s}.
\ee

Using the standard sums
\be
\sum_{s=0}^{\infty}s(s-1)x^s
=
\frac{2x^2}{(1-x)^3},
\ee
one obtains
\be
\langle S\rangle=4.
\ee

This agrees with the numerical result
\be
\langle S\rangle_{\rm num}
\simeq3.999.
\ee

\item Rescaled spacings

If the spacings are unfolded so that their mean is unity, define
\be
r=\frac{S}{\langle S\rangle}
=\frac{S}{4}.
\ee

The allowed values are therefore
\be
r=\frac12,\frac34,1,\frac54,\frac32,\ldots
\ee

with
\be
P\left(r=\frac{s}{4}\right)
=
\frac{s-1}{2^s}.
\ee

Thus, even after unfolding, the distribution remains discrete.

One can take any initial point $x_0$ on the interval  but a special interest is 
with an initial condition $x_0=\frac{1}{\sqrt{2}}$.

The map is ``chaotic" in the sense that by introducing a change $x_0\rightarrow x_0+\delta$ the difference in the $n$ step is
\be
\delta_n= 2^n\delta_0 \qquad \lambda=ln 2
\ee
where $\lambda$ is the Lyapunov exponent. 

 Strictly speaking, the Bernoulli shift   is a piecewise linear map with a derivative of 2 everywhere it is defined, and a single
discontinuity at x = 1/2. It does not have local smooth maxima but one can define ``peaks" $x_n$ 
such that $x_n> x_{n_1}$ and $ x_n>x_{n+1}$, as can be seen in figure (\ref{Bernoulii}).
The spacing $s_n$ between  the locations of successive peaks $e_n$  and $e_{n-1}$ is 
\be
s_n = e_n-e_{n-1}
\ee 
The corresponding PMF ( probability mass function),\footnote{The notion of PMF is used for discrete random variable } associated with the spacings, is given by
\be
p(s=k)=\frac{k-1}{2^k}
\ee
This is described in figure (\ref{Bernoulii})

  \begin{figure}[h!]
  \centering
  \includegraphics[width=0.48\linewidth]{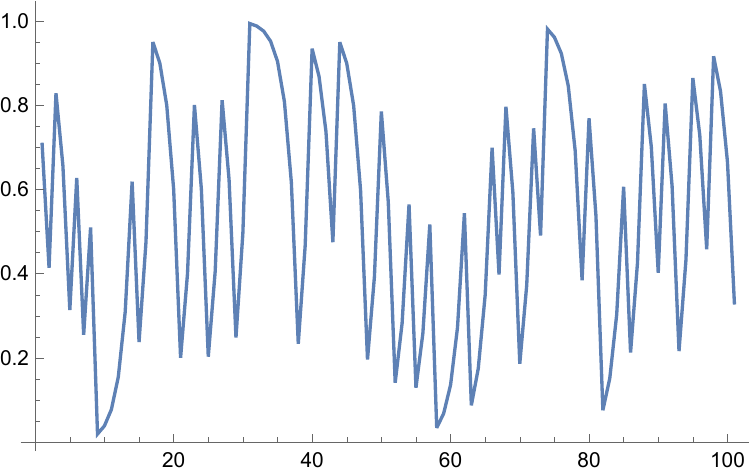}
  \includegraphics[width=0.48\linewidth]{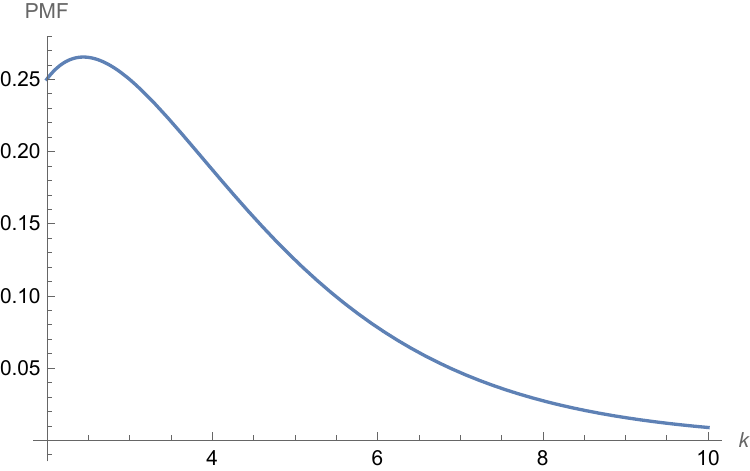}
  \caption{The Bernoulli shift map (left) and the PMF of the spacings (right).}\label{Bernoulii}
\end{figure}

\subsection{ The classical pinball scattering}
A landmark example of classical  physics chaotic behavior is the scattering of a small ball  from three disks\cite{Gaspard1998}, which is referred to as the pinball system. The system is depicted in the upper part of  figure (\ref{pinballab}) where a particle is injected with an angle $\phi$ and a location denoted by $\theta$. The figure shows two incident angles that  differ by $10^{-6}$ and yield a very different outcome angle\cite{Bianchi:2025kna}. 

This is obviously a manifestation of the butterfly effect. An approximation to the  corresponding Lyapunov exponent is given
\be
\lambda= ln \left( \frac{2 L}{R}\right )
\ee
where $R$ are the radai of the disks and $L$ is the distance between them.  
As was shown in the detailed analysis of \cite{Bianchi:2025kna} erratic functions that correspond to the pinball  classical scattering  are those of the
the out-coming  angle as function of $\theta$ and the cosine of that angle as a function of both $\theta$ and $\phi$  which are depicted on the right and left sides of figure (\ref{pinballab}) respectively.

In the former, which is a one-dimensional  function, the erratic nature is manifest in the locations of the peaks of the scattering angle. 

The latter, a two dimensional function, admits non erratic structures in the form of the three "eyes" but also erratic ones,  in the form of the parallel  lines inside the eyes which admit a repulsion and do not exist in the non-chaotic scattering from two disks\cite{Bianchi:2025kna}. 
\begin{figure}[h!]
  \centering
  \includegraphics[width=0.8\linewidth]{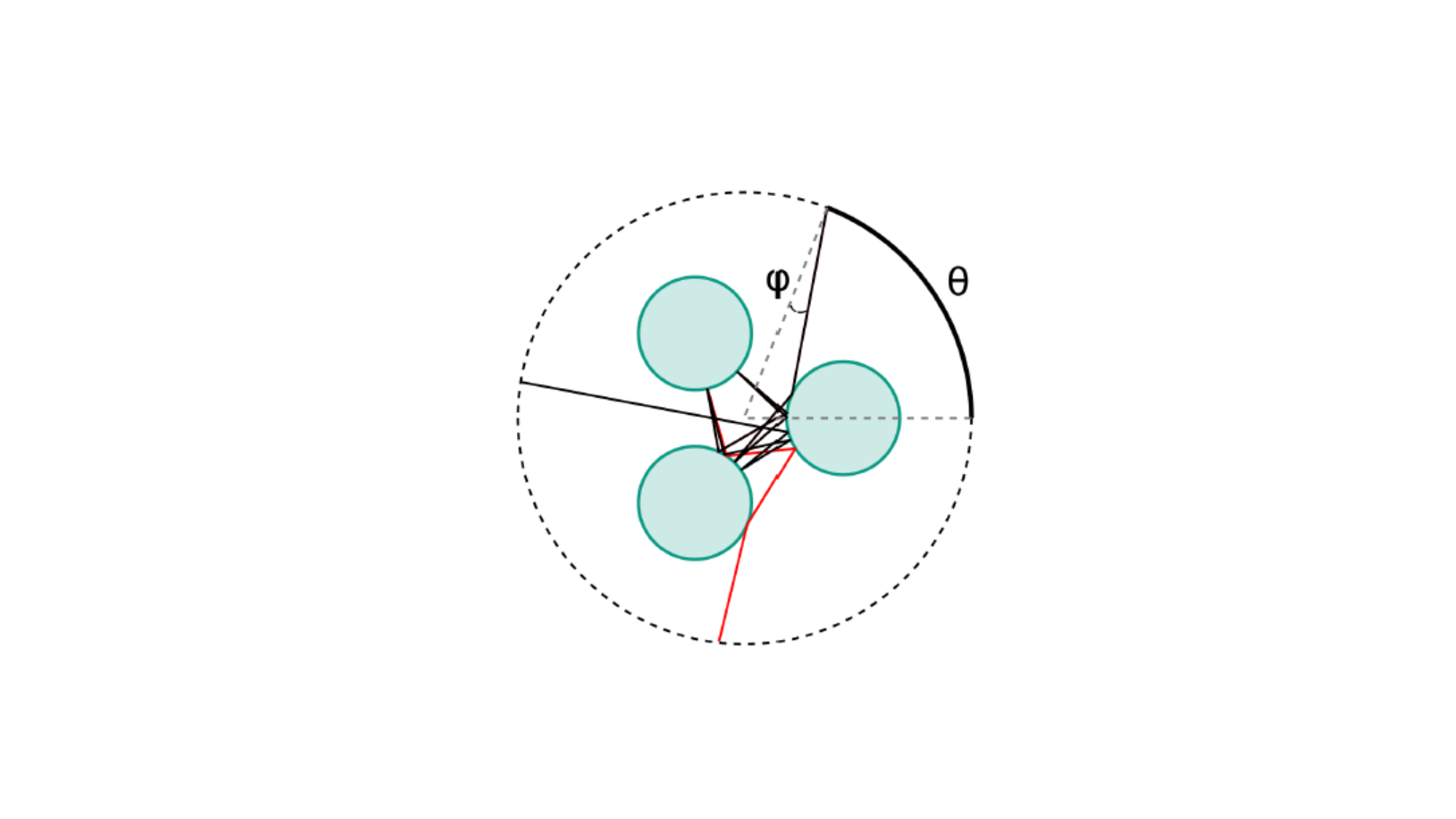}
  \includegraphics[width=0.8\linewidth]{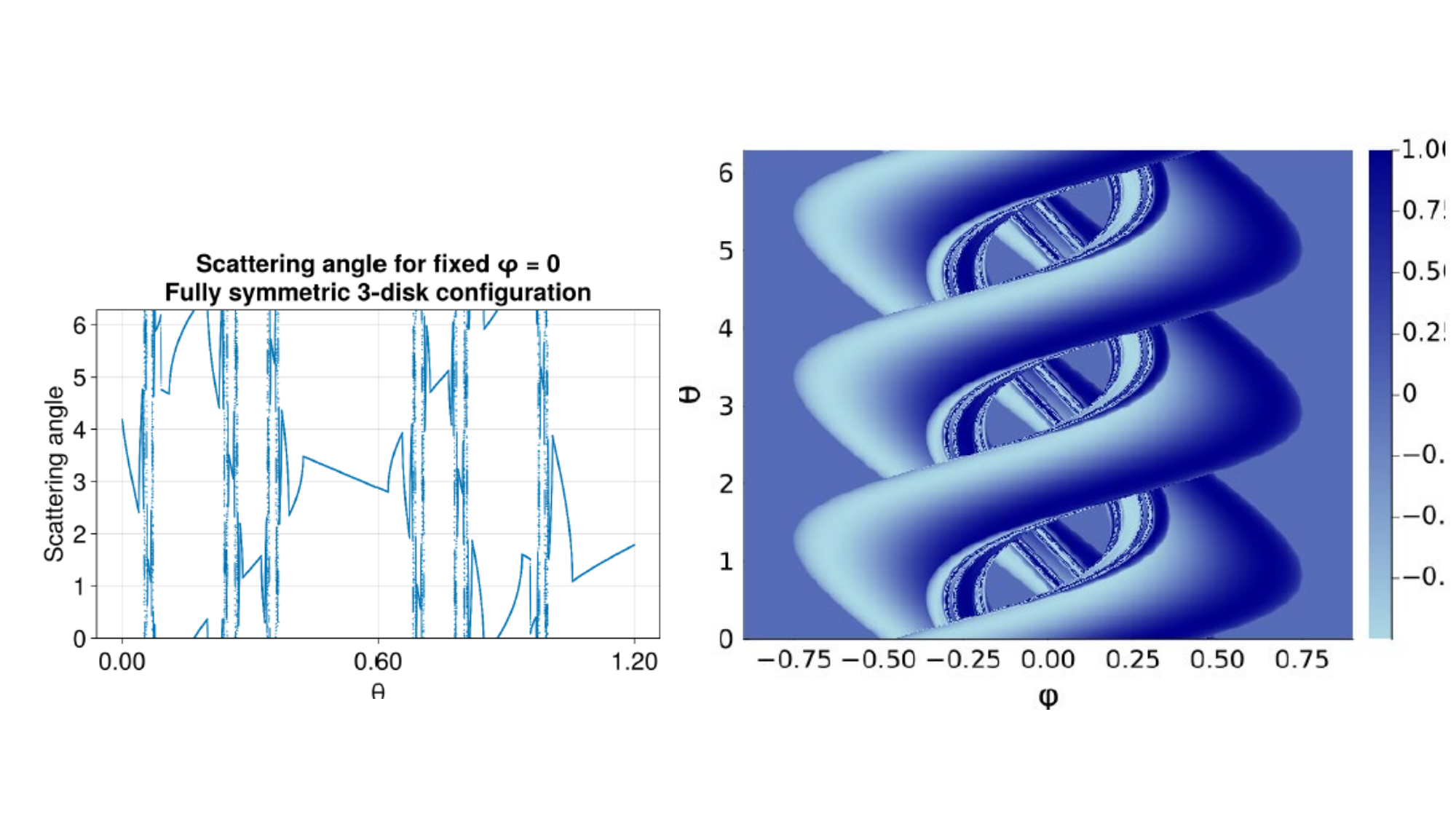}
  \caption{The scattering off a pinball system described in the upper figure. The Scattering angle as a function of $\theta$ and its cosine as a function of $\theta$ and  $\phi$ are drawn in the lower left and right plates. }\label{pinballab}
\end{figure}
It is well known that the one-dimensional erratic function is self similar. This makes the assignment of a PDF somewhat subtle. Instead we will present the PDFs associated with the quantum pinball scattering. 
\subsection{ The quantum pinball scattering}

Upon quantizing the system of the pinball, the scattering amplitude admits a cloud of peaks as depicted in the upper figures of figure(\ref{Quantpinball}) 
for two different wave numbers. The exact details of the setup are specified in \cite{Bianchi:2025kna}. The exact solution of the quantum pinball system was found in \cite{Gaspard:1989exq}

To measure the corresponding PDFs , we computed the location of the peaks of the scattering amplitude  and determined the best fits to the nearest-neighbor spacings $\delta_n$, and the spacing ratios $r_n$ along a greedy path  as discussed in (\ref{greedy}).

The distribution of nearest-neighbor spacings 
is shown  in the middle patterns in figure (\ref{Quantpinball}). The histogram  was fitted  by comparing it  to the Poisson, GOE and logistic\cite{Bianchi:2025kna}. The latter is given by 
\begin{equation}
    p_L(\delta) = \frac{1}{\sigma} \frac{e^{-(\delta-\mu)/\sigma}}{(1+e^{-(\delta-\mu)/\sigma})^2}
\end{equation}
where $\mu = 1$ (which is a choice of normalization). The best fit, as can be seen in the middle patterns in figure (\ref{Quantpinball})  is the logistic map with  $\sigma \approx 0.11$.

On the other hand, the distribution of $r_n$ along a path, which is again peaked at $r_n = 1$ was compared to the Poisson ratio PDF (\ref{Poissonrati}) and to  the GOE distribution (\ref{Atas}). It was found that its best   fit  is to  a simple distribution with
\begin{equation}
    f_B(r) = \frac{1}{2 B(a,1)} \begin{cases} r^{a-1} & 0\leq r \leq 1 \\
    \frac{1}{r^{1+a}} & 1 \leq  r
    \end{cases}
\end{equation}
which is a special case of a Beta distribution $\mathrm{B}(a,b)$ with $b=1$,\footnote{The Beta distribution owes its name to Euler Beta function and should not be confused with the $\beta$-ensemble distribution of RMT.} extended to $r > 1$ by using the inversion symmetry $r\to1/r$, implying $f(r) = \frac1{r^2}f(\frac{1}{r})$. Alternatively, one can use the normalized ratios $\tilde r_n$ which are always between 0 and 1 and fit to an ordinary Beta distribution. In our fits the parameter $b$ is fixed to 1 while $a \approx 2.9$. This is depicted in the lower plates of figure (\ref{Quantpinball}) .

\begin{figure}[ht!]
     \centering
     \includegraphics[width=\linewidth]{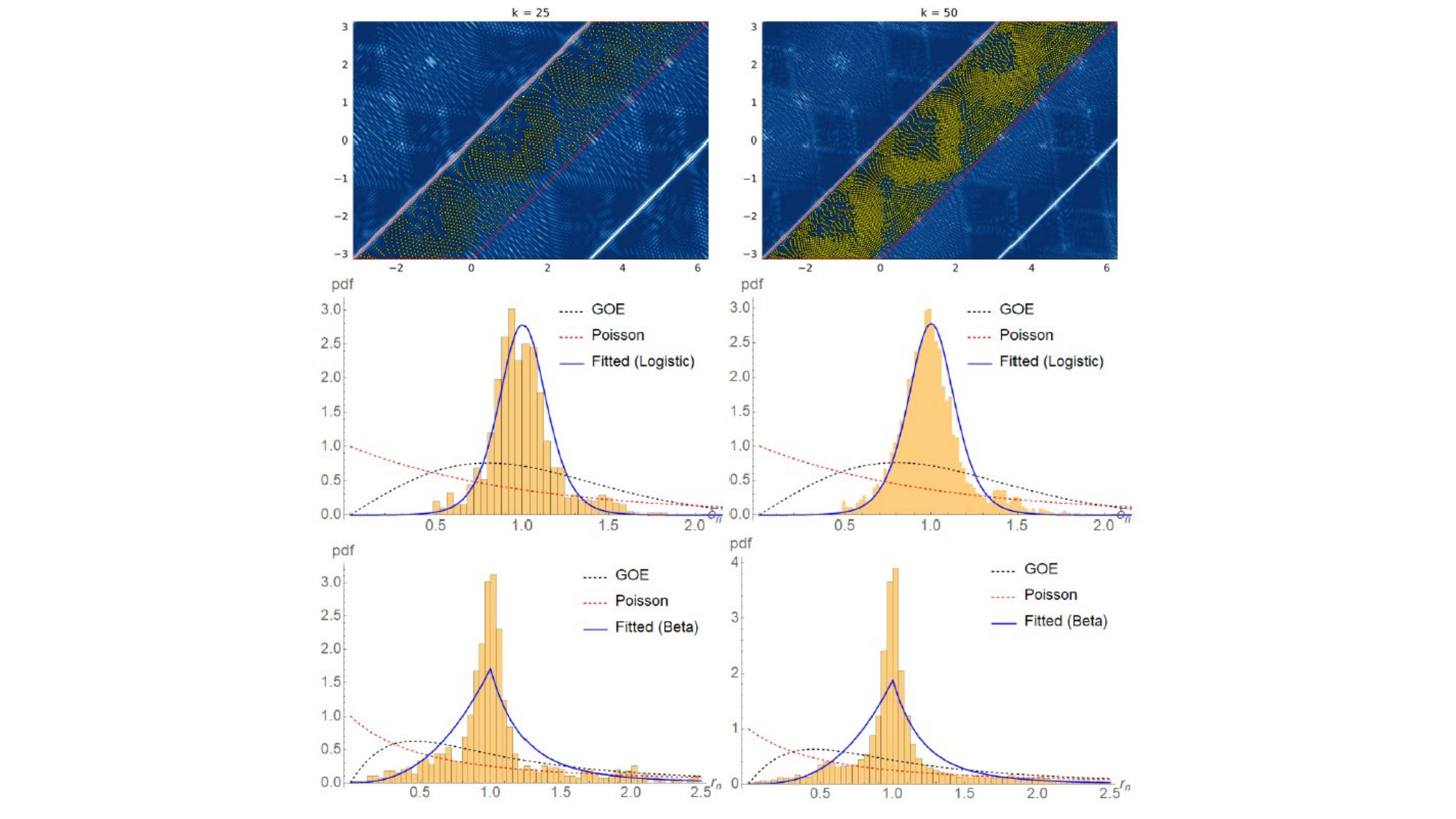}
     \caption{Peaks of the \textbf{three-disk} pinball amplitude for $k=25$ (left) and $k=50$ (right) and their distributions. Top row: The function and its peaks. Middle row: Distributions of nearest neighbor spacings. Bottom row: Distributions of path spacing ratios.}
     \label{Quantpinball}
 \end{figure}
  
  \newpage

  \subsection{Spectra of energy eigenvalues   that correspond to RMT }

One of the most powerful tools to detect quantum chaotic behavior is checking whether the energy eigenvalues of a quantum system can be faithfully mapped to the eigenvalues of a random matrix in the context of  RMT\cite{BohigasGiannoniSchmit1984,BerryTabor1977}.

The universal approach advocated in this paper is checking whether the peaks of an erratic function associated to a suspected chaotic process  admit a PDF of a form of a RMT one or a similar one.

This obviously raises the question whether also for the chaotic  energy  eigenvalues one can advice a function of the energy  which has peaks at the locations of the energy eigenvalues. 

 The answer is that indeed one  can construct mathematical functions of energy whose maxima (or more commonly, peaks, resonances, or poles) explicitly correspond to the energy eigenvalues of a quantum system.

In theoretical and computational physics, this is actually a standard technique used to find or analyze energy spectra. Here are the primary ways this is done:

\item  1. The Resolvent (Green's Function) and the Density of States-

The most common way to relate energy to the eigenvalues is through the resolvent operator (or Green's function), defined as:
\be
G(E) = \frac{1}{E - \hat{H}}
\ee

If one looks at the expectation value of this operator, or its trace (the Green's function projected onto a state or integrated over all space), one gets a function of the continuous variable $E$.

 The resolvent can be written as a sum over the exact energy eigenvalues $E_n$:

\be
\langle \psi | G(E) | \psi \rangle = \sum_n \frac{|\langle \psi | \psi_n \rangle|^2}{E - E_n}
\ee
As a function of a real number $E$, $G(E)$ goes to infinity (has poles) exactly when $E$ equals an energy eigenvalue $E_n$.

 To turn these infinite poles into finite maxima, one introduces a small damping factor $\eta$  into the imaginary part  of the Green's function (representing a tiny bit of lifetime broadening ):

\be
D(E) = -\frac{1}{\pi} \text{Im} \left[ \text{Tr}\left(\frac{1}{E - \hat{H} + i\eta}\right) \right] = \sum_n \frac{\eta}{\pi((E - E_n)^2 + \eta^2)}
\ee
This function $D(E)$ is the Density of States (DOS). It consists of a series of sharp Lorentzian peaks (maxima) whose centers are located precisely at the energy eigenvalues $E_n$.

In figure (\ref{DE50}) the density of states that corresponds to  a GOE random matrix of  size $N=50$ is plotted in blue  as a function of the energy, It  is a continuous function of energy $E$. It features multiple local maxima , and each maximum sits precisely on top of one of the GOE energy eigenvalues.
 At the bottom axis, red points mark the exact eigenvalue locations before broadening. 

\begin{figure}[h!]
  \centering
  \includegraphics[width=0.48\linewidth]{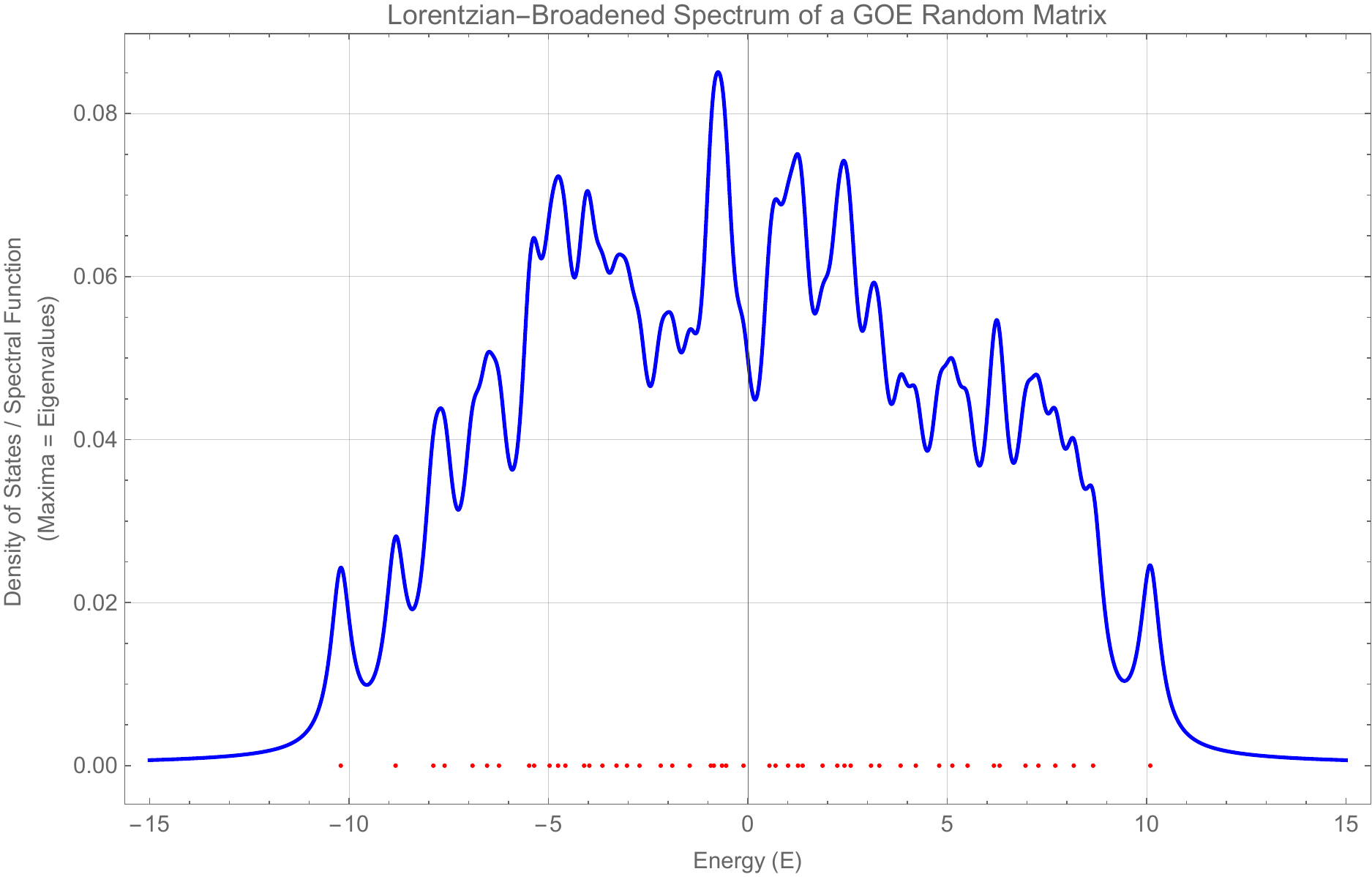}
  \includegraphics[width=0.48\linewidth]{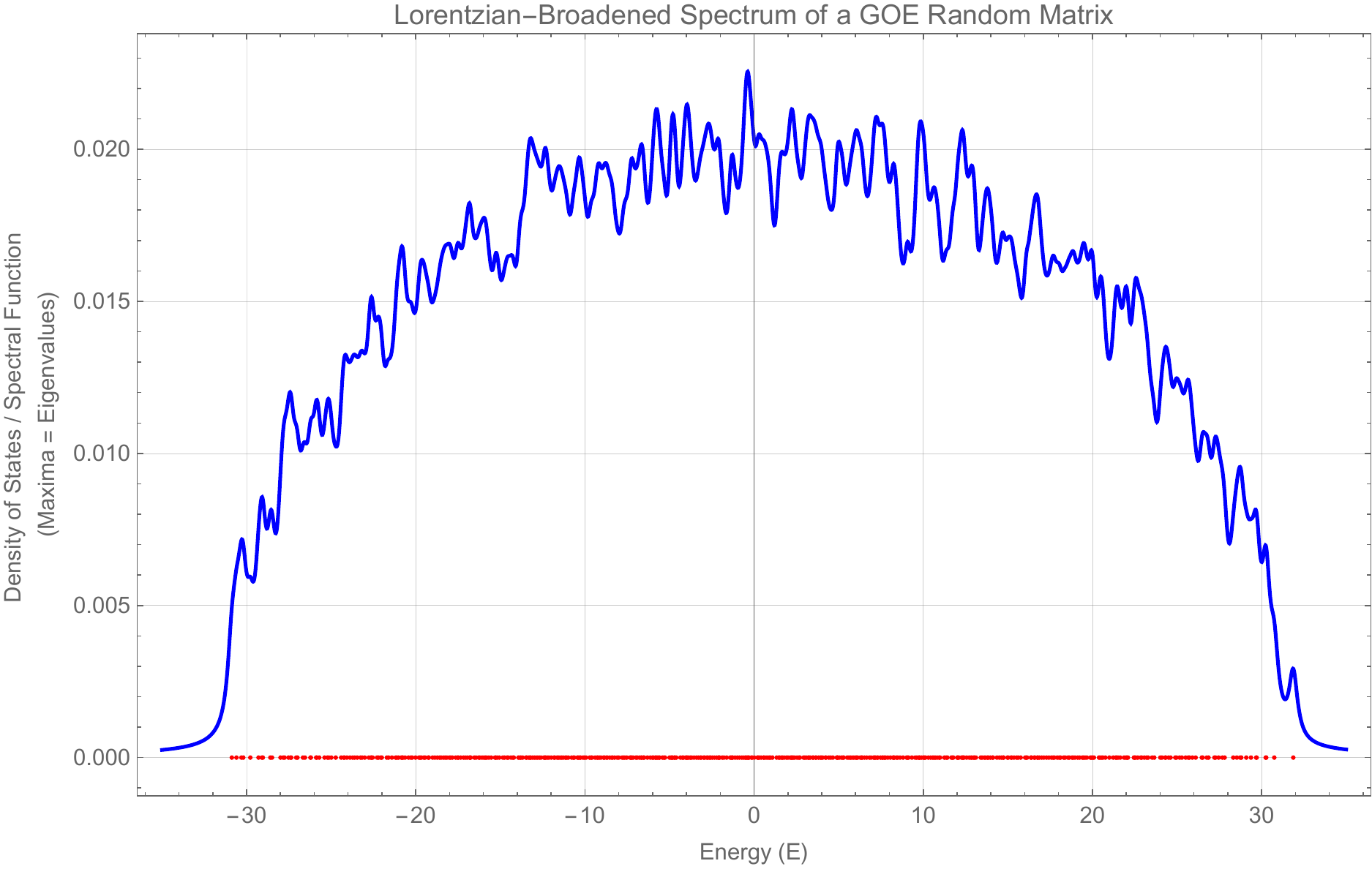}
  \caption{$D(E)$ as a function of $E$ and (blue) and the red dots are the exact locations of the GOE eigenvalues.The left figure $N=50$ and right $N=500$}\label{DE50}
\end{figure} 

 It is apparent that upon increasing  $N$ If  the individual peaks merge to form the characteristic smooth Wigner Semicircle envelope, while still retaining the local fluctuations/maxima dictated by GOE level-repulsion.

\item  2. Scattering Phase Shifts (The Breit-Wigner Formula)-

In scattering theory, one sends a particle toward a potential and measures how it scatters as a function of energy $E$.
 The scattering cross-section $\sigma(E)$ is a function of energy.
 For systems with bound states or resonances (quasibound states), the cross-section exhibits sharp peaks (resonances).

 The energies at which these maxima occur correspond directly to the energy eigenvalues of the system (or the real parts of the energies for unstable/decaying states). The cross-sections  as a function of the energy are  drawn in figure (\ref{cross}) for eigenvalues of GOE with $N=30$ and $N=100$. 
\begin{figure}[h!]
  \centering
  \includegraphics[width=0.48\linewidth]{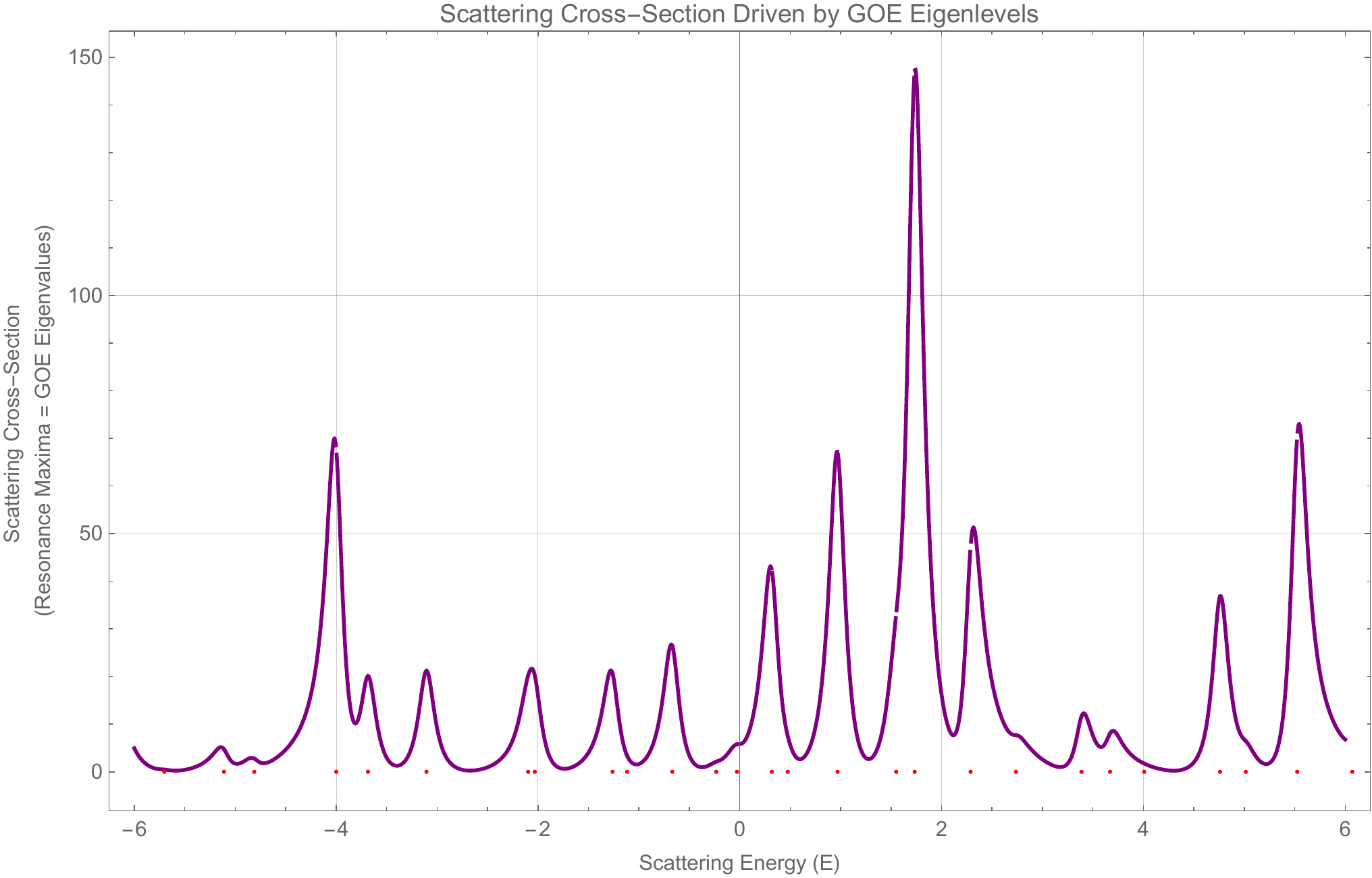}
  \includegraphics[width=0.48\linewidth]{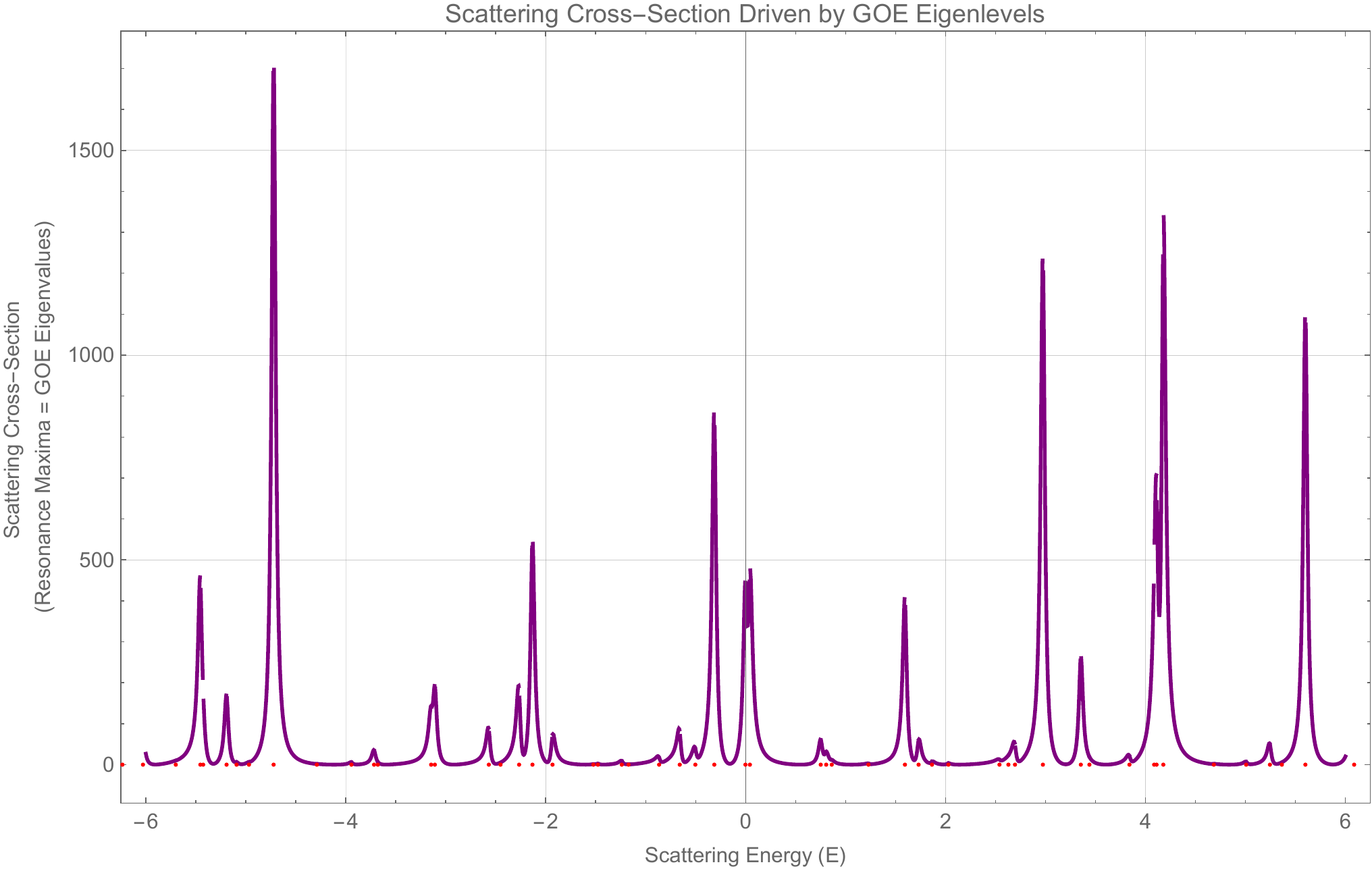}
  \caption{The cross section  as a function of $E$ and (purple) and the red dots are the exact locations of the GOE eigenvalues. The left figure $N=30$, GammaWidth = 0.2 and right $N=100$ and GammaWidth = 0.05 The locations of the eigenvalues are marked as red points on the energy axis}\label{cross}
\end{figure} 

\subsection{ The phase shifts of the leaky torus}

An extremely interesting model that displays chaotic behaviour is scattering on the leaky torus. Originally proposed by Gutzwiller \cite{Gutzwiller:1983}, the leaky torus geometry is constructed by taking the two dimensional hyperbolic space with the metric 
\begin{equation} ds^2 = \frac{dx^2+dy^2}{y^2} \label{eq:leaky_metric} \end{equation}
where we set radius to unity. One looks at the region, in the upper half plane \(y>0\), between the geodesics (i) \(x=-1\), (ii) \(x = 1\), (iii) \((x-\frac12)^2+y^2 = (\frac12)^2\), and (iv) \((x+\frac12)^2+y^2 = (\frac12)^2\).
Then, identifying boundary (i) with (iii) and (ii) with (iv), the result is a torus with a cusp point at infinity which is drawn in figure (\ref{fig:leaky_geometry}).

\begin{figure}[ht!]
\centering
\includegraphics[width=\linewidth]{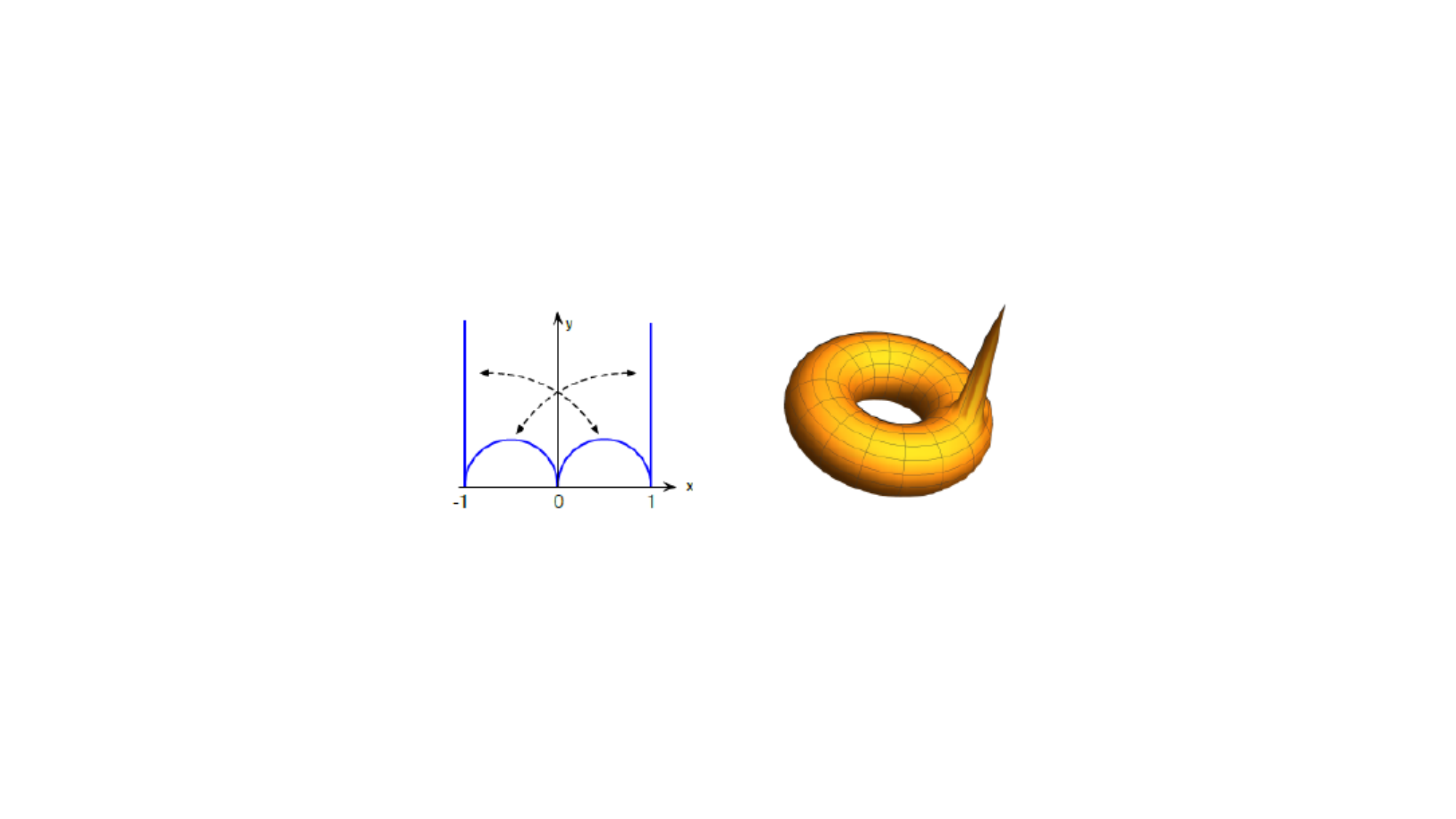} 
\caption{The leaky torus is defined by the identifications (right). A wave is sent on  a torus 
 from  the  cusp at infinity (left). 
 .}
\label{fig:leaky_geometry}
\end{figure}

The scattering in this setting involves sending an incoming free wave from \(y=\infty\) and measuring the phase shift of the outgoing wave at some finite \(y = y_0 > 0\). The phase shift is found to be exactly
\be S(k) \equiv e^{i\delta(k)} = \frac{\pi^{-ik}\Gamma(\frac12+ik)\zeta(1+2ik)}{\pi^{+ik}\Gamma(\frac12-ik)\zeta(1-2ik)} \ee
where $k = \sqrt{2E}$ is the momentum of the incoming wave.\footnote{Here we follow the analysis of \cite{Bianchi:2024fsi}}

\begin{figure}[ht!]
\centering
\includegraphics[width=0.60\textwidth]{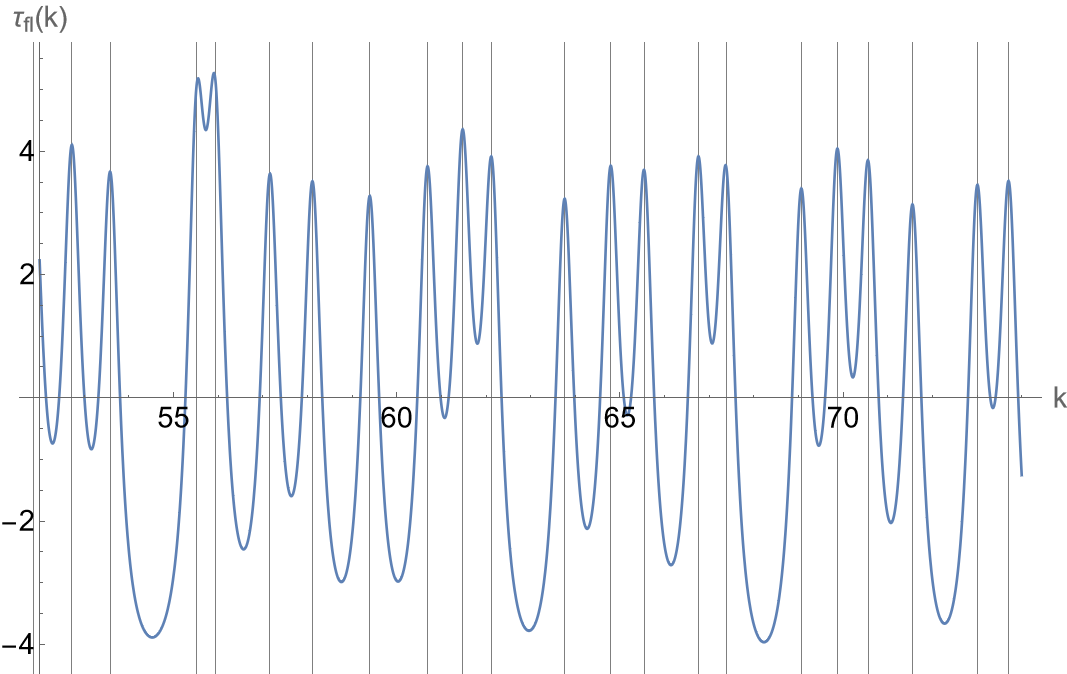} 
\caption{The function $\tau_{\mathrm{fl}}(k)$, displaying erratic behavior as a succession of resonances at zeta function zeros. One can plot any range of $k$ to see similar fluctuations, yet without ever repeating exactly. Vertical lines are located at the zeros of the Riemann zeta, $z_n = \frac12+2ik_n$.}
\label{fig:leaky_function}
\end{figure}

Looking at the fluctuating part of the function,
\be e^{i\delta_{\mathrm{fl}}(k)} \equiv \frac{\zeta(1+2ik)}{\zeta(1-2ik)}\,, \qquad  \tau_{\mathrm{fl}}(k) = \frac{d\delta_{\mathrm{fl}}(k)}{dk} \ee
one can see that it has peaks almost exactly at the zeta function zeros (see figure \ref{fig:leaky_function}), with some small deviations when two neighboring resonances overlap. 

As a result, the spacings between successive peaks in $\tau(k)$ will follow closely the Wigner-Dyson PDF of the GUE.

Let us then revisit the distribution of non-trivial zeros of the zeta function, being the solutions of
\be \zeta(\frac12+i y_n) = 0 \ee
To see the distribution of spacings, first we use the average density function, which is known to be
\be \rho(y) = \frac{1}{2\pi} \log\frac{y}{2\pi} \ee
and we define the unfolded spectrum using the cumulative density
\be \lambda_n \equiv I(y_n) = \frac{y_n}{2\pi}\left(\log\frac{y_n}{2 \pi}-1\right) \ee
It is well known that the spacings $\delta_n \equiv \lambda_{n+1}-\lambda_n$ follow the GUE distribution, and as such the zeta function was an object of interest from the early days of quantum chaos \cite{Berry:1986}.

\subsection{The  Electric potential of randomly placed charges}\label{Eporpc}
In \cite{Bianchi:2025kna} a toy model for chaotic behavior was proposed in the form of the electric potential of randomly placed charges on a plane. 

The electric potential in three-dimensional space, generated by $N$ identical point charges located on the 2D plane $z=0$, as measured at $z=z_0$. is given by 
$ V(x,y,z=z_0)= {\cal F}(x,y)  $

where 

\begin{equation}
 {\cal F}(x,y) = \sum_{n=1}^N \frac{1}{\sqrt{(x-\lambda^{(1)}_n)^2+(y-\lambda^{(2)}_n)^2+z_0^2}} \label{eq:f2d}   
\end{equation}
where $\lambda^{(1)}$ and $\lambda^{(2)}$ are the eigenvalues of two independent random matrices, $M^{(1)}$ and $M^{(2)}$, drawn from some ensemble and $z_0$ is a non-zero real constant.

The charges are assumed to be randomly distributed, and are located at the points $(x_n=\lambda^{(1)}_n,y_n=\lambda^{(2)}_n)$ with $0\le \lambda^{(1)}_n,\lambda^{(2)}_n\le N$.

This toy model \cite{Bianchi:2025kna} was introduced as a generalization of the  the phase shift of the leaky torus. Note that whereas the latter  is a quantum property, the electric potential just mentioned is classical. It is thus an example of an erratic function that has both classical and quantum interpretations.

\begin{figure}[ht!]
    \centering
    \includegraphics[width=0.33\textwidth]{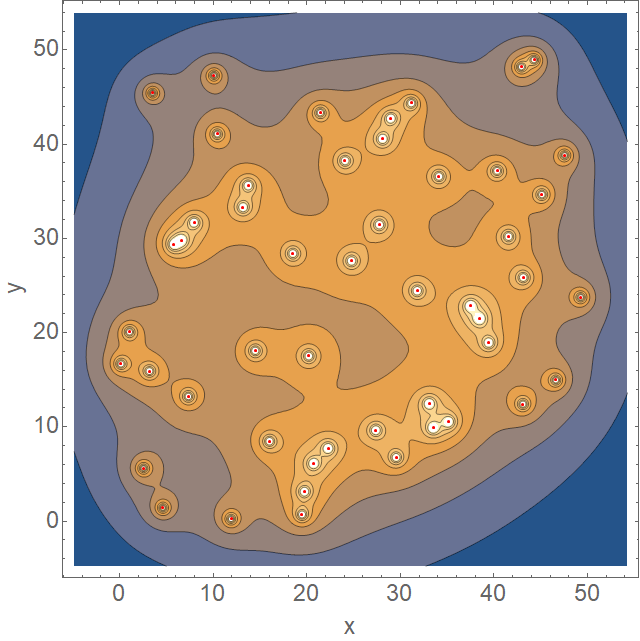}
    \includegraphics[width=0.40\textwidth]{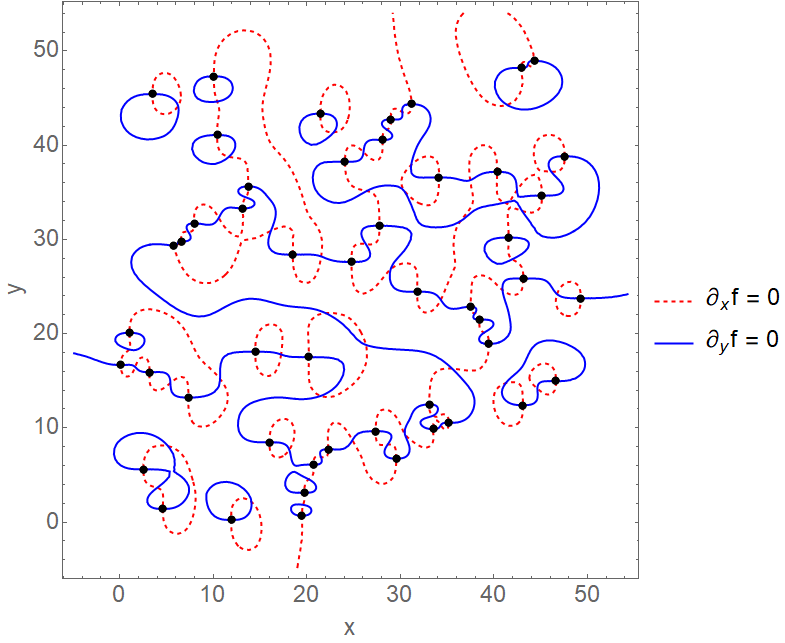}
    \caption{An example of the function \eqref{eq:f2d} with eigenvalues $\{\lambda_n^{(1)}\}$ and $\{\lambda_n^{(2)}\}$ corresponding to unfolded GOE spectra. Here $N=50$ and $z_0 = \frac14$. Left: Contour plot of the function. Lighter areas represent higher values, and the positions of peaks are marked with red points. Right: Plot of lines where the partial derivatives of $f$ are zero, with peaks marked in black. }
    \label{fig:2dpeaks}
\end{figure}

This function has randomly located peaks on the plane. If we choose $z_0$ such that it is smaller than the typical spacing between peaks, they will typically not overlap and we can see them all distinctly. For illustration we plot in figure \ref{fig:2dpeaks} an instance of this function with $\{\lambda_n^{(1)}\}$ and $\{\lambda_n^{(2)}\}$ drawn independently from the GOE. We also plot the lines along which the partial derivatives ${\partial}{\cal F}/{\partial x} $ and ${\partial}{\cal F}/{\partial y}$ are zero. The peaks are located at the intersections of these lines.

\begin{figure}[ht!]
    \centering
    \includegraphics[width=0.35\textwidth]{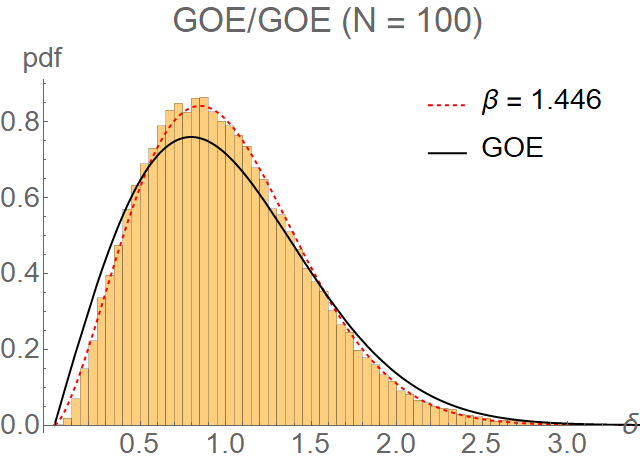}\hspace{1cm}
    \includegraphics[width=0.35\textwidth]{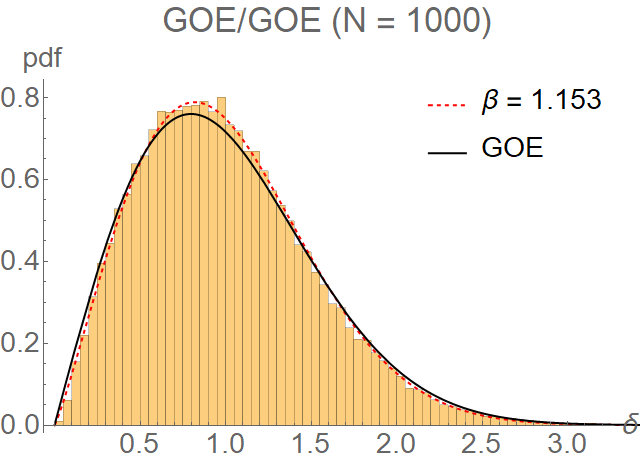} \\
    \includegraphics[width=0.35\textwidth]{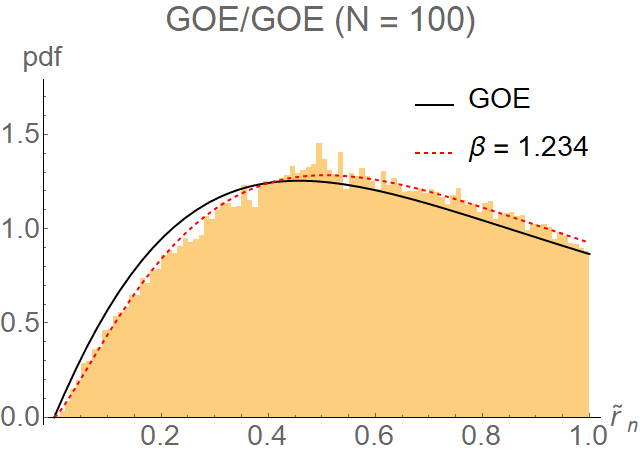}\hspace{1cm}
    \includegraphics[width=0.35\textwidth]{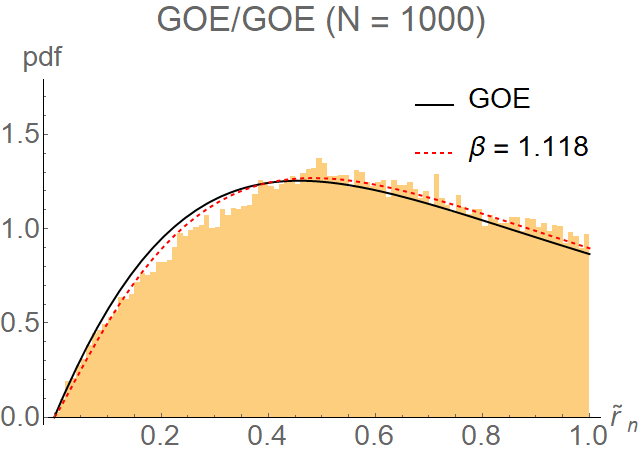}
    \caption{The distributions of $\delta_n$ (top) and $r_n$ (bottom) for the GOE-GOE case, for $N = 100$ (left) and $1000$ (right).}
    \label{fig:toymodelspacings}
\end{figure}
\newpage
\subsection{Quenched  QFT in 1+1 dimensions}
As an example of an analysis of extrema points in the context of QFT, we bring the case of 
correlation functions produced by local operator quenches in finite-volume free massive scalar field theory in 1+1  dimensions.  The basic observable is the subtracted equal-time two-point function in the locally excited state and its spatiotemporal patterns of extrema. This is shown in figure (\ref{fig:figG}). The analysis of these correlators  as one dimensional functions was done in \cite{Ageev:2025yiq} and as multi-dimensional ones in \cite{Ageev:2026jvx}.
These correlators were analyzed  these extrema by the multidimensional diagnostics recently using the methods discussed in the previous section: all-pair distance distributions, nearest-neighbor spacings, greedy-path spacing ratios, and the extrema form factor.
The results for the cloud of minima points is presented in figure (\ref{fig:minima_axis}).
 
\begin{figure}[ht!]
    \centering
    \includegraphics[width=\textwidth,height=0.21\linewidth]{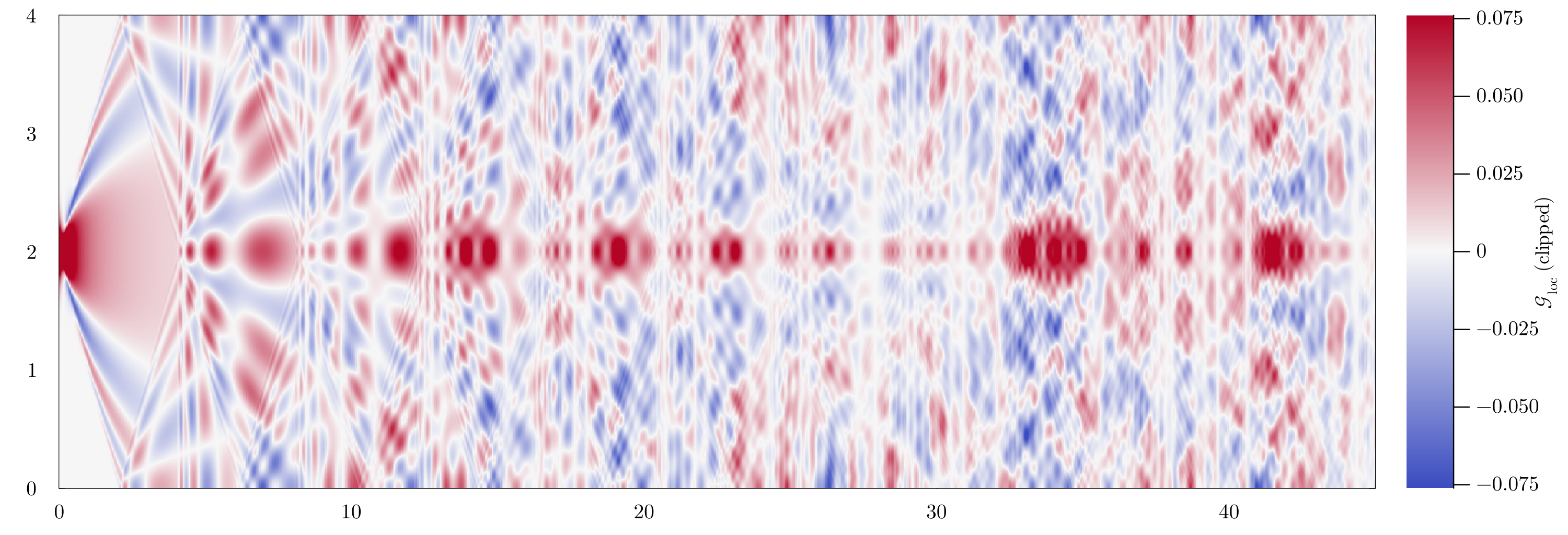}
    \caption{The two-point correlation function  for the strip size $L=4$, mass $m=10$.}
    \label{fig:figG}
\end{figure}

\begin{figure}[ht!]
\centering
 \includegraphics[width=\textwidth]{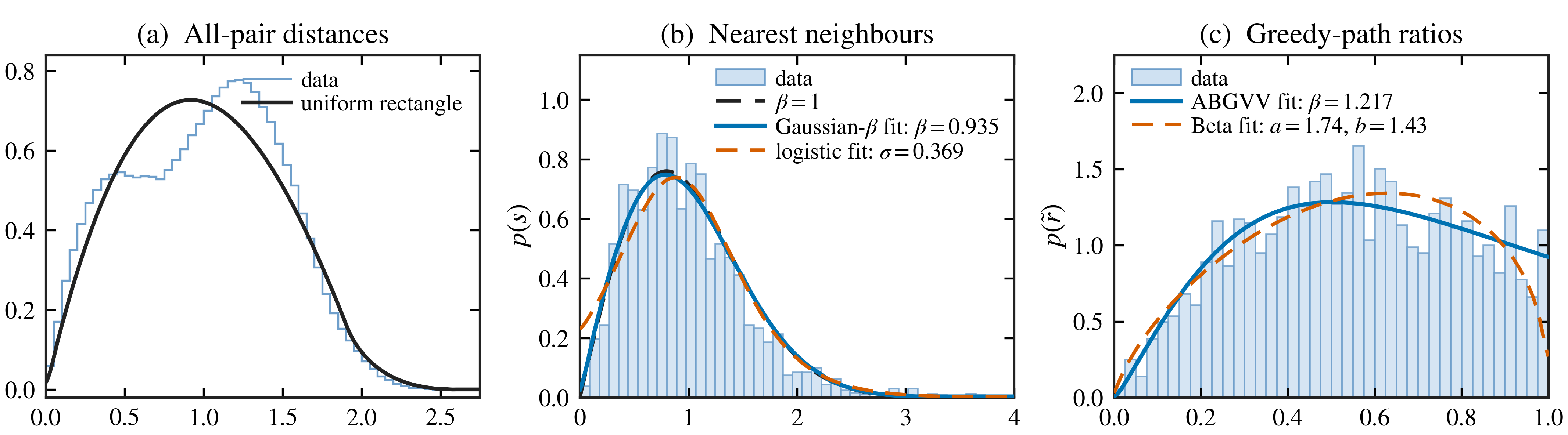}
\caption{The  analysis of the refined minima cloud. The left panel shows the histogram of all pair distances together with the analytic uniform-rectangle reference curve. The middle panel shows the nearest-neighbor spacing distribution and its Gaussian-$\beta$ and logistic fits. The right panel shows the greedy-path ratio distribution together with the ABGR-type $\beta$ fit and the ordinary Beta fit.}
\label{fig:minima_axis}
\end{figure}
\newpage
\subsection{ Decay  of $HES\rightarrow T+T$ high excited string states}\label{Dohess}
A landmark example of an erratic  behavior in string theory is 
the  decay process   of a DDF  high excited state of the  bosonic string theory into two tachyonic states drawn in figure (\ref{HEStTT}) was discovered in \cite{Gross:2021gsj}. The erratic behavior is manifested in the  decay amplitude as a function of $\alpha$  an angle related to the polarization of the DDF state. 

As discussed in (\ref{locex}) we  analyze the  scattering amplitude ${\cal A}(\alpha)$ by locating its extrema points that correspond to the zeros of the    logarithmic derivative
\be {\cal F}(\alpha) \equiv \frac{d}{d\alpha}\log {\cal A} \qquad {\cal F}(x_n) = 0.\ee
 Then we define  the spacings \(\delta_n\) and the ratios of consecutive spacings, \(r_n\) and $\tilde r_n$:
\be \delta_n =  x_{n+1}-x_n \ee
and
\be r_n \equiv \frac{x_{n+2}-x_{n+1}}{x_{n+1}-x_{n}} = \frac{\delta_{n+1}}{\delta_n}\,,\qquad \tilde r_n =min\{r_n,\frac{1}{r_n}\} \ee

In figure (\ref{HESTT}) the logarithmic derivative of the amplitude  of a state with $N=100$ made out of the  partitions $g_n(18,10,8,7,7,7,7,5,4,4,4,4,4,3,2,1,1,1,1)$  such that $N=\sum n g_n=100$ and $J=\sum n_g=19$ is drawn. 
The erratic nature of the amplitude is manifest.  The analysis of the amplitude using the ``locations"of its maxima, the spacings between adjacent peaks and their ratios was performed in \cite{Bianchi:2022mhs} and \cite{Bianchi:2023uby}.

 \begin{figure}[ht!]
  \centering
  \includegraphics[width=0.9\linewidth]{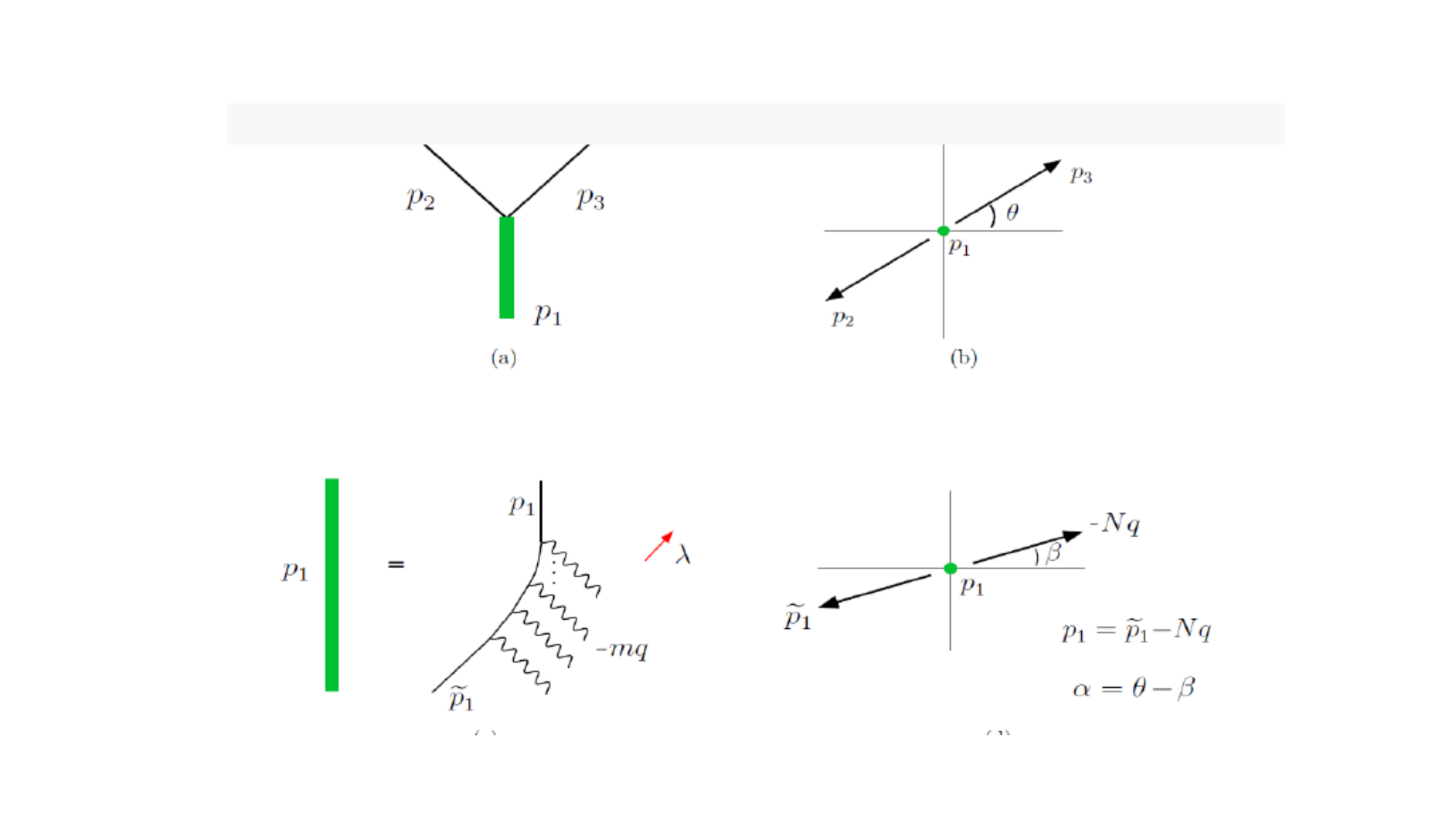}
  \caption{The logarithmic derivative of the amplitude of a decay of $HES\rightarrow T+T$ .}\label{HEStTT}
\end{figure}

   \begin{figure}[h!]
  \centering
  \includegraphics[width=0.9\linewidth]{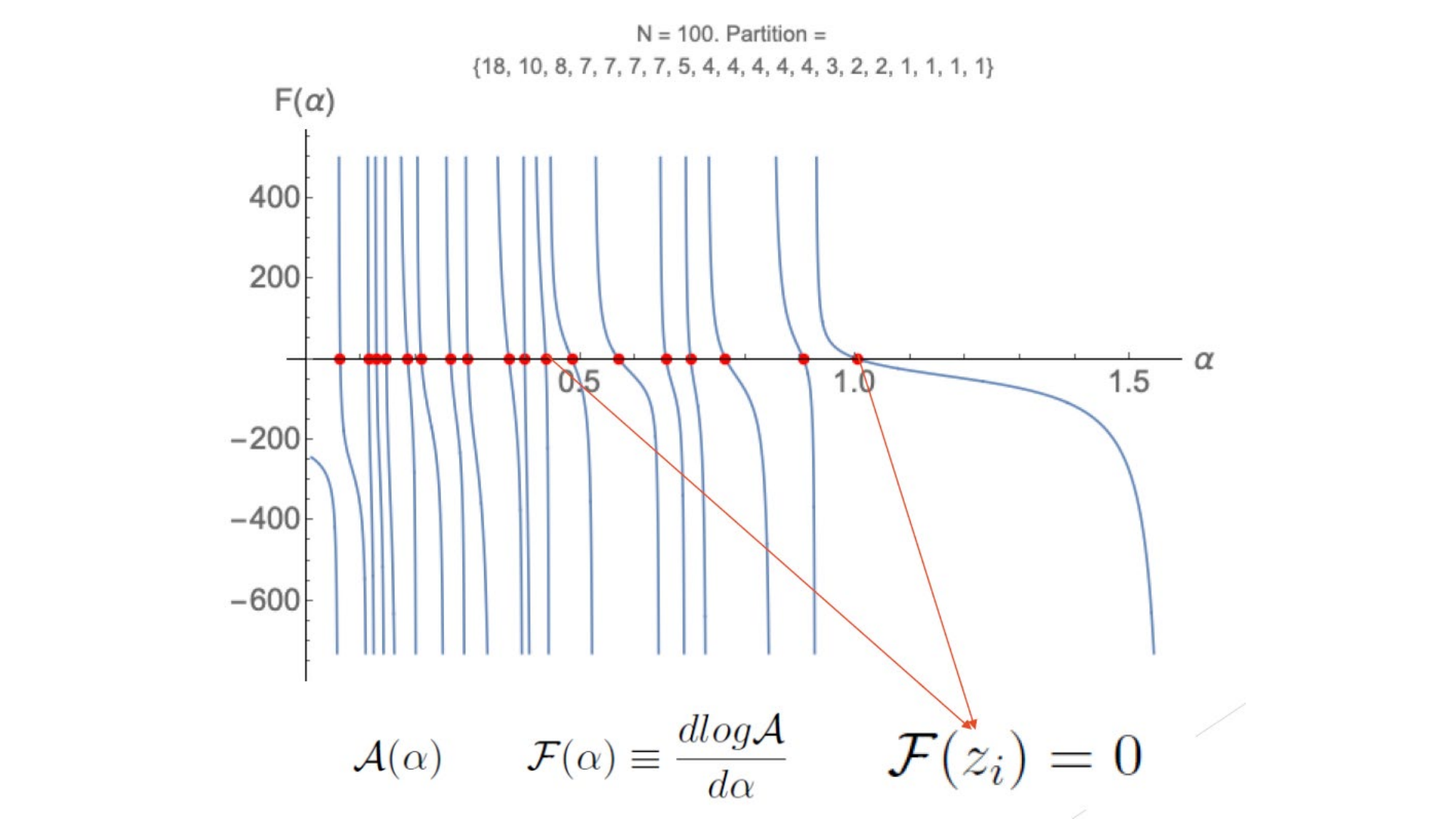}
  \caption{The logarithmic derivative of the amplitude of a decay of $HES\rightarrow T+T$ .}\label{HESTT}
\end{figure} 

The PDF associated with the ratios  $r_n$ is drawn in (\ref{HESTTr}).
\begin{figure}[h!]
  \centering
  \includegraphics[width=0.6\linewidth]{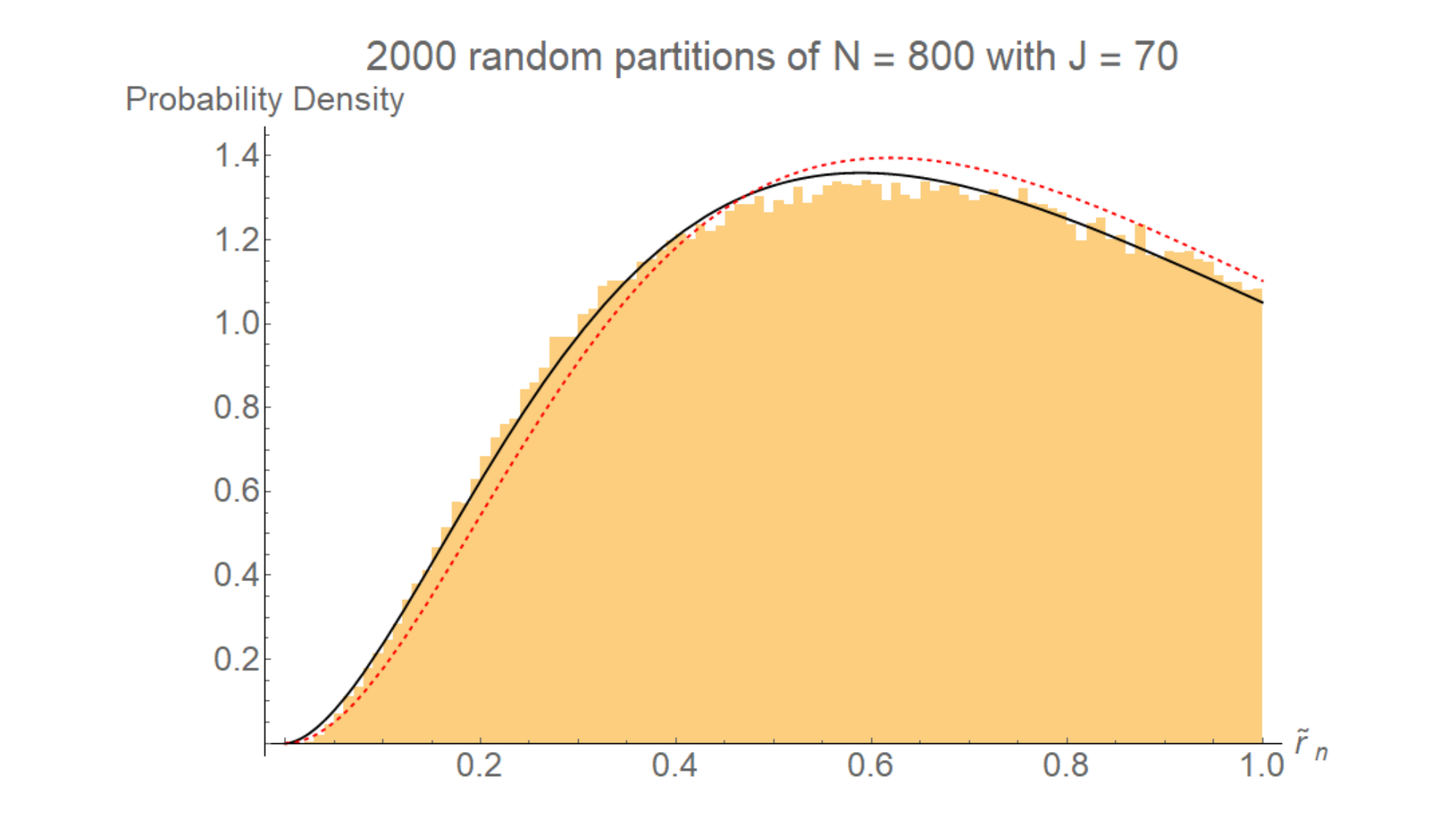}
  \caption{The measured distributions of ~r across samples of 2000 states chosen at random
with fixed N and J, fitted to a $\beta=1.72$ ensemble distribution (black line). The distribution is also
compared to the GUE distribution (red, dashed)..}\label{HESTTr}
\end{figure} 

Once the histogram of the the spacings $\delta_n$
 and of the ratios $r_n$ is calculated numerically  the questions are what PDFs fit these distributions best. 
 
  From the analogy between the spacings of the extrema points and those associated with the quantum energy eigenvlues, we are guided to check  the fitness with the results of the RMT.   The distributions for the difference of eigenvalues $\delta = \lambda_2-\lambda_1$ of a $2\times 2$ matrix of the ensemble, properly normalized such that $\avg{\delta} = 1$, are given by the well known Wigner surmises\cite{Mehta:book,Akemann:book}:
\be p_{\beta}(\delta) = {\cal N}_{\beta} \delta^\beta \exp(-c_\beta \delta^2) \label{eq:beta_delta} \ee
with
\be {\cal N}_\beta = 2\frac{[\Gamma(\frac{\beta+2}2)]^{\beta+1}}{[\Gamma(\frac{\beta+1}2)]^{\beta+2}}\,,\qquad c_\beta = \left(\frac{\Gamma(\frac{\beta+2}2)}{\Gamma(\frac{\beta+1}2)}\right)^2\ee
It was found out that this PDF is an excellent approximation of the PDF of spacings for large matrices as well.
The three classical Gaussian ensembles, orthogonal (GOE), unitary (GUE) and symplectic (GSE), have  the values $\beta$= 1, 2, and 4, respectively. ( In fact there are 10 classical ensembles)

We suspect that the case of the HES decay
follows the   the special case of $\beta=2$ GUE distribution whose PDF is 
\be
p_{\text{GUE}}(\bar \delta) = \frac{32}{\pi^2} \bar\delta^2 \exp\left(-\frac{4\bar \delta^2}{\pi}\right) \label{eq:pdfGUE}
\ee

Rather than the spacings themselves, we will focus on the ratios of consecutive spacings. In the minimal case of a $3\times3$ matrix with eigenvalues $\lambda_1 < \lambda_2 < \lambda_3$, the PDF of the ratio $r = (\lambda_3-\lambda_2)/(\lambda_2-\lambda_1)$ in the $\beta$-ensemble is \cite{Atas:2013dis}
\be f_\beta(r) = \frac{3^{\frac{3+3\beta}2}\Gamma(1+\frac\beta2)^2}{2\pi \Gamma(1+\beta)} \frac{(r+r^2)^\beta}{(1+r+r^2)^{1+\frac32\beta}} \label{eq:beta_r}\ee

Also for this quantity this PDF seems to be an excellent approximation also for larger matrices. 

With the assumption that the spacings $\delta_n$ are independent random variables drawn from (\ref{eq:pdfGUE}), then the PDF of
their ratios $r_n$ is predicted to be
\begin{equation} f_{\text{GUE}}(r) = \frac{16}{\pi} \frac{r^2}{(1+r^2)^3} \label{eq:pdfGUE_r}\end{equation}
This function is peaked at \(r = 1/\sqrt2\), with the expectation value of \(r\) being \(\avg{r} = 4/\pi \approx 1.273\). This distribution turns out to be very similar to the log-normal distribution with \(\mu = 0\), and \(\sigma = \sqrt{(\pi/2)^2-2)}\) 

The best fit in figure (\ref{HESTTr}) is with $\beta =1.72$. It seems that by considering states with higher $N$  and obviously complicated partitioning the result will approach $\beta=2$
\subsection{ String scattering  $HES+T\rightarrow T+T$}
A higher-dimensional generalization of the decay discussed in the previous section can take the form of a 
scattering of a generic,  DDF highly-excited open bosonic string state with a massive scalar (tachyon), yielding a two-scalar final state, as represented in figure \ref{fig:stringConfScattkine}. The scattering amplitude is a function of two angles $\phi$ related to the polarization vector of the DDF state and the scattering angle $\theta$ both drawn in \ref{fig:stringConfScattkine}. We denote by $x=\cos \theta$ and $y=\cos\phi$
The amplitude was computed in \cite{Bianchi:2026jui}
 following \cite{Firrotta:2024qel}.

The amplitude is drawn in 
 figure \ref{fig:areas},   for the a randomly selected HES with $N=50$. 
 The main observation is that the amplitude admit curve structure. The colored areas between the curves in the figure correspond to the ``level spacings'' $\delta_n=A_{n+1}-A_n$ (See \ref{areaspac}).   One can observe immediately that the curves are non-intersecting and  view this phenomenon as curve repulsion. 
As we increase $N$, the same overall structure remains, with a higher number of curves that grows linearly with $N$. Crucially, for generic partitions of large $N$, the spacings between curves become erratic.

\begin{figure}[h!]
\centering
\includegraphics[width=0.48\textwidth]{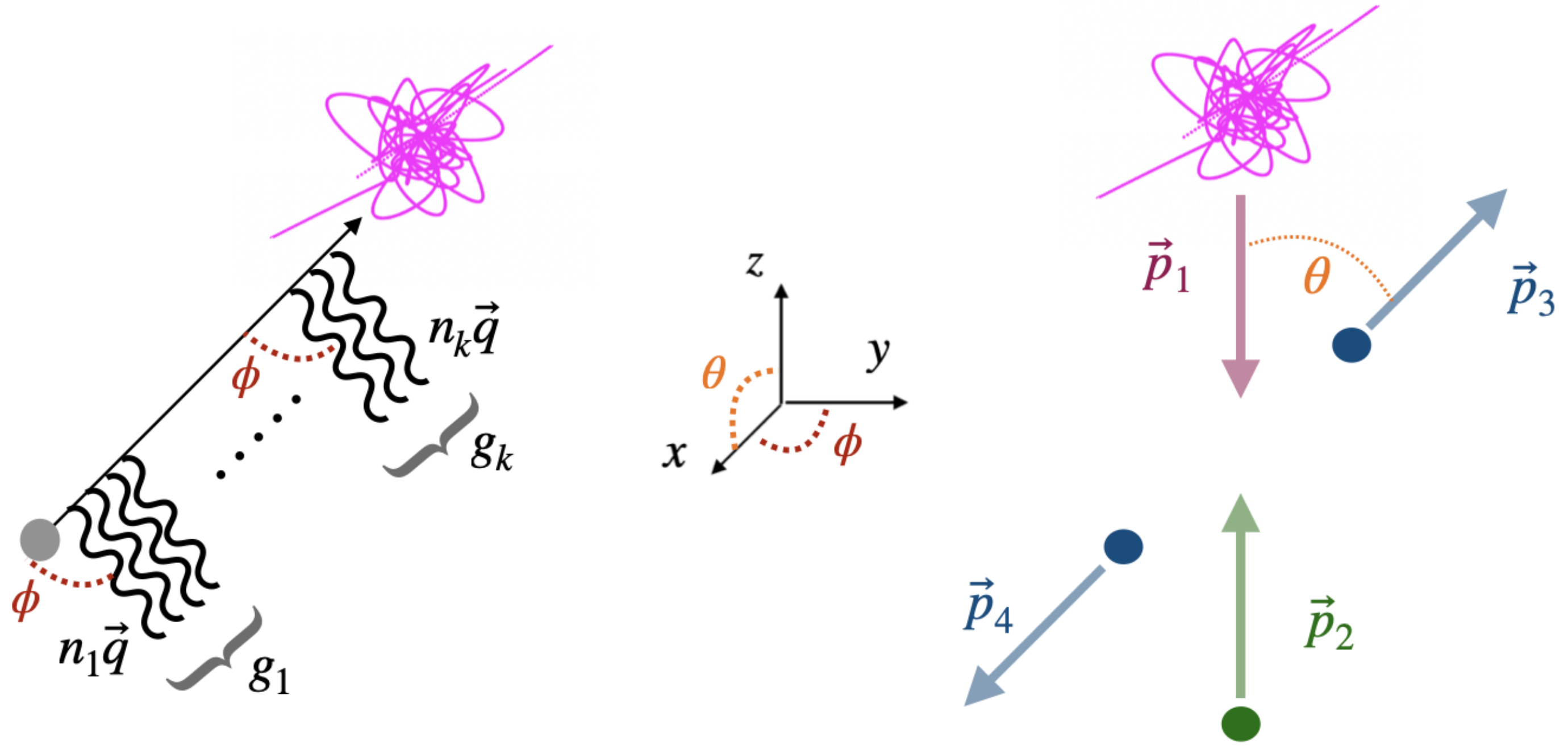}
\caption{Kinematical configuration of the formation of the DDF HES (left) and the process $HES+T\rightarrow T+T$ in the center of mass frame (right).}
\label{fig:stringConfScattkine}
\end{figure}
\begin{figure}[t!]
    \centering
    \includegraphics[width=0.48\textwidth]{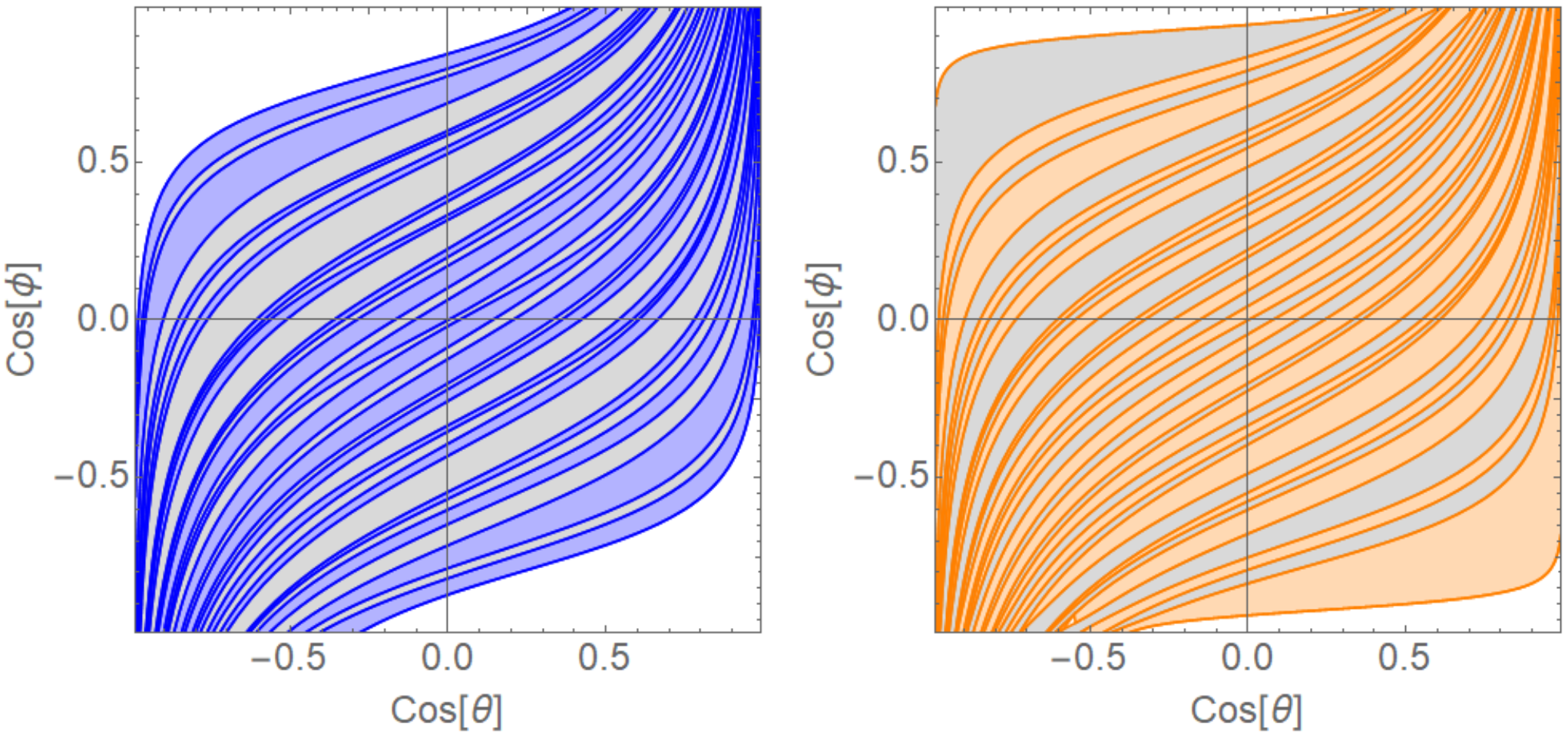}

    \caption{Curves of zeros of the $x$-derivative (left), and $y$-derivative (right) of $\log{\cal D}(x,y)$, for the state with $N = 50 = 13+10+9+5+5+3+2+1+1+1$. The areas colored in alternating colors are the spacings $\delta_n$.
    }
    \label{fig:areas}
\end{figure}
\begin{figure}[h!]
    \centering

    \includegraphics[width=0.23\textwidth]{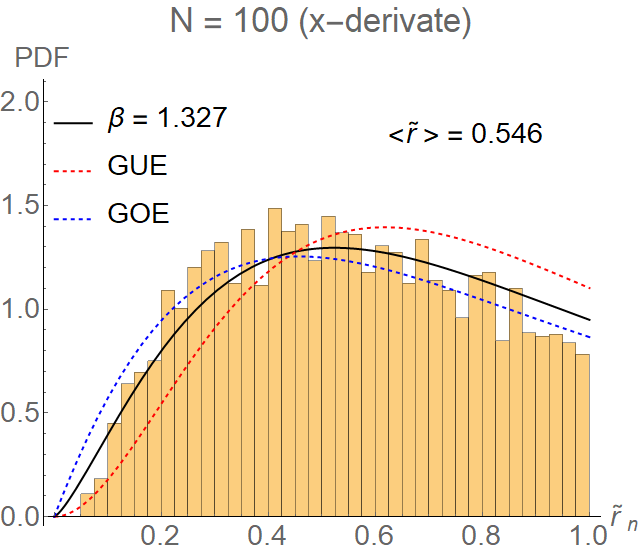}
    \includegraphics[width=0.23\textwidth]{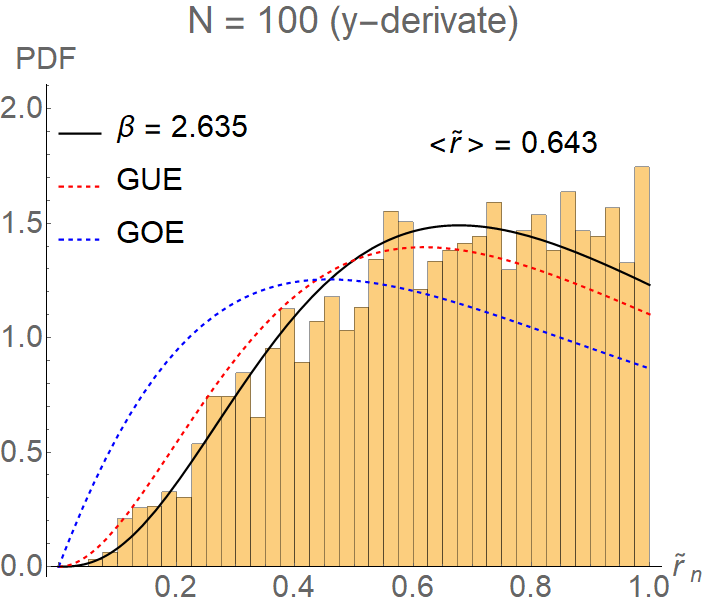}
    \caption{Distributions of ratios of adjacent areas compared to RMT (eq. \eqref{eq:beta_r}) for the GOE, GUE, and best-fitted $\beta$ values. The distribution is averaged over 100 states of $N = 100$. The left and right plots correspond to the areas associated with the $x$- and $y$-derivatives, respectively.
    }
    \label{fig:areas_spacings}
\end{figure}
To carry out a quantitative analysis, we determine the two types of curves over samples of 100 states at each level $N$, chosen by taking random integer partitions of $N$. Then, we take all the resulting spacing ratios for the areas, $\{r_{(x),n}\}$ and $\{r_{(y),n}\}$, and fit their distributions to the $\beta$-ensemble formula \eqref{eq:beta_r}. Independently of the fitting, we  compute the average value of the normalized spacing ratios $\langle \tilde r_n\rangle$, and compare it to the expected values for the non-chaotic Poisson distribution (0.386) and the expected RMT values of 0.536 for GOE and 0.603 for GUE. 

     
     

The distributions obtained are plotted in figure \ref{fig:areas_spacings} for the case of $N = 100$.  It should be noted that at larger $N$, the distributions agree better with the RMT formula. For $N = 100$, the best fit for the distribution computed from the curves where the $x$-derivative vanishes has $\beta=1.33$, while for the $y$-derivatives we find $\beta=2.63$.

This is an intriguing result. The angle $\phi$ (corresponding to $y$) is related to the polarization vector. This is similar to the angle that appeared in the decay associated with a three-point function \cite{Bianchi:2022mhs}. In that case the results admitted an extrapolation to $\beta\approx2$, as in the GUE. On the other hand $\theta$ (corresponding to $x$), which is a genuine scattering angle, seems to be invariant under a $Z_2$ symmetry and hence, as in quantum chaotic systems with time-reversal invariance, is related to the GOE distribution with $\beta=1$. We can also observe that whereas  $\beta_{(y)}$ and $\langle\tilde r_{(y)}\rangle$ are monotonically decreasing their analogs in $x$ oscillate. It is not clear why that is the case.

\subsection{Forming families of  erratic functions }\label{Ffoef}
The erratic functions and their analysis in terms of the extrema points are not related directly  to dynamical processes. It will be shown in (\ref{FPtTg}) that an attempt to map them into a thermodynamics  description  requires embedding them in families of erratic functions. 

In fact generically  this can be done. One can identify a parameter or several of them that  define a family of erratic functions such that a given one of them has a particular value of the parameter. In certain cases changing the parameters that define a family changes smoothly the erratic functions. By this type of changes one may move from erratic functions associated with chaotic behavior to non-erratic functions which obviously to not follow from chaotic processes. 

There are interesting cases where  small changes induce a major change in the erratic function. This may be viewed as  a butterfly-like event.

Let's demonstrate these ideas  in the examples given in (\ref{EfatcP}):
\item The quantum pinball scattering-

In \cite{Bianchi:2025kna} we have determined the scattering amplitudes and the corresponding PDFs  of the extrema points for a large family of the basic setup of the pinball system. The family was characterized by asymmetric parameters. For instance one member of the family has radii $(R_1=1,R_2= 1+\epsilon,R=1-\epsilon)$. The  characteristic parameter $\epsilon$, or better $k\epsilon$ where $k$ is the wave number, is assign to each scattering amplitude. In a similar way one can introduce an asymmetry parameter in the distances between the disks. Using these parameters one can interpolate between the chaotic three disks scattering and the one with two disks which in non-chaotic. \cite{Bianchi:2025kna}

\item The electric potential of randomly placed charges(\ref{Eporpc})-

The $x$ and $y$ coordinates  of the electric charges are taken to be $\lambda^{(1)},\lambda^{(2)}$  which are  eigenvalues of two independent matrices. The latter  can be associated with $\beta_1$ and $\beta_2$ of $G\beta E$. This  means that the erratic electric potential is a member of a family of potential characterized by $\beta_1$ and $\beta_2$. In this case small changes of the latter imply small changes in the PDFs associated with the erratic function before and after the change. 

\item Decays of $HES \rightarrow T+T$-

A given erratic decay amplitude(\ref{Dohess}  can be embedded in a family of them by alternating the partitioning of a given level, namely changing the set $n_g$ where $N=\sum_n ng_n$. It has been observed already in the original paper \cite{Gross:2021gsj} that a small change of the partitioning creates a large change in the amplitude. It is easy to check that for $N=n,n_g=1$ and also $N=n_1=J$ there is no erratic behavior and only for more complicated partitioning it occurs. Thus, by starting from non-erratic partitioning and changing it gradually, a transition to erratic behavior will take place.  

\newpage
 \section{From PDFs to Thermodynamics-generalities}\label{FPtTg}
Next we would like to relate the probability distribution functions    associated with  a given erratic function, which results from a chaotic process to thermodynamics.
namely, we would like to assign to them entropy, energy function, internal energy, partition function  and  temperature. 
In addition we would like to examine the  ``first law of thermodynamics" that corresponds to the erratic functions. 

This assignment of thermodynamics can be achieved in steps:
(I) The entropy is identified with Shannon entropy for discrete PDFs and differential entropy for the continuous cases. 

(II) We equate the PDF with the Gibbs thermodynamic probability distribution  which enables us to determine the energy function and the internal energy. In fact as will be seen shortly the correspondence with the Gibbs PDF enables us to determine the product $\beta E= \frac{E}{T}$

(III) To determine the temperature separately  we will need to use the assignment of a given PDF with its ``family" as was done in (\ref{Ffoef})

(iV) The last step is to write down the first law of thermodynamics.

Since we have different probability distributions associated with a given  erratic function,   correspondingly there are   various types of entropies, energies and temperatures. We will want to check if and how they are related.

Denoting the probability distribution of the locations of the extrema points, the spacings and the ratios as $p_e,p_s, p_r$
respectively we can define entropies $S_e,S_s,S_r$,
 energies $E_e,E_s,E_r$, and temperatures $T_e,T_s,T_r$

 A very interesting question is what is the added value that one can gain about the probability distributions and  the erratic functions related to them from the   the map to thermodynamics.

Thermodynamics formulations of dynamical systems were dicussed in \cite{Ruelle1978} and \cite{Bowen1975}
\subsection{From one dimensional  probability distribution to  Entropies }

Thus, the  first step  in assigning a thermodynamics is  to determine various  classical Shannon entropies associated with the probability distributions.  
    \item 

For a discrete set of variables
\be 
X :=\{x_i,p_i(x_i)\}\qquad i\in\{1,...N\} \qquad \sum_{i=1}^N p_i=1
\ee
with probability $p_i$ for the variable $x_i$,
{\bf Shannon's entropy}\cite{Shannon1948} and \cite{Shannon1948PartII} is defined as 
\be
S=-\sum_{i=1}^N p_iLog(p_i) 
\ee
\item
In a similar manner for a continuous  PDF  $p(x)$,
\be 
X :=\{x,p(x)\}\qquad x\in\{0, L\} \qquad \int_0^L p_(x) dx =1
\ee 
 the {\bf differential entropy} is defined by
\be
S(X) = -\int_{-\infty}^{\infty} p(x)\,\ln p(x)\,dx.
\ee
\item 
It is important to note that this definition is not invariant under a change of coordinates. 
For $y=ax$ the differential entropy is 
\be
S(Y)=S(X) + \ln|a|
\ee
Consequently, comparing differential entropies between PDFs is meaningful only after specifying the same normalization/unfolding convention.
\item Naturally one can generalize these computations also to the case of probability distribution of several variable.
For instance, suppose that there are two sets of random variables $X$ and $Y $ such that
\be 
(X,Y) :=\{(x_i,y_j),p_{ij}(x_i,y_j)\}\qquad i,j \in\{1,...N\} \qquad \sum_{i=1,j=1}^N p_{ij}=1
\ee
The corresponding 
Shanon's entropy is  
\be
S=-\sum_{i=1,j=1}^N p_{i,j}Log(p_{i,j}) 
\ee
\item
For a continuous variables $(x,y)$,
\be 
(X,Y) :=\{(x,y),p(x,y)\}\qquad x\in\{0, L_x\} \qquad 
y\in\{0, L_y\} \qquad\int_0^{L_x} dx \int_0^{L_y} dy p(x,y) =1
\ee
 with a probability distribution $p(x,y)$,  the differential entropy is defined by
\be
S(X) = -\int_0^{L_x} dx \int_0^{L_y} dy p(x,y)\,\ln p(x,y).
\ee

\subsubsection{The ``energy" function }

 Given the probability distribution $p(z)$ we define the  "Energy" ($E(z)$) function  by  equating it to  the 
Boltzmann-Gibbs distribution:
\be\label{Boltzmann} 
p(z) = \frac{1}{Z} e^{-E(z)/k_B T}=\frac{1}{Z} e^{-\beta E(z)}
\ee
where $\beta = \frac{1}{k_B T}$ is the inverse temperature and $Z$ the partition function is given by
\be
Z = \int e^{-\beta E(z)}dz
\ee
so that the dimensionless energy function $\beta E(z)$ (sometimes called the "surprisal" or self-information) is given by 
 \be\label{betaE}
\beta E(z) = - \ln p(z) -  \ln Z
\ee

 \subsubsection{The `` Internal energy" }

 Defining the  "Internal Energy" ($U$)
   the expectation value of the  energy function:
\be 
\beta U = \langle \beta E \rangle = \beta\int p(z) E(z) dz= -\int p(z) \ln p(z) dz-  ln Z
\ee

Using our definition of  Shanon's entropy  and defining the Helmholtz  free energy $F$  
\be
 \beta F= -ln Z
\ee
we get the well known relation 
\be
\beta F= \beta U -S \qquad F =U -TS
\ee

{\bf It is important to emphasize that at this stage there is no way to disentangle the energy $E(x)$
from the temperature. }
 
\subsubsection{The temperature}

To have a meaningful thermodynamics, and to disentangle $\beta$ and $E(x)$, the distribution $p(x)$ needs to be part of a family of distributions parameterized by some external controls (e.g., parameters $\theta$ like the mean or variance) It addition it requires an independently specified Hamiltonian or energy scale and an identifiable Gibbs family.

Instead of a single PDF $p(x)$, imagine we have a family of probability distributions parameterized by external fields $\boldsymbol{\theta} = (\theta_1, \theta_2, \dots)$:
\be
p(x; \boldsymbol{\theta}) = \frac{1}{Z(\boldsymbol{\theta})} \exp \left( -\beta E(x) - \sum_{i} \theta_i F_i(x) \right)
\ee

$E(x)$ is the "bare" or baseline energy (the internal structure derived from the original PDF).
$F_i(x)$ are external "observables" or order parameters (e.g., position, momentum, variance, skewness).
$\theta_i$ are the conjugate forces or Lagrange multipliers (which act as generalized temperatures, magnetic fields, or chemical potentials).
$Z(\boldsymbol{\theta})$ is the partition function:
$$Z(\boldsymbol{\theta}) = \int \exp \left( -\beta E(x) - \sum_{i} \theta_i F_i(x) \right)\,dx$$

How this disentangles them:
By taking derivatives of the log-partition function (the free energy / pressure function $\Phi(\boldsymbol{\theta}) = \ln Z(\boldsymbol{\theta})$), we can pull out the expectation values of the energy and the observables.
 The average energy is: $\langle E \rangle = -\frac{\partial \Phi}{\partial \beta}$
 The response to a field is: $\langle F_i \rangle = -\frac{\partial \Phi}{\partial \theta_i}$

By tuning $\theta_i$, we force the system to respond, which allows us to measure $\beta$ independently of the bare energy function $E(x)$.

This procedure of extending the PDF into a family of PDFs and extracting the thermodynamical quantities will be explicitly demonstrated  for the cases  of a Gaussian distribution (\ref{TGd}) and the $G\beta E$ PDF(\ref{TsPoGe}).

It should be noted that a family of PDFs
$p(x;\lambda)$
can provide a meaningful effective temperature if there is an independently specified energy function or an otherwise identifiable thermodynamic structure.

One should  distinguish
\[
p(x;\lambda)
=
\frac{e^{-\beta(\lambda)E(x)}}{Z(\beta(\lambda))}
\]
from
\[
p(x;\lambda)
=
\frac{e^{-\beta E(x;\lambda)}}{Z(\lambda)}.
\]

The first varies temperature at fixed energy function; the second changes the energy function itself.
These are different operations.

\subsubsection {The First Law of Thermodynamics}

In thermodynamics, the First Law relates changes in internal energy ($U$) to heat and work, commonly written as 
\be dU = T dS - dW \qquad or\qquad  dF = -S dT - \sum \langle F_i \rangle d\theta_i
\ee,

If one perturbs the distribution $p(x)$ by changing its parameters $\theta_i$, the change in the log-partition function ($\ln Z$, which acts as the thermodynamic free energy) yields a strict first-law-like accounting:
Internal energy change:
\be
dU = \sum \theta_i d\langle F_i \rangle + \sum \langle F_i \rangle d\theta_i
\ee
and entropy change:
\be
dS = \frac{1}{T} (dU - \text{Work})
\ee
It is important to realize that using the map between $p(z)$ and the Gibbs PDF a fictitious  first law can be associated also  with  non-physics chaotic processes.


\subsubsection{The constraints for a Valid Thermodynamic PDF }

A natural question to ask is whether there is a thermodynamics picture to any pdf?
It seems that 
mathematically, one can formally associate a thermodynamic picture to almost any  PDF by taking the negative logarithm to define an "energy" function :
$E(x) = -\ln p(x)$( up to a constant).
However, physically and thermodynamically, there are strict constraints if one wants that picture to make thermodynamic sense (e.g., normalizability, finite energy, and stable statistical mechanics).


To map a PDF to a standard canonical ensemble ($P(x) = \frac{1}{Z} e^{-E(x)/T}$), the distribution must satisfy a few mathematical and physical rules:
\item
1. Normalizability (The Partition Function Must Exist):

The integral $\int e^{-E(x)/T} dx$ must converge to a finite number $Z$. If $Z = \infty$ or $Z = 0$, the partition function breaks down, meaning the thermodynamic "state" cannot exist in equilibrium. For example, a uniform distribution over an infinite line cannot be normalized.
\item
2. Support and Boundaries:

The energy function $E(x)$ must be well-defined wherever $p(x) > 0$. If $p(x) = 0$, the energy goes to $+\infty$, namely,  an infinite potential barrier, which is physically allowed.  
\item
3. Extensivity and Thermodynamic Limit:

In true physical thermodynamics, energy must be extensive (scaling with  the system size $N$). For arbitrary mathematical  PDFs of a single variable $x$, we are usually looking at a "micro-level" or information-theoretic thermodynamics rather than macroscopic bulk matter, so this constraint is often relaxed.
\section{Thermodynamics of spacings -Benchmark systems  }\label{TosBs}

The benchmark systems are not merely examples of PDFs. They are systems for which independent information about chaos or statistical organization is already known. They therefore provide tests of what information about the underlying process is retained by the extrema statistics.

\subsection{The Bernoulli shift map}

Recall the standard Bernoulli shift map\ref{TBsm}
\be
x_{n+1}=2x_n \pmod 1,
\qquad
0\leq x_n<1.
\ee
The probability mass function of the peak spacings is
\be
p_s=
\frac{s-1}{2^s},
\qquad s\geq2.
\ee
The mean spacing is
\be
\langle S\rangle
=
\sum_{s=2}^{\infty}
s\,\frac{s-1}{2^s} =4.
\ee

\item Entropy of the spacing distribution

Because \(S\) is discrete, the appropriate entropy is the Shannon
entropy is found to be 

\be
S
=
4\ln2
-
\sum_{s=2}^{\infty}
\frac{s-1}{2^s}\ln(s-1).
\ee
Numerical evaluation gives
\be
S\simeq1.8794469\ .
\ee
In fact the Bernoulli shift map serves as an landmark  example of  the dynamics of an chaotic process and its corresponding extrema analysis.  
\be
\text{known dynamical chaos}
\longrightarrow
\text{extrema process}
\longrightarrow
\text{information retained/lost}.
\ee

\item{The energy function}

Using the map to the Gibbs PDF we get 
\be
\beta E_s
=
s\ln2-\ln(s-1)-\ln Z.
\ee

The additive constant $-\ln Z$ can be made arbitrary because adding a constant to
all energies does not change a canonical probability distribution.

Thus a natural dimensionless effective energy is
\be
\mathcal{E}(s)
\equiv
\beta E(s)
=
s\ln2-\ln(s-1).
\ee
This may be viewed as  the basic energy landscape associated with the spacing
distribution.

\item{Interpretation of the effective energy}

The effective energy consists of two terms:
\be
\mathcal{E}(s)
=
s\ln2-\ln(s-1).
\ee

The first term, $s\ln2,$
increases linearly with the spacing. It therefore suppresses large gaps.

The second term, $-\ln(s-1),$
lowers the effective energy because there are \(s-1\) possible
configurations associated with a gap of length \(s\).

Thus one may write
\be
\mathcal{E}(s)
=
\underbrace{s\ln2}_{\text{cost of a long spacing}}
-
\underbrace{\ln(s-1)}_{\text{degeneracy/entropy contribution}}.
\ee

This observation is important. The quantity
\be
-\ln p_s
\ee
is naturally closer to an effective \emph{free energy} than to a
microscopic energy if the factor \(s-1\) represents a multiplicity of
microscopic configurations.

\item{A more physical thermodynamic decomposition}

Indeed, write
\be
p_s=(s-1)2^{-s}.
\ee
This has precisely the form
\be
p_s
=
\frac{g_s e^{-\beta E_s}}{Z},
\ee
where \(g_s\) is a degeneracy.
Choose
\be
g_s=s-1
\ee
and
\be
E_s=\epsilon s,
\ee
where \(\epsilon\) is an elementary energy scale.
Then
\be
p_s
=
\frac{(s-1)e^{-\beta\epsilon s}}{Z}.
\ee

Comparison with
\be
p_s=\frac{s-1}{2^s}
\ee
gives
\be
e^{-\beta\epsilon}=\frac12.
\ee

Consequently,
\be
\beta\epsilon=\ln2.
\ee

This gives a particularly clean thermodynamic interpretation of the
Bernoulli spacing distribution.

\item{Partition function}

The partition function is
\be
Z(\beta)
=
\frac{e^{-2\beta\epsilon}}
{\left(1-e^{-\beta\epsilon}\right)^2}.
\ee
Since for the Bernoulli map,
\be
e^{-\beta\epsilon}=\frac12,
\ee therefore
\be
Z=1.
\ee
This is consistent with the fact that
\be
\sum_{s=2}^{\infty}\frac{s-1}{2^s}=1.
\ee
\item{Internal energy}

The thermodynamic internal energy is
\be
U
=
\langle E\rangle
=
\sum_{s=2}^{\infty}p_sE_s= U=4\epsilon= U=4\frac{\ln2}{\beta}.
\ee

Notice, however, that this latter equation applies at the particular
temperature corresponding to the Bernoulli distribution.

For the full canonical family,
\be
U(\beta)
=
-\frac{\partial}{\partial\beta}\ln Z.
\ee
Since
\be
\ln Z
=
-2\beta\epsilon
-
2\ln\left(1-e^{-\beta\epsilon}\right),
\ee
we obtain
\be
U(\beta)
=
\frac{2\epsilon}
{1-e^{-\beta\epsilon}}.
\ee
At
\be
e^{-\beta\epsilon}=\frac12,
\ee
this gives
\be
U=4\epsilon,
\ee
as expected.

\item{Entropy}

There are now two different entropies that should be distinguished.

The entropy of the observed spacing variable is
\be
S
=
-\sum_s p_s\ln p_s\simeq1.8794469 .
\ee

However, if \(g_s=s-1\) represents a genuine degeneracy of
microstates, the canonical thermodynamic entropy is
\be
\frac{S_{\mathrm{th}}}{k_B}
=
-\sum_s p_s
\ln\left(\frac{p_s}{g_s}\right).
\ee

Since
\be
\frac{p_s}{g_s}=2^{-s},
\ee
we obtain
\be
\frac{S_{\mathrm{th}}}{k_B}
=
\sum_s p_s\,s\ln2.
\ee

Using
\be
\langle S\rangle=4,
\ee
gives
\be
S_{\mathrm{th}}
=
4k_B\ln2.
\ee

The relation between the two entropies is
\be
\frac{S_{\mathrm{th}}}{k_B}
=
S+\langle\ln g_S\rangle.
\ee

Here
\be
g_S=S-1,
\ee
so
\be
\frac{S_{\mathrm{th}}}{k_B}
=
H_S+\langle\ln(S-1)\rangle.
\ee

Numerically,
\be
4\ln2\simeq2.7725887,
\ee
and hence
\be
\langle\ln(S-1)\rangle
\simeq
2.7725887-1.8794469
\simeq0.8931418.
\ee

Thus
\be
H_S\simeq1.87945
\ee
is the entropy of the coarse-grained spacing variable, whereas
\be
\frac{S_{\mathrm{th}}}{k_B}=4\ln2\simeq2.77259
\ee
includes the assumed internal degeneracy \(g_s=s-1\).

\item{Helmholtz free energy}

The Helmholtz free energy is
\be
F=-k_BT\ln Z.
\ee

For the general canonical family,
\be
Z
=
\frac{e^{-2\beta\epsilon}}
{\left(1-e^{-\beta\epsilon}\right)^2},
\ee
and therefore
\be
F
=
2\epsilon
+
2k_BT
\ln\left(1-e^{-\epsilon/(k_BT)}\right).
\ee

At the Bernoulli value
\be
e^{-\beta\epsilon}=\frac12,
\ee
we have
\be
Z=1,
\ee
so
\be
F=0.
\ee

This is consistent with
\be
F=U-TS_{\mathrm{th}},
\ee
because
\be
U=4\epsilon
\ee
and
\be
TS_{\mathrm{th}}
=
T(4k_B\ln2)
=
4\epsilon.
\ee

Therefore
\be
F=4\epsilon-4\epsilon=0.
\ee

The value \(F=0\) here depends on the chosen energy normalization and
partition-function convention; it should not by itself be assigned
independent physical significance.

\item{Can the temperature be determined?}

This is the crucial point.

From the spacing distribution we find
\be
\beta\epsilon=\ln2.
\ee

Equivalently,
\be
\frac{\epsilon}{k_BT}=\ln2.
\ee

The PDF therefore determines the \emph{dimensionless inverse
temperature}
\be
\beta\epsilon,
\ee
but it does not determine \(\epsilon\) and \(T\) separately.

Thus the absolute temperature cannot be obtained from the spacing
distribution alone.

If we choose units in which
\be
\epsilon=1,
\qquad
k_B=1,
\ee
then
\be
T_{\mathrm{eff}}
=
\frac{1}{\ln2}
\simeq1.442695.
\ee

This is a perfectly legitimate dimensionless effective temperature,
but it is a choice of units, not an independently measured physical
temperature.

Therefore,
\be
T_{\mathrm{eff}}=\frac{1}{\ln2}
\ee
should not be interpreted as an intrinsic physical temperature of the
Bernoulli map.

What is intrinsically determined is only
\be
\beta\epsilon=\ln2.
\ee

\item{Heat capacity}

Once the canonical family is introduced, one may also define a heat
capacity
\be
C
=
\frac{\partial U}{\partial T}.
\ee

Since
\be
U
=
\frac{2\epsilon}
{1-e^{-\epsilon/(k_BT)}},
\ee
we find
\be
C
=
2k_B
\left(\frac{\epsilon}{k_BT}\right)^2
\frac{
e^{-\epsilon/(k_BT)}
}{
\left[
1-e^{-\epsilon/(k_BT)}
\right]^2
}.
\ee

At the Bernoulli point,
\be
\frac{\epsilon}{k_BT}=\ln2,
\ee
and therefore
\be
C
=
4k_B(\ln2)^2.
\ee

Thus
\be
C_{\mathrm{Bernoulli}}
=
4k_B(\ln2)^2
\simeq1.92181\,k_B.
\ee

Notice that quantities such as the heat capacity require the
\emph{extended family} \(p_s(\beta)\). They cannot be obtained from a
single PDF without specifying how the distribution is to change with
temperature.

\subsection{The  Poisson spacing PDF}

For the Poisson spacing distribution (which corresponds to regular, integrable quantum systems), the probability density function is:
\be
P(s) = e^{-s}
\ee
(Note: This is already normalized such that the mean spacing $\langle s \rangle = \int_0^\infty s e^{-s} ds = 1$).

Here are its corresponding thermodynamic
properties.  

 \item 1. Shannon Entropy ($H$)
The differential Shannon entropy for the Poisson distribution is calculated as:

\be
S = -\int_0^\infty e^{-s} (-s) \, ds   =1
\ee

 \item 2. Internal Energy ($E(s)$)

From comparing the Poisson and Gibbs PDFs we get 
\be
\beta E(s) = s
\ee
The internal energy (the ensemble average of this energy) is:
\be
\langle \beta E \rangle = \int_0^\infty E(s) P(s) \, ds = \int_0^\infty s e^{-s} ds = 1
\ee

 \item 3. Effective Temperature ($T$)-
 
The partition function $Z$ for the Poisson distribution is:
$$Z = \int_0^\infty e^{-E(s)/T} ds = \int_0^\infty e^{-s/T} ds = T$$

If we impose the  normalization $\int_0^\infty P(s) ds = 1$, this fixes the baseline temperature of the standard Poisson process at:
\be
T = 1.
\ee

Alternatively if we generalize the distribution for an arbitrary temperature $T$, it becomes:
\be
P(s; T) = \frac{1}{T} e^{-s/T}
\ee

 If we then  impose a unit mean spacing
\be
<s>=1
\ee
selects $T=1$ under the convention that $E(s)=s$.

\subsection{ The  spacing PDF of the Gaussian$\beta$ ensemble}\label{TsPoGe}

To find the thermodynamic properties (Entropy, Energy and  Temperature) for the spacing distribution of eigenvalues in the Gaussian $\beta$-Ensemble (GBE)\cite{Dyson1962I,Dyson1962II,Dyson1962III}, we have to shift from the multi-particle joint PDF to the distribution of the distance between adjacent eigenvalues, $s = x_{i+1} - x_i$.

While the exact exact analytical spacing distribution for all $\beta$ is famously complex (involving Painlevé transcendents or Fredholm determinants), we can use the famous Wigner Surmise approximations, which capture the exact thermodynamic behavior with incredible accuracy.

Here is how the thermodynamic dictionary applies to the eigenvalue spacing PDF.

\item  The entropy  of the Gaussian beta ensemble

For the Gaussian \(\beta\) ensemble (G\(\beta\)E), the nearest–neighbor
spacing distribution is well approximated by the Wigner surmise
\be
p_\beta(s)
= a_\beta\, s^{\beta}\,
\exp\!\left(-b_\beta s^{2}\right),
\qquad s>0,
\ee
where the constants \(a_\beta\) and \(b_\beta\) are fixed by normalization
and unit mean spacing:
\be
b_\beta
= \left[
\frac{\Gamma\!\left(\frac{\beta+2}{2}\right)}
{\Gamma\!\left(\frac{\beta+1}{2}\right)}
\right]^2,
\qquad
a_\beta
= \frac{2\, b_\beta^{(\beta+1)/2}}
{\Gamma\!\left(\frac{\beta+1}{2}\right)}.
\ee

A direct calculation  of the differential entropy yields
\be
\begin{aligned}
S_\beta
&= -\ln a_\beta
- \beta\,\mathbb{E}[\ln s]
+ b_\beta\,\mathbb{E}[s^2]
\\[6pt]
&= -\ln a_\beta
- \frac{\beta}{2}
\left[
\psi\!\left(\frac{\beta+1}{2}\right)
- \ln b_\beta
\right]
+ \frac{\beta+1}{2},
\end{aligned}
\ee
where \(\psi(x)\) is the digamma function.

Substituting \(a_\beta\) explicitly, this can be written as
\be
\begin{aligned}
S_\beta
&=
\ln \Gamma\!\left(\frac{\beta+1}{2}\right)
- \ln 2
- \frac{1}{2}\ln b_\beta
&\quad
- \frac{\beta}{2}\,
\psi\!\left(\frac{\beta+1}{2}\right)
+ \frac{\beta+1}{2}
\end{aligned}
.
\ee

The Special cases are:
\be
\beta=1\;(\text{GOE}),\qquad
\beta=2\;(\text{GUE}),\qquad
\beta=4\;(\text{GSE}),\qquad
\ee
 
Notice that 

$\beta=1,2,4$
 are the classical symmetry classes, while the 
$\beta$-ensemble extends continuously to $\beta>0$.
The differential entropy  
takes the following numerical values:
\be
\begin{aligned}
\beta = 1 \; (\text{GOE}): \quad & h_1 \;\approx\; 0.716243, \\[6pt]
\beta = 2 \; (\text{GUE}): \quad & h_2 \;\approx\; 0.528798, \\[6pt]
\beta = 4 \; (\text{GSE}): \quad & h_4 \;\approx\; 0.276758 .
\end{aligned}
\ee 
The entropy decreases as \(\beta\) increases, reflecting stronger level
repulsion and increased spectral rigidity in the Gaussian ensembles:
\be
h_{\text{GOE}} > h_{\text{GUE}} > h_{\text{GSE}}.
\ee
\item  {The Energy $E(s)$}

Following our information-theoretic dictionary, we define the effective energy of a spacing $s$ by inverting the Boltzmann factor $p(s) \propto \exp(-E(s)/T)$:
\be
\beta E(s)  = -\ln\left(\frac{\pi}{2}\right) - \ln(s) + \frac{\pi}{4} s^2 -\ln Z
\label{EGOEspacing}\ee
for the GOE spacing distribution.
This reveals a very interesting effective potential energy for the spacing $s$:

1. Repulsive Term ($-\ln s$): This comes from level repulsion (Wigner's avoids crossing theorem). When $s \to 0$, the energy goes to $+\infty$, meaning zero spacing has an exponentially small probability.

2. Confinement Term ($\frac{\pi}{4} s^2$): This acts like a harmonic spring pulling large spacings back down. Extremely large gaps are energetically penalized.

\item 3. The temperature

To find the temperature, we map the generalized spacing distribution $p(s) = a_\beta s^\beta \exp(-b_\beta s^2)$ directly into the canonical Gibbs form:

$$p(s) = \frac{1}{Z} \exp\left( -\frac{E(s)}{T} \right)$$

Let's rewrite the exponent of $p(s)$:
$$p(s) = \exp\left( \beta \ln s - b_\beta s^2 + \ln a_\beta \right)$$

To match this to $\exp(-E(s)/T)$, let's define the fundamental physical energy of the spacing as a harmonic-like well with a logarithmic correction:
$$E_{phys}(s) = s^2 - \frac{\beta}{b_\beta} \ln s$$

With the convention that the coefficient of \(s^2\) in the effective energy is normalized to unity, one may define the dimensionless effective temperature
\be
T = \frac{1}{b_\beta}
\ee

\begin{figure}[h!]
  \centering
  \includegraphics[width=0.4\linewidth]{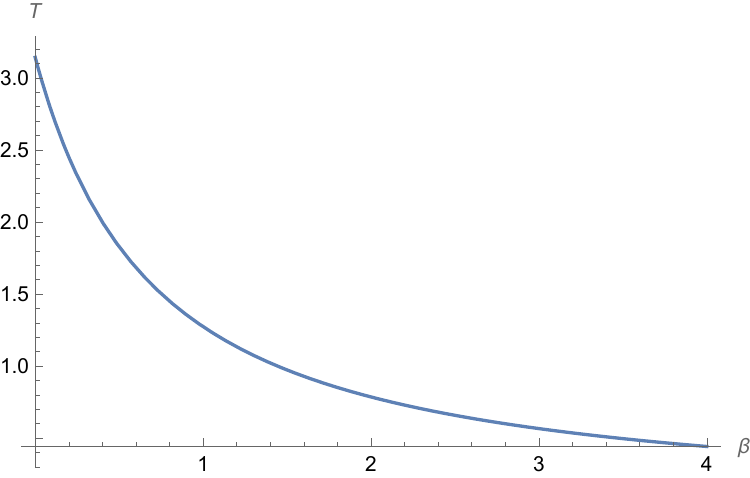}
  \caption{The temperature as a function of $\beta$ for    Gaussian
  $\beta$–ensemble .}\label{Tbeta}
\end{figure}
The temperature as a function of $\beta$ for    Gaussian
  $\beta$–ensemble is drawn in figure (\ref{Tbeta}).
  
For the GOE  ($\beta = 1$), the normalization constants give $b_1 = \frac{\pi}{4}$. Therefore, the temperature of the spacing distribution is:

\be
T_{\text{spacing}} = \frac{4}{\pi} \approx 1.273
\ee

\section{Thermodynamics for joint distribution of all extrema or all spacings.}\label{Tfjdoaeoas}

In this setting, the thermodynamic structure should be interpreted
somewhat differently. The natural objects are not physical particles
with a literal Hamiltonian and temperature. They are the extrema
\be
x_1<x_2<\cdots<x_N
\ee
of an erratic function \({\cal A}(x)\), and the spacings
\be
s_i=x_{i+1}-x_i.
\ee

The question is whether the statistics of the point process
\(\{x_i\}\), or equivalently of the spacings \(\{s_i\}\), possess
enough additional structure to support a useful \emph{effective}
thermodynamics.

The important point is that a one-spacing PDF \(P(s)\) is only a
marginal description. The richer thermodynamic structure, if it
exists, will generally live at the level of the joint distribution of
all extrema or all spacings.

\subsection{What would \(H\) mean?}

From a single spacing distribution one can always define
\be
V_{\mathrm{eff}}(s)=-\log P(s)+C.
\ee
Thus
\be
P(s)=\frac{1}{Z}e^{-V_{\mathrm{eff}}(s)}.
\ee

For example, suppose
\be
P(s)=A s^\alpha e^{-B s^\gamma}.
\ee
Then
\be
V_{\mathrm{eff}}(s)
=
B s^\gamma-\alpha\log s+C.
\ee

The term
\be
-\alpha\log s
\ee
produces repulsion as \(s\rightarrow0\), whereas
\be
B s^\gamma
\ee
confines large spacings.

For a single \(P(s)\), however, this remains mainly a useful
representation rather than a genuine thermodynamics.

The more interesting candidate for \(H\) is instead a function of
\emph{all} spacings,
\be
H(s_1,\ldots,s_N),
\ee
such that
\be
P(s_1,\ldots,s_N)
=
\frac{1}{Z}
e^{-\beta H(s_1,\ldots,s_N)}.
\ee

Then \(H\) measures statistical interactions among the spacings.

For instance, one might find approximately
\be
H
=
\sum_i V(s_i)
+
\sum_i W(s_i,s_{i+1})
+\cdots.
\ee

Here \(V(s_i)\) describes the statistics of an individual gap, while
\(W(s_i,s_{i+1})\) describes correlations between neighboring gaps.
This contains information that is completely absent from the
one-spacing PDF \(P(s)\).

\subsection{What would temperature mean?}

One should not initially introduce a physical temperature \(T\) in
this problem. Unless the erratic function itself arises from a thermal
system, there is no reason for \(T\) to have units of Kelvin.

Instead, write
\be
P(\{s_i\})
=
\frac{1}{Z(\beta)}
e^{-\beta H(\{s_i\})}.
\ee

Only the combination
\be
\beta H
\ee
is initially meaningful.

An effective \(\beta\) becomes useful if one has a \emph{family} of
erratic functions controlled by some parameter \(\lambda\), for
example their roughness, noise amplitude, disorder strength, spectral
bandwidth, correlation length, or another parameter controlling
irregularity.

One could then ask whether
\be
P_\lambda(\{s_i\})
\simeq
\frac{1}{Z(\beta(\lambda))}
e^{-\beta(\lambda)H(\{s_i\})}
\ee
with essentially the same \(H\).

If this occurs, \(\beta(\lambda)\) plays a role analogous to inverse
temperature.

This would be nontrivial: changing the erraticness of the underlying
function would change mainly the effective temperature of the extrema
process rather than changing the effective interaction itself.

\subsection{Constraints have a natural meaning for spacings}

Suppose the extrema occupy an interval of length \(L\). Then
necessarily
\be
\sum_{i=1}^{N}s_i=L.
\ee

This is a genuine geometric constraint.

Equivalently,
\be
\frac{1}{N}\sum_i s_i=\bar{s}=\frac{L}{N}.
\ee

After unfolding the extrema to unit mean spacing, one normally has
\be\langle s\rangle=1.
\ee

Thus the fixed total length is analogous to a thermodynamic
constraint.

There may be additional constraints generated by the underlying
erratic function. For example,
\be
\left\langle s^2\right\rangle,
\qquad
\left\langle s_i s_{i+1}\right\rangle,
\qquad
\left\langle \log s\right\rangle
\ee
could be constrained by properties of the process generating the
extrema.

Maximum entropy subject to such constraints produces distributions
of Gibbs form.

For example, maximizing
\be
S[P]
=
-\int_0^\infty P(s)\log P(s)\,ds
\ee
subject to
\be
\int_0^\infty P(s)\,ds=1
\ee
and
\be
\langle s\rangle=1
\ee
gives
\be
P(s)=e^{-s}.
\ee

Thus the Poisson spacing distribution follows from maximum entropy
when the mean spacing is the only relevant constraint.

This provides a natural baseline.

If instead one also imposes, for example,
\be
\langle\log s\rangle=c,
\ee
then maximum entropy gives
\be
P(s)
\propto
s^\alpha e^{-\lambda s}.
\ee

Thus a power-law behavior
\be
P(s)\sim s^\alpha,
\qquad s\rightarrow0,
\ee
can be interpreted in a maximum-entropy framework as additional
statistical structure beyond the mean-spacing constraint.

\subsection{Dynamics of extrema}

There are two useful notions of dynamics in this setting.

Suppose the erratic function belongs to a parameterized family
\be
{\cal A}(x;\tau).
\ee

As \(\tau\) changes, the extrema move:
\be
x_i=x_i(\tau),
\ee
and therefore
\be
s_i(\tau)
=
x_{i+1}(\tau)-x_i(\tau).
\ee

Then \(\tau\) becomes an effective time, and one can study the
stochastic dynamics of the extrema.

One might obtain something schematically like
\be
\frac{ds_i}{d\tau}
=
-\frac{\partial H}{\partial s_i}
+\eta_i(\tau),
\ee
or a Fokker--Planck equation
\be
\frac{\partial P}{\partial\tau}
=
\sum_i
\frac{\partial}{\partial s_i}
\left[
P\frac{\partial H}{\partial s_i}
+
D\frac{\partial P}{\partial s_i}
\right].
\ee

The stationary distribution could then be
\be
P_{\mathrm{st}}(\{s_i\})
\propto
e^{-H/D}.
\ee

In this situation \(D\), which measures the strength of the
stochastic motion of the extrema, would play a role analogous to an
effective temperature.

\subsection{Dynamics hidden in the spatial sequence}

Even if there is no external parameter \(\tau\), there is still the
sequence
\be
s_1,s_2,s_3,\ldots
\ee
and one may regard the index \(i\) as a discrete ``time'' variable.

One can then ask whether
\be
P(s_{i+1}\mid s_i)
\ee
depends on \(s_i\).

If
\be
P(s_{i+1}\mid s_i)=P(s_{i+1}),
\ee
the neighboring spacings are statistically independent.

If instead
\be
P(s_{i+1}\mid s_i)\neq P(s_{i+1}),
\ee
the extrema contain correlations invisible in the one-spacing
distribution \(P(s)\).

For example, one may calculate
\be
C(k)
=
\left\langle
(s_i-\bar{s})(s_{i+k}-\bar{s})
\right\rangle.
\ee

This is one of the first quantities worth examining in such a
problem.

Notice     that  $s_i$ are  conceptually different
from the $k$-spacing
\begin{equation}
\delta_{k,n}=x_{n+k}-x_n.
\end{equation}

Let  introduce the notation
\begin{equation}
P_m(\delta_1,\ldots,\delta_m)
\end{equation}
for the joint probability density of $m$ consecutive nearest-neighbor
spacings. More precisely, assuming stationarity,
\begin{equation}
P_m(s_1,\ldots,s_m)
=
P\left(
\delta_n=s_1,\,
\delta_{n+1}=s_2,\ldots,
\delta_{n+m-1}=s_m
\right),
\end{equation}
with the corresponding interpretation as a probability density for
continuous variables.

This notation avoids possible confusion between a joint distribution of
consecutive spacings and the $k$-spacing $\delta_{k,n}$.

The two quantities contain different information.

The $k$-spacing
\begin{equation}
\delta_{k,n}
=
\delta_n+\delta_{n+1}+\cdots+\delta_{n+k-1}
\end{equation}
contains only the sum of $k$ consecutive nearest-neighbor spacings.

In contrast,
\begin{equation}
P_k(s_1,\ldots,s_k)
\end{equation}
retains the complete statistical organization among these $k$
consecutive spacings.

For example, for $k=2$,
\begin{equation}
\delta_{2,n}
=
\delta_n+\delta_{n+1}.
\end{equation}
Therefore, the PDF of the second-neighbor spacing is obtained from the
two-spacing joint PDF according to
\begin{equation}
P_{\delta_2}(S)
=
\int_0^\infty ds_1
\int_0^\infty ds_2\,
P_2(s_1,s_2)
\delta_{\rm D}(S-s_1-s_2),
\end{equation}
where $\delta_{\rm D}$ denotes the Dirac delta function.

Equivalently,
\begin{equation}
P_{\delta_2}(S)
=
\int_0^S
P_2(s,S-s)\,ds.
\end{equation}

More generally,
\begin{equation}
P_{\delta_k}(S)
=
\int_0^\infty
\prod_{j=1}^{k}ds_j\,
P_k(s_1,\ldots,s_k)
\,
\delta_{\rm D}
\left(
S-\sum_{j=1}^{k}s_j
\right).
\end{equation}

Thus, the $k$-spacing PDF is a projection of the full $k$-spacing
joint PDF and consequently contains less information.

If consecutive nearest-neighbor spacings are statistically independent,
then
\begin{equation}
P_k(s_1,\ldots,s_k)
=
\prod_{j=1}^{k}P_1(s_j).
\end{equation}

For $k=2$ this gives
\begin{equation}
P_{\delta_2}(S)
=
\int_0^S
P_1(s)P_1(S-s)\,ds.\label{s1s2}
\end{equation}
Another example is with following our last discussion, is the \(k=3\) Gaussian \(\beta\)-ensemble.
For three ordered eigenvalues,
\[
x_1<x_2<x_3,
\]
define
\[
s_1=x_2-x_1,\qquad
s_2=x_3-_2.
\]
For a Gaussian joint density of the form
\[
P(x_1,x_2,x_3)
\propto
e^{-a\sum_i x_i^2}
\prod_{i<j}|x_i-x_j|^\beta,
\]
one obtains
\[
P_2(s_1,s_2)
\propto
s_1^\beta s_2^\beta(s_1+s_2)^\beta
\exp\left[
-\frac{2a}{3}
(s_1^2+s_1s_2+s_2^2)
\right].
\]
This immediately shows
\[
P_2(s_1,s_2)\neq P_1(s_1)P_1(s_2).
\]

Going back to (\ref{s1s2}),this is simply the convolution
\begin{equation}
P_{\delta_2}=P_1*P_1.
\end{equation}

For general $k$,
\begin{equation}
P_{\delta_k}
=
\underbrace{
P_1*P_1*\cdots*P_1
}_{k\ {\rm times}}.
\end{equation}

This observation suggests a useful test for correlations in an erratic
function. One can compare the measured PDF of the $k$-spacing,
$P_{\delta_k}(S)$, with the $k$-fold convolution constructed from the
measured nearest-neighbor PDF $P_1(s)$.

If
\begin{equation}
P_{\delta_k}^{\rm measured}(S)
\neq
\underbrace{
P_1*P_1*\cdots*P_1
}_{k\ {\rm times}},
\end{equation}
then the sequence of nearest-neighbor spacings cannot be described as
independent random variables.

This may provide an additional diagnostic for distinguishing structured
erraticity from simple uncorrelated stochastic irregularity.

\subsection{Why the extrema themselves may be more fundamental}

Suppose the extrema have joint density
\be
P(x_1,\ldots,x_N).
\ee

One could ask whether it takes approximately the form
\be
P(x_1,\ldots,x_N)
\propto
e^{-\beta H(x_1,\ldots,x_N)}.
\ee

For example,
\be
H
=
\sum_i V(x_i)
+
\sum_{i<j}U(|x_i-x_j|).
\ee

If
\be
U(r)=-\alpha\log r,
\ee
then
\be
P(x_1,\ldots,x_N)
\propto
e^{-\beta\sum_iV(x_i)}
\prod_{i<j}|x_i-x_j|^{\beta\alpha}.
\ee

This is precisely the type of structure that appears in random-matrix
Coulomb-gas descriptions.

For extrema of erratic functions, a crucial question is therefore
whether the extrema exhibit an effective interaction analogous to
\be
-\log|x_i-x_j|,
\ee
or perhaps some short-range interaction.

If they do, then the thermodynamic interpretation is not merely
cosmetic: it tells us something about the collective organization of
the extrema.

One may want  to display explicitly the hierarchy
\begin{equation}
P(x_1,\ldots,x_N)
\quad\longrightarrow\quad
P(\delta_1,\ldots,\delta_{N-1})
\quad\longrightarrow\quad
P_1(\delta).
\end{equation}

Information is progressively discarded as one moves from left to right.

This provides a useful connection between the point-process description
of the extrema and the one-spacing PDFs studied extensively in the
paper.

\subsection{The underlying function \({\cal A}(x)\) and the origin of the
effective Hamiltonian}

This is particularly important for the present problem.

The extrema satisfy
\be
{\cal A}'(x_i)=0.
\ee

Therefore the point process of extrema is determined by the
statistical properties of
\be
{\cal A}'(x).
\ee

If \({\cal A}(x)\) is a random process, for example a Gaussian random
function with correlation function
\be
C(r)=\langle {\cal A}(x){\cal A}(x+r)\rangle,
\ee
then derivatives of \(C(r)\) determine statistical properties of the
zeros of \(f'\), and hence of the extrema of \(f\).

There is therefore a chain
\be
\text{statistics of }{\cal A}
\longrightarrow
\text{statistics of }{\cal A}'
\longrightarrow
\text{point process of zeros of }{\cal A}'
\longrightarrow
\{x_i\}
\longrightarrow
\{s_i\}.
\ee

Ideally, the effective thermodynamics should emerge from this chain
rather than being imposed afterward on \(P(s)\).

In particular, the correlation structure of \({\cal A}\) may generate
effective repulsion or attraction between its extrema.

One can  extend this chain by one additional step:
\begin{equation}
(x)
\longrightarrow
{\cal A}'(x)
\longrightarrow
\{x_i\}
\longrightarrow
\{\delta_i\}
\longrightarrow
\left\{
P_1(\delta),
P_2(\delta_i,\delta_{i+1}),
P_3(\delta_i,\delta_{i+1},\delta_{i+2}),
\ldots
\right\}.
\end{equation}

This makes explicit that the ordinary spacing PDF $P_1(\delta)$ is only
the first member of a hierarchy of statistical descriptions.

The hierarchy can be written schematically as
\begin{equation}
P_1
\longrightarrow
P_2
\longrightarrow
P_3
\longrightarrow\cdots,
\end{equation}
where higher-order joint PDFs reveal correlations that are necessarily
invisible in the marginal distribution $P_1$.

This distinction may become particularly important for the central
question of the paper. A fit of the nearest-neighbor spacing PDF to a
particular distribution, such as a Gaussian $\beta$-ensemble
distribution, logistic distribution, log-normal distribution, or other
PDF, characterizes only the one-spacing statistics. It does not by
itself determine whether the erratic function originates from a chaotic
process.

Stronger evidence can be obtained by examining whether the sequence of
extrema possesses nontrivial higher-order statistical organization. In
particular, one may examine
\begin{equation}
P_2(\delta_i,\delta_{i+1}),
\qquad
P_3(\delta_i,\delta_{i+1},\delta_{i+2}),
\qquad
C(k),
\end{equation}
and related information-theoretic quantities.

For example, the mutual information between spacings separated by $k$
steps is
\begin{equation}
I(k)
=
\int ds\,ds'\,
P(\delta_i=s,\delta_{i+k}=s')
\ln
\left[
\frac{
P(\delta_i=s,\delta_{i+k}=s')
}{
P_1(s)P_1(s')
}
\right].
\end{equation}

For statistically independent spacings,
\begin{equation}
I(k)=0.
\end{equation}

Consequently, the combination of the $k$-spacing PDFs and the joint PDFs
of consecutive spacings provides substantially more information than
the nearest-neighbor spacing PDF alone.

\subsection{A concrete hierarchy for the analysis}

A natural organization is\cite{DaleyVereJones2003}
\be
{\cal A}(x)
\quad\longrightarrow\quad
\{x_i\}
\quad\longrightarrow\quad
\{s_i\}.
\ee

At the lowest level one has the nearest-neighbor spacing distribution
\be
P_1(s).
\ee

The next objects to calculate are
\be
P_2(s_i,s_{i+1})
\ee
and
\be
C(k)
=
\langle s_i s_{i+k}\rangle-\langle s\rangle^2.
\ee

One should then compare
\be
P_2(s,s')
\ee
with
\be
P_1(s)P_1(s').
\ee

A particularly revealing quantity is
\be
W(s,s')
=
-\log
\frac{P_2(s,s')}
{P_1(s)P_1(s')}.
\ee

If
\be
W(s,s')=0,
\ee
neighboring spacings are independent.

If it is nonzero, \(W\) provides a natural measure of the statistical
interaction between neighboring spacings.

One might then attempt to construct an effective Hamiltonian of the
form
\be
H_{\mathrm{eff}}(\{s_i\})
=
\sum_i V(s_i)
+
\sum_i W(s_i,s_{i+1})
+\cdots.
\ee

This is the point at which a thermodynamic description of an
extrema-spacing problem starts becoming genuinely useful rather than
merely formal.


For extrema of an erratic function, the correspondence may be
summarized schematically as
\be
\begin{array}{ccl}
\text{thermodynamic concept}
&\longleftrightarrow&
\text{extrema-spacing problem}
\\[4pt]
H
&\longleftrightarrow&
\text{effective statistical interaction among extrema or gaps},
\\[4pt]
\beta
&\longleftrightarrow&
\text{parameter controlling the strength of fluctuations},
\\[4pt]
\text{constraint}
&\longleftrightarrow&
\displaystyle \sum_i s_i=L,\quad
\langle s\rangle=1,\quad
\text{and possible correlation constraints},
\\[6pt]
\text{dynamics}
&\longleftrightarrow&
\text{motion of extrema under variation of the underlying function},
\\[4pt]
\text{interaction}
&\longleftrightarrow&
\text{correlations between }s_i\text{ and }s_j.
\end{array}
\ee

The central distinction is therefore between merely writing
\be
P_1(s)=e^{-V_{\mathrm{eff}}(s)}
\ee
and demonstrating that the complete set of spacings has a structured
joint distribution
\be
P(s_1,\ldots,s_N)
\propto
e^{-\beta H_{\mathrm{eff}}(s_1,\ldots,s_N)}.
\ee

The latter can contain genuine information about collective
organization, correlations, constraints, and possibly the dynamics of
the extrema.



\subsection{What can thermodynamics add?}

The thermodynamic language becomes more informative if one studies a
family of PDFs
\[
P(s;\mu)
\]
depending on a parameter \(\mu\).

If the distributions can be represented approximately as
\[
P(s;\mu)
=
\frac1{Z(\beta)}
e^{-\beta(\mu)V(s)}
\]
with approximately the same potential \(V(s)\), then
\[
\beta(\mu)
\]
becomes a meaningful effective control parameter.

One may then investigate quantities analogous to
\[
U(\beta),
\qquad
S(\beta),
\qquad
F(\beta),
\qquad
C(\beta).
\]

But for a single arbitrary PDF,
\[
P(s),
\]
the thermodynamic construction alone cannot establish whether the
underlying function is chaotic.

Thermodynamic quantities derived from extrema PDFs characterize
statistical structure,
not chaos by themselves.

\section{Summary of thermodynamics descriptions}\label{Sotd}
In table (\ref{summary}) we summarize the theormodynamics description of the three different levels. Level 0 the pdf  of  locations of the extrema points, level 1 the pdf of the spacings and level 2 the pdf of the ratios of spacings. For each case teh pdf,entropy, energy and temperature are specified.
\begin{sidewaystable}[htbp]
\centering
\begin{tabular}{|c|c|c|c|c|c|}
\hline 
Extrema & &PDF & entropy& energy&  temperature \\ \hline  & Bernoulli & $ \frac{(s-1)}{2^s}$   & 1.88 & $s \ln(2) - \ln(s-1) $ & 1 \\ \hline
 & Flat & $\frac{1} {N}$ or $\frac{1}{(b-a)}$ & $ln N$ or $ln(b-a)$  & $E=E_0$ & undefined   \\ \hline
 & Delta & $\delta(x-x_0)$ &$-\infty$& $E(x\neq x_0)=+\infty$ & 0\\ \hline
 & Surmise  & $\frac{2}{\pi R^2} \sqrt{R^2 - x^2} $  & $\frac{1}{2} + \ln\left(\frac{\pi R}{2}\right)$  & -$\ln\left(\frac{2}{\pi R^2}\right) - \frac{1}{2}\ln(R^2 - x^2)$ & $2R^2$ \\ \hline \hline
 Spacings &Poison & $ e^{-s}$ & $1$ & $ T(s-ln Z)$ & 1 \\ \hline
 & $G\beta E$ &  $a_\beta\, s^{\beta}\,
\exp\!\left(-b_\beta s^{2}\right)$
 & $-\ln a_\beta
- \frac{\beta}{2}
\left[
\psi\left(\frac{\beta+1}{2}\right)
- \ln b_\beta
\right]
+ \frac{\beta+1}{2}$
 & $-\ln\left(\frac{\pi}{2}\right) - \ln(s) + \frac{\pi}{4} s^2$ &   $\frac{1}{b_\beta}$ \\ \hline
 & Log normal & $\frac{1}{x \sigma \sqrt{2\pi}} \exp \left( - \frac{(\ln x - \mu)^2}{2\sigma^2} \right)$ & $\ln(\sigma \sqrt{2\pi e}) + \mu$ & $ \frac{1}{2 \sigma^2} (\ln x)^2 + \left( 1 - \frac{\mu}{ \sigma^2} \right) \ln x$  &\\
 \hline
 & Logistic  & $\frac{e^{-\frac{x-\mu}{s}}}{s \left( 1 + e^{-\frac{x-\mu}{s}} \right)^2}$& $  \ln(s) + 2=\ln(T) + 2$&
$-\ln \left( \frac{e^{-x/s}}{s \left( 1 + e^{-x/s} \right)^2} \right) - \ln Z$&   \\
 \hline\hline
 ratios & Poison  &$\frac{1}{(1+r)^2}$ & $2$ & $2\ln(1+r)$ & 1  \\
 \hline
 &  $G\beta E$  & $\frac{1}{Z_\beta}\,
\frac{(r+r^2)^{\beta}}{(1+r+r^2)^{1+\frac{3}{2}\beta}}$ & & & \\ \hline
 \end{tabular}
 \caption{Summary}
 \label{summary}
 \end{sidewaystable}
\subsection{ Comparison between locations, spacings and spacing ratios}

As we explained in (\ref{Feftcp}) 
for any given process suspected to be a chaotic one,  we first identify one or several erratic functions of the underlying variables. For instance for the pinball scattering we have analyzed\cite{Bianchi:2025kna} both the outcome angle and the number of bouncing as a function of either the incident angle or also of the ``impact parameter angle (\ref{pinballab}).

For any erratic function we have
the PDFs of the locations of the extrema points, the spacings and the ratios as $p_e,p_s, p_r$
respectively as well  entropies $S_e,S_s,S_r$,
 energies $E_e,E_s,E_r$, and temperatures $T_e,T_s,T_r$.

 We would like to understand  for a given erratic function if and how these quantities are related and whether they are redundant or add additional information.

As an example we take the string decay amplitude for $HES\rightarrow T+T$

\subsection{The added value of the thermodynamics description}

Applying a thermodynamic description (Energy, Temperature, Entropy, and the First Law) to probability density functions (PDFs) associated with the locations of the extrema points, the spacings between adjacent ones  the ratios  between  the spacings  etc. might seem like an abstract mathematical exercise at first, but  in fact it provides several  added values. 

Here are the primary added values one gains by viewing PDFs through a thermodynamic lens:

\item
 1. A Universal ``Language" for Competing Forces (Energy vs. Entropy)-
 
In standard probability, we just have a formula like for spacings in the GOE case
\be p(s) = \frac{\pi}{2}s e^{-\frac{\pi}{4}s^2}\ee 
and we calculate moments or normalization.
The Thermodynamic View splits that formula into a direct battle between Energy ($E$) and  Entropy ($S$)  via the free energy expression ($F = U - TS$).
 For example, in the Wigner surmise(\ref{EGOEspacing}), the $-\ln s$ term is an energetic penalty (level repulsion), while the $s^2$ term is an entropic/confinement constraint. Seeing it this way tells us why the distribution looks the way it does: it is a physical trade-off between packing states tightly and obeying repulsion.

\item 2. A rigorous Way to define ``temperature" (The Shape Parameter)

In many distributions, parameters like $b$ in $\exp(-bs^2)$ are treated as arbitrary fitting constants.
 By mapping the PDF to a Boltzmann factor $\exp(-E/T)$, that parameter is instantly promoted to a physical temperature.
 This gives the parameter immediate physical intuition. 

\item  3. The first law as a sensitivity and response tool.

When you write down the First Law ($dU = TdS - dW$) for a PDF, one is  no longer just looking at a static snapshot of a distribution— one is  looking at how it responds to perturbation.
 If you change a control parameter (like the radius $R$ of a semicircle or the mean spacing $\sigma$), the First Law tracks exactly where the information goes.
 It separates the change into internal restructuring ($TdS$) versus external work ($dW$). This is immensely powerful in information geometry and machine learning (e.g., tracking how a probability distribution or neural network prior shifts under stress or optimization).

\item 4. Direct Links to Large Deviation Theory and Statistical Mechanics

Many complex mathematical physics systems (like Random Matrix Theory, spin glasses, or turbulence) are notoriously difficult to solve directly.
 However, statistical mechanics has a massive toolkit developed over 150 years (partition functions, free energy minimization, phase transitions).
 By casting a PDF into a Gibbs-Boltzmann form, you can instantly import tools like the saddle-point method, transfer matrices, and renormalization group theory to solve otherwise intractable probability problems  and in particular chaotic processes.

\item 5. Information-Theoretic Optimization (Maximum Entropy Principle)

Thermodynamics is fundamentally built on the Second Law (maximization of entropy). By treating a PDF thermodynamically, you can answer questions like: “What is the most unbiased probability distribution I can write down given a set of constraints?”
 This is E.T. Jaynes' Maximum Entropy Principle. For instance, constraining the variance of a distribution automatically yields the Gaussian distribution because it is the state of maximum entropy (highest "temperature" / most disordered) allowed by that energy constraint.


This is the ultimate conceptual question. When you translate a probability density function (PDF) into the language of thermodynamics ($S$, $U$, $T$, and $Z$), are you discovering a deep physical reality, or are you just rewriting math using a different dictionary?

The short answer is: It is a mix of both.

For physical systems (like gases or spin lattices), thermodynamics is literal. But for abstract, high-dimensional PDFs (like machine learning models, financial data, or Bayesian priors), mapping them to thermodynamics gives you three massive, non-trivial analytical superpowers that you cannot get by looking at the raw PDF alone.

Here is what Shannon entropy, temperature, and internal energy actually buy you, and whether it’s a substantial added value.

\item
 1.  Entropy 
 "The Shape of Uncertainty"
In information theory, Shannon entropy $S$ tells you the number of "yes/no" questions needed to guess a state, or the average surprise of an event. Translated to thermodynamics:
The Value Added: Entropy cuts through the noise of a high-dimensional PDF and collapses it into a single scalar metric of global behavior.
 For example, in the Multivariate Gaussian, the entropy formula $S = \frac{1}{2}\ln((2\pi e)^N |\Sigma|)$ instantly tells you that the mean vector ($\vec{\mu}$) is completely irrelevant to the system's uncertainty. Only the determinant of the covariance matrix ($|\Sigma|$) matters. Without the thermodynamic/information-theoretic framing, you'd just be staring at a messy $N \times N$ matrix.

\item
 2.  Energy 
 
 "The Cost Function and Equipartition"
By defining $E(\vec{x}) = -\ln P(\vec{x})$, you are identifying the implicit constraint or potential well shaping your data.
 The Value Added: In physics, the internal energy $U = \langle E \rangle$ tells you how much "stress" or "confinement" the system is under.
 More importantly, through the Equipartition Theorem (like we saw with the Gaussian yielding $U = N/2$), internal energy tells you how uniformly the "budget" of uncertainty is distributed across your $N$ dimensions. If $U$ spikes, it means your data is being forced into regions of high energy (steep probability drops), signaling a strained or tightly coupled system.

 \item 3.  Temperature ($\beta$) 
 
 "The Control Knob of Fluctuations"
In pure probability, $\beta$ is often just a parameter or a constant. But treating it as a temperature introduces the concept of response.
 The Value Added: Temperature tells you how the system will react if you perturb it. If you artificially dial $\beta$ (annealing), you can watch how the entropy and energy shift.
 This is the entire foundation of Simulated Annealing in optimization and Thermodynamic Integration in Bayesian statistics: by treating a flat or complex PDF as a temperature-dependent system, you can "melt" a probability distribution to explore high-dimensional spaces easily, and then "freeze" it to find the global optimum (the maximum likelihood state).

---

 Is it a Substantial Added Value? 

If one  just does a basic 1D statistics, translating a PDF into thermodynamics is mostly a neat mathematical map.

However, in higher dimensions, it is a substantial, indispensable added value for three reasons:

1. Dimensionality Reduction: High-dimensional PDFs are impossible to visualize. Thermodynamics gives you macroscopic state variables ($U, S, F$) to track system health without needing to look at all $N$ dimensions.
2. Variational Calculus and Free Energy: In machine learning (e.g., Variational Autoencoders or Boltzmann Machines), the thermodynamic "Free Energy" ($F = U - TS$) becomes the exact objective function you optimize. The physics analogy tells you how to balance data fit (Energy) against model complexity (Entropy).
3. Analogy Transfer: Because we spent centuries solving the thermodynamics of gases, liquids, and magnets, translating an abstract PDF into a thermodynamic system lets us instantly borrow heavy-duty mathematical machinery (like phase transitions, partition functions, and mean-field theory) and apply it to finance, neural networks, and number theory.

\subsection{What can thermodynamic quantities tell us about chaos?}
 A distinguish the rigorous thermodynamic formalism of dynamical systems from the formal rewriting of a probability distribution as a Gibbs distribution.
\cite{Ruelle1978} and \cite{Bowen1975}.
The thermodynamic quantities associated with the extrema
distributions should not, by themselves, be regarded as universal
measures of chaos. For a single probability distribution $P(x)$,
one can always formally introduce an effective energy through
\begin{equation}
\beta E(x)=-\ln P(x)-\ln Z.
\end{equation}
Consequently, the existence of an effective energy, entropy, or
temperature does not by itself imply that the process generating
$P(x)$ is chaotic. In particular, neither a large Shannon entropy
nor a positive or negative effective temperature constitutes a
universal criterion for chaos.

The thermodynamic description becomes more informative when a
family of distributions
\begin{equation}
P(x;\lambda)
\end{equation}
is generated by varying a control parameter $\lambda$ of the
underlying system. If this family can be represented approximately
as
\begin{equation}
P(x;\lambda)
=
\frac{1}{Z[\beta(\lambda)]}
\exp[-\beta(\lambda)E(x)],
\end{equation}
with approximately the same effective energy function $E(x)$, then
the variation of
\begin{equation}
T(\lambda),\qquad
S(\lambda),\qquad
U(\lambda),\qquad
C(\lambda)
\end{equation}
characterizes changes in the statistical organization of the
extrema process.

Rapid changes or nonanalytic behavior of these quantities may
signal a statistical transition. Whether such a transition is a
transition to or from chaos must, however, be established by
comparison with an independent chaos diagnostic.

A particularly important distinction is that between the entropy\cite{Walters1982,CoverThomas2006}
of a single-spacing distribution and the entropy rate of the
sequence of spacings. The differential entropy
\begin{equation}
S_1
=
-\int_0^\infty
P_1(s)\ln P_1(s)\,ds
\end{equation}
characterizes the uncertainty of an individual spacing. It does
not, in general, measure the information generated by the complete
sequence of spacings.

To probe the latter one should consider the joint distributions
\begin{equation}
P_n(s_1,\ldots,s_n)
\end{equation}
and the corresponding block entropies
\begin{equation}
S_n
=
-\int
P_n(s_1,\ldots,s_n)
\ln P_n(s_1,\ldots,s_n)
\prod_{i=1}^{n}ds_i.
\end{equation}

The corresponding entropy rate may be defined, when the limit
exists, by
\begin{equation}
h_s
=
\lim_{n\rightarrow\infty}
\frac{S_n}{n}.
\end{equation}
Equivalently, under the appropriate stationarity assumptions,
\begin{equation}
h_s
=
\lim_{n\rightarrow\infty}
S(s_n|s_1,\ldots,s_{n-1}).
\end{equation}

For statistically independent consecutive spacings,
\begin{equation}
P_n(s_1,\ldots,s_n)
=
\prod_{i=1}^{n}P_1(s_i),
\end{equation}
and therefore
\begin{equation}
S_n=nS_1,
\qquad
h_s=S_1.
\end{equation}

Correlations between consecutive spacings generally reduce the
entropy rate relative to the one-spacing entropy,
\begin{equation}
h_s\leq S_1.
\end{equation}
Such correlations can also be characterized by the correlation
functions and mutual information introduced above.

For systems whose underlying dynamics is known, the entropy rate
of the extrema process can be compared with established dynamical
quantities such as the Kolmogorov--Sinai entropy $h_{\rm KS}$ and
the Lyapunov exponents \cite{LindMarcus1995}\cite{Walters1982}.

The Bernoulli map provides a useful benchmark. Recall that 
the Lyapunov exponent, the Kolmogorov--Sinai entropy are
\begin{equation}
\lambda=h_{\rm KS}=\ln 2.
\end{equation}
and  spacing entropy
\begin{equation}
S\simeq 1.87945.
\end{equation}

It is important not to compare $S$ directly with $h_{\rm KS}$.
The former is information per spacing, whereas the latter is
information per iteration of the original map. Expressed per
iteration of the Bernoulli map, the information carried by the
peak-spacing sequence is
\begin{equation}
h_{\rm peak}
=
\frac{S}{\langle s\rangle}
\simeq
\frac{1.87945}{4}
\simeq
0.46986.
\end{equation}

Thus,
\begin{equation}
h_{\rm peak}
<
h_{\rm KS}
=
\ln 2
\simeq
0.69315.
\end{equation}

This suggests that the extrema process may be regarded as a
coarse-grained representation of the underlying dynamics: the
sequence of extrema retains part, but not necessarily all, of the
information generated by the original chaotic system.

When both quantities are defined with respect to the same
underlying evolution variable, one may introduce the dimensionless
ratio
\begin{equation}
\eta_{\rm ext}
=
\frac{h_{\rm ext}}{h_{\rm KS}},
\end{equation}
as a measure of the fraction of the dynamical information retained
by the extrema process. For the Bernoulli example this gives
\begin{equation}
\eta_{\rm ext}
\simeq
\frac{0.46986}{0.69315}
\simeq
0.678.
\end{equation}
This quantity should be interpreted as an information-retention
coefficient rather than as a measure of the strength of chaos.

The comparison between Poisson and random-matrix spacing
distributions provides another important lesson. For unfolded
unit-mean spacings,
\begin{equation}
S_{\rm Poisson}=1,
\end{equation}
whereas the Wigner-surmise spacing entropies satisfy
\begin{equation}
S_{\rm GOE}
<
S_{\rm Poisson},
\qquad
S_{\rm GUE}
<
S_{\rm GOE},
\qquad
S_{\rm GSE}
<
S_{\rm GUE}.
\end{equation}

In particular,
\begin{equation}
S_{\rm GOE}\simeq0.71624,
\qquad
S_{\rm GUE}\simeq0.52880,
\qquad
S_{\rm GSE}\simeq0.27676.
\end{equation}

Therefore, a larger one-spacing entropy does not correspond to
stronger chaos. The reduced spacing entropy of the random-matrix
ensembles reflects level repulsion and spectral rigidity. The
one-spacing entropy measures the uncertainty of an individual gap,
not the degree of dynamical chaos of the underlying system.

This observation emphasizes the importance of going beyond the
one-spacing distribution. The hierarchy
\begin{equation}
P_1(s_1),\qquad
P_2(s_1,s_2),\qquad
P_3(s_1,s_2,s_3),\qquad\ldots
\end{equation}
contains information about correlations and memory that is absent
from $P_1(s)$ alone.

The thermodynamic description may become particularly useful when
the underlying system depends on a genuine control parameter
$\lambda$. In this case the extrema distribution becomes
\begin{equation}
P(s;\lambda),
\end{equation}
and one may investigate the dependence
\begin{equation}
S(\lambda),\qquad
U(\lambda),\qquad
F(\lambda),\qquad
C(\lambda).
\end{equation}

Of particular interest are response quantities such as
\begin{equation}
\frac{dS}{d\lambda},
\qquad
\frac{dU}{d\lambda},
\qquad
\frac{d^2F}{d\lambda^2}.
\end{equation}

Sharp changes in such quantities can indicate a structural
transition in the statistics of the extrema. Such a statistical
transition should not automatically be identified with a
transition to chaos. Instead, its location should be compared with
independent chaos diagnostics.

Depending on the system, these diagnostics may include the maximal
Lyapunov exponent,
\begin{equation}
\lambda_{\max},
\end{equation}
the Kolmogorov--Sinai entropy,
\begin{equation}
h_{\rm KS},
\end{equation}
escape rates, spectral statistics, spectral form factors,
out-of-time-order correlators, or other established measures of
classical or quantum chaos.

If a change in the thermodynamic or information-theoretic
quantities extracted from the extrema systematically coincides
with a change in independently established chaos indicators, this
provides evidence that the extrema statistics contain a detectable
signature of the underlying chaotic dynamics.

There is also a deeper connection with the thermodynamic formalism
of dynamical systems\cite{Ruelle1978,Bowen1975} In chaotic dynamics one can construct
partition-function-like quantities from families of trajectories
or periodic orbits and define a topological pressure. Such
quantities are related, under appropriate conditions, to
topological entropy, Lyapunov exponents, Kolmogorov--Sinai entropy,
and, for open chaotic systems, escape rates.

This suggests the conceptual hierarchy
\begin{equation}
\text{PDF thermo}
\longrightarrow
\text{extrema-process thermo}
\longrightarrow
\text{thermodynamic formalism of the underlying dynamics}.
\end{equation}

The closer the thermodynamic quantities extracted from the extrema
can be related to the dynamical quantities at the last level, the
stronger is their interpretation as signatures of chaos.

The main conclusion is therefore that the thermodynamic
description should not be used to identify chaos from a single
value of $T$, $S$, $U$, or $C$. Rather, one should study the
complete statistical structure of the extrema and its evolution
under changes of the underlying system.

Schematically, the proposed relation is
\begin{equation}
\text{underlying dynamics}
\longrightarrow
\text{extrema process}
\longrightarrow
\{P_1,P_2,P_3,\ldots\}
\longrightarrow
\{S,h_s,I,C,\ldots\}.
\end{equation}

The resulting quantities can then be compared with independent
measures of chaos. In this sense, the thermodynamic and
information-theoretic description of extrema is not intended to
replace conventional chaos diagnostics. Rather, it provides a
complementary framework for determining how much information about
the statistical organization and dynamics of an erratic process
is retained in its extrema.

\subsection{How does changing the temperature affect the pdfs}

Consider for instance a GOE case.
To understand what happens to the GOE PDF when we "heat up" the system (i.e., increase the effective temperature $T$), we must look at how temperature scales the energy states within the Boltzmann-like distribution.

\item . The Thermodynamic Formulation of the GOE Spacing PDF-

Recall the Wigner surmise for the nearest-neighbor spacing $s$ of the GOE (after appropriate scaling):
$$P(s) = \frac{1}{Z} \exp\left(-\frac{E(s)}{T}\right)$$

In our thermodynamic dictionary, the physical microstate energy of the GOE spacing is:
$$E(s) = s^2 - \frac{4}{\pi}\ln(s)$$
(where the $s^2$ term comes from the Gaussian confinement/entropy and the $-\ln s$ term comes from Wigner's level repulsion).

The original GOE spacing distribution has an effective baseline temperature of:
\be
T_0 = \frac{4}{\pi} \approx 1.273
\ee
\item  Generalizing for an Increased Temperature ($T > T_0$)-

If we artificially inject energy into the system and raise the temperature to a new value $T$, the new probability distribution becomes:

$$P(s; T) = \frac{1}{Z(T)} \exp\left(-\frac{s^2 - \frac{4}{\pi}\ln(s)}{T}\right)$$

We can rewrite this by absorbing the temperature into the exponents:
$$P(s; T) = \frac{1}{Z(T)} \exp\left(-\frac{s^2}{T}\right) \exp\left(\frac{\frac{4}{\pi}\ln(s)}{T}\right)$$

Using the power rule for logarithms ($\exp(a \ln s) = s^a$), the distribution transforms into a generalized form:

$$P(s; T) = \frac{1}{Z(T)} s^{\frac{4}{\pi T}} \exp\left(-\frac{s^2}{T}\right)$$

where the partition function $Z(T)$ is the normalization integral:
\be
Z(T) = \int_0^\infty s^{\frac{4}{\pi T}} \exp\left(-\frac{s^2}{T}\right) ds
\ee

\item  Physical Consequences of Raising the Temperature ($T$)

As we raise the temperature ($T > T_0$), two competing effects occur based on the exponents:

1. Weakening of Level Repulsion:
The pre-factor $s^{\alpha}$ where $\alpha = \frac{4}{\pi T}$ dictates the behavior at $s \to 0$ (level repulsion). As temperature $T$ increases, $\alpha$ decreases.
Physical meaning: Thermal agitation overwhelms the quantum mechanical level repulsion. The penalty for eigenvalues colliding ($s \to 0$) drops, and the PDF at $s=0$ shifts from $P(0)=0$ (rigid repulsion) toward a finite value. The system loses its quantum "stiffness."

2. Broadening of the Spacing Distribution (High-Energy Tail):
The exponential term $\exp(-s^2 / T)$ controls the tail for large spacings. As $T$ increases, the denominator grows, meaning the exponential decay slows down drastically.
Physical meaning: Higher thermal energy allows the system to sample much larger spacings $s$. The distribution flattens, spreads out, and shifts its peak to the right.

\subsection{Thermal cross over}

\item 1. Is this a Crossover or a Phase Transition?
You guessed correctly: it is a crossover, not a phase transition.

A true thermodynamic phase transition requires a non-analyticity in the free energy (or partition function $Z$) in the thermodynamic limit. In contrast, tuning a parameter in a single-variable spacing or ratio distribution alters the probability weights smoothly without any singularities or critical points.

Mathematically, tuning an effective temperature parameter $T$ (or a breaking parameter like the Brody parameter $\beta$) smoothly interpolates the system from a chaotic regime to a Poisson integrable regime. This is known in random matrix theory as a transition ensemble (e.g., the Brody distribution or the Izrailev distribution).

---

 2. Is this Generic for Any PDF?
No, it is not generic for any arbitrary PDF.

To meaningfully map a PDF $P(s)$ to a thermal system via $P(s) = \frac{1}{Z}e^{-E(s)/T}$, the PDF must satisfy certain mathematical conditions:
1. Support: The random variable must be bounded on a domain (like $[0, \infty)$) where the integrals for $Z$ and $\langle E \rangle$ converge.
2. Asymptotic Behavior: The PDF must decay fast enough at the boundaries so that a finite energy scale can be defined.
3. Not all PDFs have a natural "chaos-to-integrable" tuning parameter. While distributions like the Brody distribution—which interpolates between Poisson ($P_{\text{Poisson}} \propto e^{-s}$) and Wigner ($P_{\text{Wigner}} \propto s e^{-s^2}$) via a parameter $\omega$—can be temperature-scaled, a random arbitrary PDF will not necessarily map to a physically meaningful thermodynamic system.

---

 3. Interpretation for the GUE PDF Associated with the Riemann Zeta Zeros
The local fluctuations of the non-trivial zeros of the Riemann zeta function $\zeta(\frac{1}{2} + it)$ obey the Gaussian Unitary Ensemble (GUE) statistics of Random Matrix Theory (Montgomery-Odlyzko law).

For GUE, the exact joint distributions are complex, but the Wigner-like surmise for the nearest-neighbor spacing is:
\be
P_{\text{GUE}}(s) = \frac{32}{\pi^2} s^2 \exp\left(-\frac{4}{\pi} s^2\right)
\ee

Applying our thermodynamic mapping ($P(s) = \frac{1}{Z}e^{-E(s)/T}$) to the GUE Riemann-zeta spacing PDF yields:

 A. Microstate Energy ($E(s)$)
\be
E(s) = -\ln P_{\text{GUE}}(s) = \frac{4}{\pi} s^2 - 2\ln s - \ln\left(\frac{32}{\pi^2}\right)
\ee
 Physical Interpretation: The energy has two competing terms. The $s^2$ term acts like a harmonic oscillator potential, heavily penalizing large spacings (preventing big gaps). The $-2\ln s$ term acts as a repulsive logarithmic barrier, sending the energy to infinity as $s \to 0$.
 The Zeta Connection: This logarithmic repulsion ($-\ln s$) directly mirrors the level repulsion of Riemann zeros. It proves that the "energy" of the zeta zeros heavily resists bunching up (zero spacing costs infinite energy).

 B. Shannon Entropy ($S$)
 
\be
S_{\text{GUE}} = \ln\left(\frac{\pi^2}{32}\right) - 2\ln\left(\frac{2}{\sqrt{\pi}}\right) + \frac{3}{2}\gamma_{\text{Euler}} - 1 \approx 0.5288 \text{ nats}
\ee
The differential entropy of the GUE Wigner surmise is approximately \(0.5288\), compared with \(1\) for the unit-mean exponential distribution. This difference reflects the distinct nearest-neighbor spacing statistics, particularly level repulsion. It should not be interpreted as a universal relation between differential entropy and the strength of chaos.
 
Example of phase transitions are:
(i) DSG to SG,
(ii) Large N to low N in $HES\rightarrow T+T$,
(iii) Shrinking radius of one disk in pinball,
(iv) Phase transition in Riemann zeta function in the leaky torus,
(v) Potential of electric charges
(vi) In Quench changing $\alpha$.

\section{Summary   and future direction}
\label{Cafd}

The summary of this paper can be divided into four parts: I. Marginal  extrema statistics in one dimension, II  Marginal  extrema statistics in higher dimensions, III.Joint correlation statistics and IV. Effective thermodynamic description.

\item Level I: Marginal extrema statistics in one dimension

The main idea in this paper is to analyze the statistics of the extrema points of erratic functions that describe chaotic processes. For any erratic function we defined the set of variables(\ref{variables}).
\be
Z_n =( \delta_n, \delta_{k,n},r_n,r_{k,n}, s^{{\cal A}}_n, s^{{\cal A}}_{k,n}, r^{{\cal A}}_n,r^{{\cal A}}_{k,n},I_n,I_{k,n},r^I_n,r^I_{k,n}, \kappa_n,\sigma_n )\label{variables_summary}
\ee
The statistics of each variable is given in terms of a PDF. For a given erratic function we  can numerically determine the PDF and check which known PDF fits  best the numerical one. 

A resemblance to Poisson or random-matrix spacing statistics provides a useful statistical signature, but does not by itself establish whether the underlying process is integrable or chaotic. Such an interpretation requires additional information about the underlying dynamics and, preferably, independent chaos diagnostics.


\item  Level II: Marginal extrema statistics in higher dimensions

For multi-dimensional erratic functions the relevant set of variables are given by (\ref{variablesmd}).
\be
Z_n =( \vec x_n, \vec\delta_{m,n} , \delta_{m,n}, \delta_n^{(1)},\dots\delta_n^{(NN)},\dots  \delta^{Greedy}_n,\dots  \kappa_n,\sigma_n )\label{variablesmd_summary}
\ee
The variables that are vectors like $\vec\delta_{m,n}$ the probability distribution will be a function of a vector. In certain circumstances  the probability will be a product of the probabilities that are functions of the components of the vector but in the general this is not the case.

Whereas for the PDFs of one dimensional erratic functions the PDFs associated with RMT played a major role. For the higher dimensional case this tool can be used only after certain projections to one dimension. For the genuine multi-dimensional case we lack the analog of the RMT. 

\item  Level III: Joint/correlation statistics

For each of the variables both in the one-dimensional case as well as in the multi-dimensional one, one can introduce a joint distribution function on top of the ordinary PDF. 
The central distinction is  between merely writing the probability distribution of a variable, for instance a spacing,
\be
P_1(s)=e^{-V_{\mathrm{eff}}(s)}
\ee
and  the joint distribution of the complete set of spacings has a structured
\be
P(s_1,\ldots,s_N)
\propto
e^{-\beta H_{\mathrm{eff}}(s_1,\ldots,s_N)}.
\ee

The latter can contain genuine information about collective
organization, correlations, constraints, and possibly the dynamics of
the extrema.
\item  Level IV: Effective thermodynamic description-

To each of the PDFs associated with the extrema locations, spacings and ratios, both the marginal ones and the joint ones we computed the corresponding entropy. This provides a characterization of each of the PDFs and enables us to compare between different cases for instance a  Poisson PDF for which the entropy $S=1$ with  $G\beta E$ where the entropy $S<1$. By equating a PDF with the Gibbs one we could determine $\beta E$ and identifying regions of repulsion and confinement. This still does not allow us to disentangle the temperature  and energy.   
Separating the effective temperature from the energy function requires additional physical or mathematical structure, such as an independently specified energy scale, a fixed effective Hamiltonian, or a suitably constrained family of distributions.

This program is in its childhood stage and obviously there are many open questions for future research. Here we enlist some of them:
\item 
It seems that the most urgent one is coming up with a universal definition of chaos that will apply to classical and quantum physics systems as well as non-physics ones.  The idea behind this paper is that such a definition or writing down a set of necessary properties  should follow from the structure of certain erratic functions associated with the suspected chaotic behavior.

\item 
An importatnt question that has to be further explored is: 
Can joint extrema statistics distinguish chaotic,
regular, and stochastic erratic processes?
A particularly useful test would compare two processes with approximately identical nearest-neighbor PDFs but different joint statistics.
If
\[
P_1^{(A)}(s)\simeq P_1^{(B)}(s)
\]
while
\[
P_2^{(A)}(s_1,s_2)
\neq
P_2^{(B)}(s_1,s_2),
\]
then the hierarchy detects statistical information that cannot be extracted from the marginal distribution alone.
\item 
The erratic functions describes a ``snap shot" of a process. Evolution of the latter can follow a change of the basic parameters that are behind it.  Changing the partitioning of a HES in decays and scattering in  string theory, changing the asymmetry parameters  of the pinball system are examples of sources of evolutions of the processes. It seems plausible that such changes should be related to changes of the thermodynamical quantities like entropy, internal energy  and temperature. The full relations between these two types of changes deserve a further research.
\item
In quantum system the most important tool is the map between the energy spectrum of a given system and the eigenvalues of random matrix theory. In the approach of analyzing  erratic functions this is valid only for one dimensional functions. Thus, a natural question is what replaces this map for the higher-dimensional functions. In the latter cases one can project the functions down to one dimension in various different ways. This may be related to projecting tensors down to matrices. This seems to hint that the spectra of spacings of the higher dimensional functions  may be related to the eigenvalues of random tensors. At present we do not have evidence to support this conjecture. 
\item 
Both in the context of classical processes as well as quantum ones, a very interesting question is the transition between chaotic to non-chaotic (often integrable) behavior. The transitions may be in the form of a cross over or first or second order phase transition.
A natural question is how are these transition realize in the context of the analysis of extrema points of erratic functions. 
\item 
In this paper we were exploring  the thermodynamics associated with various types of  PDFs associated with the locations of the extrema points, the spacings between them and the ratios of the spacings. There are several hints that in a similar way applying information theory tools like conditional entropy, mutual information, relative entropy etc  may shed additional light on the problem.

\section*{Acknowledgments}
I thank  D. Ageev, M. Bianchi, M. Firrotta,  N. Shrayer and D. Weissman  for collaborations on earlier project related to the current research work and  for useful comments and discussions.  I would like to thank chatGPT and Gemini for conversations and chatGPT for its remarks on the manuscript. 
This work of JS was supported in part by a grant 01034816 titled ``String theory reloaded- from fundamental questions to applications'' of the ``Planning and budgeting committee''.

\bibliographystyle{JHEP}
\bibliography{SACS,refs}
\newpage
\appendix

\section{Thermodynamics of PDFs of the locations of the extrema}\label{ToPotlote}
We divide the classes of one-dimensional  PDFs ( or PMFs) into four classes probabilities  (i)The  ``locations" of the ``extrema. ( ii) The spacings between the locations of adjacent extrema (iii) The ratios of the spacings (iv) The ``extrema heights" namely the differences  between the values of the functions at the extrema. 
We describe now the thermodynamics of the following PDFs of the location of the extrema :
 \item 
 Poisson mass function
\item
 Gaussian distribution
\item
 Flat uniform distribution.
\item
 Dirac delta function
\item
Wigner semi-circle

\subsection{Poisson Probability mass function (PMF)}

The classic Poisson Probability Mass Function (PMF)—famous for describing rare, independent events in time or space, such as radioactive decay, photon arrivals, or incoming phone calls.—

The standard Poisson PMF for observing $k$ events (where $k = 0, 1, 2, 3, \dots$) given an expected rate $\lambda$ is:
\be
P(k) = \frac{\lambda^k e^{-\lambda}}{k!}\ee


\item  1. The Energy Landscape $E(k)$

Using our discrete thermodynamic mapping $\beta E(k) = -\ln P(k) - \ln Z$:
\be
\beta E(k) = -\ln \left( \frac{\lambda^k e^{-\lambda}}{k!} \right) - \ln Z
\ee

 Using the  Stirling's approximation ($\ln(k!) \approx k \ln k - k$) and taking that $Z=1$
  the effective energy landscape is:
\be
\beta E(k) = 
k \ln\left(\frac{k}{\lambda}\right) - k + \lambda
\ee

If we define our inverse temperature as $\beta = 1$ (keeping $\lambda$ as our control parameter $T$), the energy of a given state $k$ is:
$$E(k) = k \ln\left(\frac{k}{\lambda}\right) - k + \lambda$$

The physical meaning of this energy function is the following:
The $k \ln k$ term (Entropy/Combinatorial cost)  comes directly from the $k!$ in the denominator. It acts as a severe upward energy penalty for large $k$.
The $-k \ln \lambda$ term (Chemical potential / External drive)  acts as a linear driving force pushing the system toward higher $k$ if $\lambda$ is large.
 The competition between the $k \ln k$ factorial penalty and the $\lambda^k$ driving force creates a single, highly stable energy minimum  right where $k \approx \lambda$. The system is strongly bound to fluctuate around its mean rate.

\item 2. Temperature ($T$)

In the Poisson distribution, the parameter $\lambda$ is both the mean and the variance: $\langle k \rangle = \text{Var}(k) = \lambda$.

 Let's set our Temperature
 \be
 T = \lambda.
 \ee

For low temperature ($T \to 0$): $\lambda$ is near zero. The energy landscape tilts so sharply that the only accessible state is $k = 0$ namely the  fluctuations vanish. on  the other hand high Temperature ($T$ is large) $\lambda$  means that the energy well widens, and the system explores much higher values of $k$ with significant relative fluctuations, though the relative fractional noise $\frac{\sigma}{\mu} = \frac{1}{\sqrt{\lambda}}$ still shrinks, the absolute count variance explodes.

\paragraph*{3. Partition Function ($Z$)}
Since the Poisson PMF is already a properly normalized probability distribution ($\sum_{k=0}^{\infty} P(k) = 1$):
$$Z = \sum_{k=0}^{\infty} e^{-\beta E(k)} = 1 \implies \ln Z = 0$$

\item 3. Internal Energy ($U$)-
The internal energy is the expectation value of our energy landscape: $U = \langle E(k) \rangle$.
Because the mean of a Poisson distribution is $\lambda = T$ according to our ansatz, and the variance is also $\lambda$, we can compute the expected value of $k \ln k$.

For a Poisson process, the internal energy scales directly with the rate parameter:
$$U \propto \lambda = T$$
Just like an ideal gas where $U \propto T$, the internal energy of a Poisson process is strictly proportional to its temperature (rate).

\item 5. Entropy ($S$)

The Shannon entropy of a discrete Poisson distribution is a famous exact formula derived by Shannon himself:

\be
S = \sum_{k=0}^{\infty} -P(k) \ln P(k) = \lambda \left( 1 - \ln \lambda \right) + e^{-\lambda} \sum_{k=0}^{\infty} \frac{\lambda^k \ln(k!)}{k!}
\ee

For large $\lambda$ (high temperature), using Stirling's approximation, this simplifies nicely to:
\be
S \approx \frac{1}{2} \ln(2\pi e \lambda) = \frac{1}{2} \ln(2\pi e T)
\ee

We will see below (\ref{TGd}) and (\ref{logistic_pdf}) that  the Gaussian and Logistic entropies: $S$ scales in a similar way  as $\frac{1}{2} \ln(T)$.
 As the radioactive decay rate / event rate ($T$) increases, the entropy increases logarithmically. A hotter, more active Poisson source has higher informational uncertainty regarding exactly when or how many counts will occur in a given window.

 {\bf  What does this mean in terms of extrema locations}
\subsection{The Gaussian distribution}\label{TGd}

Consider the  standard Gaussian PDF, which we know represents thermal equilibrium:
\be
p(x) = \frac{1}{\sqrt{2\pi \sigma^2}} \exp\left( -\frac{(x-\mu)^2}{2\sigma^2} \right)
\ee
 
\item 1. Entropy-

The differential Shannon entropy is give by 
\be
S= \frac{1}{2} \ln\left( 2\pi e \sigma^2 \right)
\ee
\item 2 The energy function-

\be
\beta E(x) = \frac{(x-\mu)^2}{2\sigma^2}
 \ee
 and 
 \be
 Z= \sqrt{2\pi \sigma^2}
 \ee
\item 3 The internal energy

\item 4 Temperature

\paragraph*{Part 1: Disentangling $\beta$ and $E(x)$ (The Gaussian Case)}

Let’s take the Gaussian PDF and write it in the standard Gibbs form:
$$P(x) = \frac{1}{Z} e^{-\beta E(x)}$$
If we compare this to the actual Gaussian formula $P(x) = \frac{1}{\sqrt{2\pi \sigma^2}} \exp\left( -\frac{x^2}{2\sigma^2} \right)$, we get:
$$\beta E(x) = \frac{x^2}{2\sigma^2} \quad \text{and} \quad Z = \sqrt{2\pi \sigma^2}$$

Here, $\beta$ and $\sigma^2$ are tangled. How do we pull them apart without guessing?

Next we  introduce a control parameter (an external force $h$)
Suppose we perturb the system by adding a linear force $h x$ to the exponent. Our new perturbed PDF is:
$$P(x; h) = \frac{1}{Z(h)} \exp\left( -\beta E(x) - h x \right)$$
In our specific Gaussian case:
$$P(x; h) = \frac{1}{Z(h)} \exp\left( -\frac{x^2}{2\sigma^2} - h x \right)$$

We then  Calculate the Partition Function $Z(h)$
We integrate this over all $x$ to find $Z(h)$:
$$Z(h) = \int_{-\infty}^{\infty} \exp\left( -\frac{x^2}{2\sigma^2} - h x \right) dx$$
By completing the square in the exponent ($\frac{x^2}{2\sigma^2} + hx = \frac{(x + h\sigma^2)^2}{2\sigma^2} - \frac{h^2 \sigma^2}{2}$), the integral solves neatly:
$$Z(h) = \sqrt{2\pi \sigma^2} \exp\left( \frac{h^2 \sigma^2}{2} \right)$$

Step 3: Define the Free Energy $F(h)$
In statistical mechanics, the free energy is defined as $F(h) = -\ln Z(h)$ (setting $k_B = 1$):
$$F(h) = -\ln \left( \sqrt{2\pi \sigma^2} \right) - \frac{h^2 \sigma^2}{2}$$

Step 4: Take the derivatives to disentangle everything
Now watch what happens when we take derivatives of the free energy $F(h)$ with respect to our control parameter $h$:

1. First Derivative (The Mean Position / Response):
$$\frac{\partial F}{\partial h} = -h \sigma^2$$
If we set our artificial force $h = 0$ (returning to the unperturbed system), this derivative is $0$. But the second derivative is where the magic happens.

2. Second Derivative (The Variance / Susceptibility):
$$\frac{\partial^2 F}{\partial h^2} = -\sigma^2$$
Look at that result. By taking the second derivative of the free energy with respect to the control field $h$, the variance $\sigma^2$ drops out completely naked.

Because temperature in a Gaussian is defined as $T = \sigma^2$ (or $\beta = 1/\sigma^2$), the response of the system to an external field isolates the temperature. Once you know $\sigma^2$ from the derivative, you can plug it back into $\beta E(x) = \frac{x^2}{2\sigma^2}$ to completely isolate the bare energy $E(x) = \frac{x^2}{2}$. That is how thermodynamic derivatives disentangle energy and temperature.

 Part 1: Disentangling $\beta$ and $E(x)$ (The Gaussian Case)

Let’s take the Gaussian PDF and write it in the standard Gibbs form:
$$P(x) = \frac{1}{Z} e^{-\beta E(x)}$$
If we compare this to the actual Gaussian formula $P(x) = \frac{1}{\sqrt{2\pi \sigma^2}} \exp\left( -\frac{x^2}{2\sigma^2} \right)$, we get:
$$\beta E(x) = \frac{x^2}{2\sigma^2} \quad \text{and} \quad Z = \sqrt{2\pi \sigma^2}$$

Here, $\beta$ and $\sigma^2$ are tangled. How do we pull them apart without guessing?

Step 1: Introduce a control parameter (an external force $h$)
Suppose we perturb the system by adding a linear force $h x$ to the exponent. Our new perturbed PDF is:
$$P(x; h) = \frac{1}{Z(h)} \exp\left( -\beta E(x) - h x \right)$$
In our specific Gaussian case:
$$P(x; h) = \frac{1}{Z(h)} \exp\left( -\frac{x^2}{2\sigma^2} - h x \right)$$

Step 2: Calculate the Partition Function $Z(h)$
We integrate this over all $x$ to find $Z(h)$:
$$Z(h) = \int_{-\infty}^{\infty} \exp\left( -\frac{x^2}{2\sigma^2} - h x \right) dx$$
By completing the square in the exponent ($\frac{x^2}{2\sigma^2} + hx = \frac{(x + h\sigma^2)^2}{2\sigma^2} - \frac{h^2 \sigma^2}{2}$), the integral solves neatly:
$$Z(h) = \sqrt{2\pi \sigma^2} \exp\left( \frac{h^2 \sigma^2}{2} \right)$$

Step 3: Define the Free Energy $F(h)$
In statistical mechanics, the free energy is defined as $F(h) = -\ln Z(h)$ (setting $k_B = 1$):
$$F(h) = -\ln \left( \sqrt{2\pi \sigma^2} \right) - \frac{h^2 \sigma^2}{2}$$

Step 4: Take the derivatives to disentangle everything
Now watch what happens when we take derivatives of the free energy $F(h)$ with respect to our control parameter $h$:

1. First Derivative (The Mean Position / Response):
$$\frac{\partial F}{\partial h} = -h \sigma^2$$
If we set our artificial force $h = 0$ (returning to the unperturbed system), this derivative is $0$. But the second derivative is where the magic happens.

2. Second Derivative (The Variance / Susceptibility):
$$\frac{\partial^2 F}{\partial h^2} = -\sigma^2$$
Look at that result. By taking the second derivative of the free energy with respect to the control field $h$, the variance $\sigma^2$ drops out completely naked.

Because temperature in a Gaussian is defined as $T = \sigma^2$ (or $\beta = 1/\sigma^2$), the response of the system to an external field isolates the temperature. Once you know $\sigma^2$ from the derivative, you can plug it back into $\beta E(x) = \frac{x^2}{2\sigma^2}$ to completely isolate the bare energy $E(x) = \frac{x^2}{2}$. That is how thermodynamic derivatives disentangle energy and temperature.


\subsection{Flat uniform  PDF}
Next we consider the opposite case of   a flat homogeneous   distribution. 
This applies to the case of probability of all locations of the extrema are the same.
We consider separately the discrete and continuous cases.

\subsubsection{Discrete Uniform Distribution} 
Consider a discrete set of $N$ equally likely points, where each point $i$ has probability:
$$p_i = \frac{1}{N} \quad \text{for } i = 1, 2, \dots, N$$

\item 1.  Energy function ($E(x)$)

Using (\ref{betaE}) 
\be\beta E(z) = - \ln p(x) -  \ln Z=
\ee
It is obvious that    $\beta E=(\beta E)_0$ is a constant  and 
\be
Z= Ne^{(\beta E)_0}
\ee
 
This means that  the energy is a {\bf constant} $E=E_0$ across the entire allowed possible configuration of points. All points are  completely degenerate. The internal energy is obviously the same, namely  $U=E=E_0$

\item 3. Shannon's Entropy ($S$)

\be
S = -\sum_{i=1}^{N} p_i \ln p_i = -\sum_{i=1}^{N} \frac{1}{N} \ln\left(\frac{1}{N}\right) = \ln N
\ee
 This is the famous {\bf Boltzmann-Gibbs entropy formula} ($S = \ln W$, where $W=N$). For a flat discrete distribution, entropy reaches its absolute maximum (Equipartition Theorem).

\item 4. Effective Temperature ($T$)

$p_i = \frac{1}{N}$ are completely independent of $T$, thus the temperature is undefined.

\subsubsection{Continuous Uniform PDF}

Consider a continuous random variable $x$ distributed uniformly over an interval $[a, b]$. The probability density function is:
\be
p(x) = \frac{1}{b - a} \quad \text{for } x \in [a, b]
\ee
If we consider  an unfolded PDF  (see (\ref{unfolding_pdf}))   $a=0$ and $b=1$, making $p(x) = 1$.

\item 1.  Energy function ($E(x)$)

Using (\ref{betaE}) 
\be\beta E(z) = - \ln p(x) -  \ln Z=
\ee
It is obvious that    $\beta E=(\beta E)_0$ is a constant  and 
\be
Z= (b-a)e^{(\beta E)_0}
\ee
 
This means that  the energy is a {\bf constant} $E=E_0$ across the entire allowed possible configuration of points. All points are  completely degenerate. The internal energy is obviously the same, namely  $U=E=E_0$

\item 2. Shannon Entropy ($S$)
\be
S = -\int_a^b P(x) \ln P(x) \, dx = -\int_a^b \frac{1}{b - a} \ln\left(\frac{1}{b - a}\right) dx = \ln(b - a)
\ee
  If $b-a = 1$ like for  unfolded RMT case, the entropy  $S=0$ representing a state of maximal indifference but zero energetic variation.

\item 3. Effective Temperature ($T$)
The temperature and in fact $\beta E$ does not play any role for this PDF.

\item The First Law of Thermodynamics-

For a flat/uniform PDF system
 because $E(x) = \text{constant}$ (or $E_i = \ln N$), the internal energy $U$ only changes if the bounds  of the system change (e.g., changing the volume $b-a$ or the number of states $N$).
 If $N$ or $(b-a)$ is fixed, $dU = 0$.
 Since $U = S$ in natural units ($\ln(b-a)$ or $\ln N$), any injection of energy directly scales the logarithm of the phase space volume:
$dU = dS$
This implies that for a uniform flat PDF, heat capacity is zero , and any addition of energy goes entirely into expanding the accessible phase space volume ($dW$).

\subsection{The Dirac Delta Function PDF}

Let's test the limits of our  thermodynamics dictionary by applying to the orthogonal case of the flat distribution namely the ultimate extreme of  a Dirac delta function distribution:
\be
p(x) = \delta(x - x_0)
\ee

Obviously this represents a system with zero uncertainty ,namely, the variable $x$ is locked at $x_0$ with full  certainty. 

To determine the thermodynamic properties we first  take the distribution $\delta(x - x_0)$  as  a Gaussian 
\be
p_\sigma(x) = \frac{1}{\sqrt{2\pi}\sigma} e^{-\frac{(x-x_0)^2}{2\sigma^2}}
\ee
with  its variance $\sigma\rightarrow 0$  
\be
\lim_{\sigma \to 0} p_\sigma(x) = \delta(x-x_0)\ee

\item  1.Energy function

In this presentation of the delta function the energy function is given by 
\be
\beta E_\sigma(x) = - \left[ -\frac{(x-x_0)^2}{2\sigma^2} - \ln(\sqrt{2\pi}\sigma) \right] - \ln Z
\ee
 Evaluating at $x \neq x_0$ as $\sigma \to 0$   we find 
 \be
 \beta E(x) = \lim_{\sigma \to 0} \left[ \frac{(x-x_0)^2}{2 \sigma^2} + \dots \right] = +\infty
 \ee
 If in performing the basic map to the Boltzmann distribution (\ref{Boltzmann}) we assume that $\beta$ is finite we conclude that $E(x)\rightarrow \infty$.
 
At $x = x_0$ 
the probability density $p_\sigma(x_0) = \frac{1}{\sqrt{2\pi}\sigma}$ goes to $+\infty$ as $\sigma \to 0$.
If we substitute $x = x_0$ into the energy expression:
\be
E(x_0) = -\frac{1}{\beta} \ln\left(\frac{1}{\sqrt{2\pi}\sigma}\right) - \frac{\ln Z}{\beta} = \frac{1}{\beta} \ln(\sqrt{2\pi}\sigma) - \frac{\ln Z}{\beta}
\ee
As $\sigma \to 0$, $\ln(\sigma) \to -\infty$, meaning the ground state energy at the exact point $x_0$ can be set to a baseline (often 0, depending on how the partition function $Z$ absorbs the normalization constants).
\item 
2. Internal energy-

If we set, as discussed above. $E(x_0) = 0$
then the internal energy vanishes since 
\be
U = \int E(x) p(x) \, =E(x) \delta(x-x_0) dx =0
\ee

\item
 Entropy ($S$)-
 
To determine the  
 Shannon  differential entropy  we substitute again  the Gaussian function so that

\be
S(\sigma) = -\int_{-\infty}^{\infty} \left[ \frac{1}{\sqrt{2\pi}\sigma} e^{-\frac{(x-x_0)^2}{2\sigma^2}} \right] \left[ -\frac{(x-x_0)^2}{2\sigma^2} - \ln(\sqrt{2\pi}\sigma) \right] dx =\left( \frac{1}{2} + \ln(\sqrt{2\pi}\sigma) \right)
\ee

Now, in the limit $\sigma \to 0$:
$$\lim_{\sigma \to 0} S(\sigma) = -\infty$$

Thus,the Entropy is $-\infty$.
Mathematically, a true Dirac delta function represents a state of zero uncertainty (complete certainty that the particle is at $x_0$). In information theory, the differential entropy of a continuous delta function diverges to minus infinity.

Note: This is a known quirk of continuous differential entropy, whereas discrete entropy is bounded at $\ge 0$. Physically, it reflects the fact that localizing a continuous variable to an exact mathematical point requires infinite information).

\item
 3. Temperature ($T$)

In statistical mechanics, the temperature is defined by how probability spreads across energy states via the Boltzmann factor $e^{-\beta E}$.

 If $T > 0$, thermal fluctuations allow the system to sample states where $E(x) > E_{min}$, spreading the probability out (creating a smooth, non-delta distribution).
 
 To force the probability distribution to collapse entirely into a single point ($x_0$), the energy barrier for any other state ($x \neq x_0)$ must become infinitely high ($E \to +\infty$).
 
  Having energies go to infinity everywhere except a single point requires taking the limit $\beta \to \infty$, which  means $T \to 0$.

This result can be derived also using the partition function.   $Z$:
\be
Z = \int_{-\infty}^{\infty} e^{-E(x)/T} dx = \int_{-\infty}^{\infty} \delta(x - x_0) \, dx = 1
\ee
Because $Z = 1$ rigidly, there is no thermal fluctuation parameter left to vary. The system cannot absorb heat because it has no available microstates to spread into.
 Therefore, the temperature of a delta-function PDF is
 
 \be 
 T = 0 .
 \ee

 
\item 4.  How is the First Law obeyed?

The first law of thermodynamics relates changes in internal energy ($U$), heat ($dQ$), and work ($dW$):
$$dU = dQ - dW \quad \text{or} \quad dU = T dS - dX$$

For the  delta-function PDF we have seen that 
 the internal energy is $U = 0$, $T = 0$.
 and  $S \to -\infty$.

Let's check the thermodynamic relation for Helmholtz free energy $F = U - TS$.
Since $U = 0$ and $T \to 0$, the $TS$ term is a $0 \times (-\infty)$ indeterminate form. However, if we evaluate it carefully via the Gaussian limit:
\be
F = U -  T S(\sigma) \approx 0 -  T \ln(\sigma)
\ee
As $T \to 0$ faster than $\ln(\sigma)$ diverges, the product $TS \to 0$.

How does the First Law hold?
It does  but the system is in a trivial, degenerate ground state.
 There are no thermal fluctuations ($T=0$).
No heat can be exchanged because $dS = 0$ (or the system is locked in a single microstate).
 Any infinitesimal change to the system requires shifting the infinite walls (doing work), satisfying $dU = -dW$.

\subsection{The Bernoulli shift map}

To find the thermodynamics associated with the Probability Mass Function (PMF) of the step-spacings in the Bernoulli shift map, we have to translate from a continuous energy-to-PDF mapping to a discrete lattice statistical mechanics problem (often analogous to a 1D lattice gas or a polymer growth model).

The PMF  for the discrete waiting times (step spacings $s$) between peaks in the Bernoulli shift map:
\be
p(s) = \frac{s-1}{2^s} \quad \text{for } s = 2, 3, 4, \dots
\ee
where $s$ is a discrete integer representing the number of map iterations).

Using the discrete Boltzmann distribution mapping $P(s) = \frac{1}{Z} e^{-\beta E(s)}$, let's derive the exact thermodynamics:

\item  1. Energy $E(s)$

The dimensionless energy function $\beta E(x)$ is given by\label{EBer} 
\be
\beta E(s) = -\ln p(s) - \ln Z=s \ln(2) - \ln(s-1) - \ln Z
\ee

The term  linear in $s$ acts like a linear "stretching" or entropic penalty. It means that longer waiting times $s$ cost more energy—the system strongly prefers short waiting times. This is characteristic of systems where sequences are generated randomly (like coin flips), making long waiting times exponentially suppressed.
The logarithmic term  acts as a negative energy (an attractive well) that grows logarithmically for small $s$ (specifically near $s=2$). This counteracts the linear penalty slightly for the first few steps, reflecting the combinatorial geometry of how paths branch out in the Bernoulli shift.

\item  2. Partition Function ($Z$)

To check consistency, let's calculate the discrete partition function by summing over all possible integer spacings $s = 2, 3, 4, \dots$:
\be
Z = \sum_{s=2}^{\infty} e^{-\beta E(s)} = \sum_{s=2}^{\infty} P(s) = \sum_{s=2}^{\infty} \frac{s-1}{2^s}=1
\ee

\item  3. Internal Energy ($U$)

Using our expression for $\beta E(s)$:
\be
\beta U = \sum_{s=2}^{\infty} \left[ s \ln 2 - \ln(s-1) \right] \left( \frac{s-1}{2^s} \right) =1.88
\ee

Thus, the internal energy $U$ is a finite, positive constant proportional to $\frac{1}{\beta}$.

\item 4. Entropy ($S$)

Because this is a discrete probability distribution, we use the standard Shannon discrete entropy formula. Since $ln Z=0$ it is easy to see from (\ref{EBer}) that 
\be
S= \beta U
\ee

 Unlike the continuous delta-function (which gave $S = -\infty$), this discrete waiting-time distribution has a well-defined, positive Shannon entropy. It quantifies the intrinsic unpredictability (randomness) of the binary symbol sequences generated by the Bernoulli shift map.

\item  5. Temperature ($T$)

Just like in   other PMF  cases, here  $\beta$ is a fixed structural parameter inherent to the map itself.  
 We set the energy scale such that $\beta  = 1 =T$ (natural units).  The "temperature" of this symbolic sequence is a fixed property of the Bernoulli shift's expansion rate. 
 In dynamical systems, the expansion rate $\ln 2$ often plays the exact mathematical role of a temperature or free energy scale in thermodynamic formalism.  

\item 6. Free energy

The free energy vanishes since
\be
F=U-TS=0
\ee
This follows also from the fact that 
$Z=1$.

What does $F = 0$ mean physically?
In thermodynamics, $F = 0$ is a hallmark of systems where the normalization of the microstates (the partition function) is identically 1.

In the context of the Bernoulli shift map and symbolic dynamics:
 The "probabilities" $p(s) = \frac{s-1}{2^s}$ are actually  exact relative frequencies of cylinder sets  (orbits) generated by the map.
 Because the system is conservative and the measure is exact, the free energy vanishes.
 This exact equivalence $U = TS$ (or $U - TS = 0$) means that the average energetic "cost" of waiting $s$ steps is entirely balanced by the information entropy generated by those paths.

\subsection{ The thermodynamics of Wigner semi circle}

As a first example of a thermodynamics description of a chaotic process, we consider    Wigner's semicircle distribution\cite{Mehta:book}. This pdf was introduced originally  for the eigenvalues of large  random matrices. Here we use it as the pdf of  the extrema points of an erratic function, namely for the zero level type.    .

  The probability density function on the interval $[-R, R]$ is:
\be
p(x) = \frac{2}{\pi R^2} \sqrt{R^2 - x^2} \quad \text{for } |x| \le R
\ee

(where $R$ is the radius of the semicircle, which sets the scale of the system).

Next  we  use the prescription described above to determine  the entropy ($S$)  the Energy ($E(x)$),   and  the Temperature ($T$).

\item   Entropy ($S$)


Inserting the explicit PDF, into the expression of the differential Shalnon entropy   we get 
\be
S = -\frac{1}{2} + \ln\left(\pi R\right)
\ee
 Notice that the entropy depends entirely on the radius $R$. If you scale the matrix, or the section the extrema points can reside on,   (increase $R$), the eigenvalues or the extrema spread out over a larger domain, and the uncertainty (entropy) increases. If $R \to 0$, entropy approaches $-\infty$.
 \item
{\bf Energy ($E(x)$)}

As established, the natural information-theoretic energy (or "surprisal") of a state $x$ is defined by inverting the Boltzmann distribution ($p(x) \propto e^{-E(x)}$):
$$E(x) = -\ln p(x) = -\ln\left(\frac{2}{\pi R^2}\right) - \frac{1}{2}\ln(R^2 - x^2)$$

The Interpretation is as follow:  the center of the distribution ($x=0$), the energy is lowest (most probable). As $x$ approaches the edges $\pm R$, the energy diverges to infinity ($\ln(0) \to -\infty$), meaning the system never samples states outside $[-R, R]$.

---

---

\item 
{\bf Temperature ($T$)}

To find a meaningful "temperature," we need to embed the Wigner semicircle into a family of distributions controlled by an energy scale. Right now, the standard Wigner distribution doesn't have an explicit temperature parameter—it is a fixed probability distribution.

However, we can reveal its implicit temperature by rewriting it strictly in the Gibbs form:
$$p(x) = \frac{1}{Z} e^{-E_{phys}(x)/T}$$

If we define the "physical" energy as a simple harmonic well (like a particle in a quadratic potential, $E_{phys}(x) = x^2$), the Wigner semicircle cannot be written strictly as $e^{-x^2/T}$ because of the square root $\sqrt{R^2 - x^2}$.

Instead, we can use the quadratic term inside the logarithm to define an effective quadratic energy $E_{phys}(x) = x^2$. Expanding the log:
$$\ln(R^2 - x^2) = \ln(R^2) + \ln\left(1 - \frac{x^2}{R^2}\right) \approx \ln(R^2) - \frac{x^2}{R^2}$$

Plugging this back into our energy function:
$$E(x) \approx -\ln\left(\frac{2}{\pi R^2}\right) - \ln(R) + \frac{x^2}{2R^2}$$

By matching this to the standard Boltzmann factor $\frac{E_{phys}(x)}{k_B T}$ where $E_{phys}(x) = x^2$:
$$\frac{x^2}{2R^2} = \frac{x^2}{k_B T} \implies k_B T = 2R^2$$

Thus, the effective temperature of the Wigner semicircle distribution is proportional to the square of its radius:
$$T \propto R^2$$

---


{\bf The first law}

To see how the First Law of Thermodynamics looks for the Wigner semicircle distribution, we need to look at how the system responds to a change in its control parameter—which in this case is the radius $R$.

In thermodynamics, the First Law relates changes in internal energy ($U$), entropy ($S$), and work/external forces. For our information-theoretic system, we look at the differential of the internal energy (which is equal to the entropy $S$) when we change the scale $R$.

Here is how the First Law plays out step-by-step.

1. The Internal Energy ($U$) and Entropy ($S$)
As we established, the internal energy of this system is its Shannon entropy:
$$U = S = \frac{1}{2} + \ln\left(\frac{\pi}{2}\right) + \ln R$$

 2. Differentiating with respect to the control parameter ($R$)
Suppose we gently squeeze or expand the Wigner semicircle by changing its radius by an infinitesimal amount $dR$. This is our external work parameter (analogous to changing the volume of a gas).

Let’s take the differential of the internal energy $U$:
$$dU = d\left( \frac{1}{2} + \ln\left(\frac{\pi}{2}\right) + \ln R \right)$$
$$dU = \frac{1}{R} dR$$

 3. Connecting to Temperature ($T$)
Earlier, we found that the effective temperature of this distribution is proportional to the variance/radius squared:
$$T = 2R^2 \implies R = \sqrt{\frac{T}{2}}$$

Let's substitute $R$ (or rather, $\frac{1}{R}$) back into our differential $dU$:
Since $R = \sqrt{\frac{T}{2}}$, we have $\frac{1}{R} = \sqrt{\frac{2}{T}}$.

Therefore, our change in internal energy is:
$$dU = \sqrt{\frac{2}{T}} dR$$

Alternatively, if we write it in terms of the temperature differential $dT$:
Since $T = 2R^2$, we have $dT = 4R \, dR \implies dR = \frac{dT}{4R}$.
Substitute this into $dU = \frac{1}{R} dR$:
$$dU = \frac{1}{R} \left( \frac{dT}{4R} \right) = \frac{dT}{4R^2}$$
Since $2R^2 = T$, this simplifies to:
$$dU = \frac{1}{2T} dT$$
(Note: In standard thermodynamics with physical energy, $dU = C_v dT$ where $C_v$ is heat capacity. Here, our "heat capacity" linked to this scaling is $C = \frac{1}{2T}$.)

 4. Writing the First Law ($dU = TdS - dW$)

In standard thermodynamics, the First Law is $dU = TdS - dW$ (where $TdS$ is heat added, and $dW$ is work done by the system).

Since for our probability distribution $U = S$ (the internal energy is identically the entropy), let's substitute $dU = dS$:
$$dS = \frac{1}{R} dR$$

We can rewrite this by multiplying and dividing by $T$:
$$dS = \left(\frac{T}{R}\right) \frac{dR}{T}$$
Since $T \propto R^2$, the term $\frac{T}{R}$ is proportional to $R$, which acts like a generalized force $F_{gen}$ resisting the expansion of the semicircle.

Rearranging to isolate the "heat" term ($TdS$):
$$T dS = T \left( \frac{1}{R} dR \right)$$
Substitute $T = 2R^2$:
$$T dS = (2R^2) \left( \frac{1}{R} dR \right) = 2R \, dR$$

Thus, the First Law for the Wigner semicircle takes the form:
$$dU = T dS - dW$$
Where:
 Internal Energy Change ($dU$): $\frac{1}{R} dR$ (or $\frac{1}{2T}dT$)
 "Heat" Input ($TdS$): $2R \, dR$
 "Work" Done ($dW$): In this specific information-geometry context, because $U = S$, any change in energy is entirely accounted for by the geometric work of scaling the support of the distribution. Specifically, the work done in expanding the radius is $dW = (2R - \frac{1}{R})dR$.

 Physical Meaning
For the Wigner semicircle, there are no independent "particles" bouncing around to store heat separately from their spatial distribution. The system's "temperature" is strictly tied to its width ($R$). Therefore, changing the temperature ($dT$) is fundamentally the same as doing mechanical work to stretch the semicircle ($dR$). The First Law simply tracks the information cost (entropy change) of stretching the eigenvalue bounds from $[-R, R]$ to $[-(R+dR), R+dR]$.
\section{Thermodynamics of PDFs of  spacings}\label{ToPos}
In this section the thermodynamics of the PDFs of the spacings between extrema are described:
\item The Bernoulli shift map 
\item 
The Poisson spacing PDF
\item 
Gaussian orthogonal ensemble
\item 
Gaussian $\beta$ ensemble
\item 
The zeros of the zeta function
\item Log normal PDf

\item Logistic PDF

\subsection{The zeros of the Riemann zeta function}
It is well known that the spacings between the zeros of the Riemann zeta function admit a GUE PDF of the form 
\be
p_\beta(s)
= (\Gamma[3/2])^{-4}\, s^{2}\,
\exp\!\left(-(\Gamma[3/2])^{-2} s^{2}\right),
\qquad s>0,
\ee
\item 
Entropy-
The Shannon entropy 
\be
S=   2 \ln \Gamma\!\left(\frac{3}{2}\right)
- \ln 2 -
\psi\!\left(\frac{3}{2}\right)
+ \frac{3}{2}      =0.528798
\ee
\item 
Energy function-
From equating $p(s)$ and the Gibbs distribution function we get 
\be
\beta E(s) = -ln P(s) - ln Z =
 -\beta \ln s + \frac{s^2}{(\Gamma[3/2])^2} +4 \ln \Gamma[3/2] -ln 2 - ln Z
\ee

\item 
Temperature-
For the GUE case the temperature is
\be
T=\frac{1}{b_2}= \Gamma[3/2]^2=\frac{\pi}{4}
\ee

\subsection{ Log Normal PDF}

Let’s take the Log-Normal PDF —famous for describing turbulence, financial asset prices, particle sizes, and internet traffic—and run it through our thermodynamic dictionary.

A standard log-normal probability distribution for a variable $x > 0$ is:
\be
p(x) = \frac{1}{x \sigma \sqrt{2\pi}} \exp \left( - \frac{(\ln x - \mu)^2}{2\sigma^2} \right)
\ee

Here, $\mu$ is the scale parameter (mean of $\ln x$) and $\sigma^2$ is the variance of $\ln x$. 

\item  1. The Energy Landscape $E(x)$

Using our mapping $\beta E(x) = -\ln P(x) - \ln Z$ the energy can be written as  
 
\be
\beta E(x) = \frac{1}{2 \sigma^2} (\ln x)^2 + \left( 1 - \frac{\mu}{ \sigma^2} \right) \ln x
\ee

This is a quadratic well in logarithmic space ($\ln x$).
The $(\ln x)^2$ term acts like a harmonic trap, but not for $x$ itself—it traps the order of magnitude ($\ln x$). It severely penalizes values of $x$ that are exponentially too large or exponentially too small.
The $\ln x$ term acts like a constant "tilt" or external force in log-space, shifting the most probable value (the mode) away from zero.

\item 2. Temperature ($T$)

If we define our thermodynamic inverse temperature as $\beta = \frac{1}{\sigma^2}$, then:
 The variance $\sigma^2$ of the log-normal distribution is directly proportional to the Temperature ($T$).
 As $T \to 0$, the log-normal distribution collapses into a sharp Dirac delta function. The fluctuations vanish, and the system freezes into a single deterministic value ($x = e^\mu$).
 As $T$ increases, the  PDF broadens dramatically, developing massive "fat tails." This is the exact mathematical signature of intermittency in turbulence —higher temperature means wilder, more extreme fluctuations where rare, massive values of $x$ become much more common.

---

\item 4. Internal Energy ($U$)

Because $E(x)$ is essentially composed of $\langle \ln x \rangle$ and $\langle (\ln x)^2 \rangle$, and for a log-normal distribution we know:
 $\langle \ln x \rangle = \mu$
 $\langle (\ln x)^2 \rangle = \sigma^2 + \mu^2$

We can plug these directly into our energy function to find $U$:
 \be
 U = \frac{1}{2}\sigma^2 + \sigma^2 \mu - \frac{1}{2}\mu^2
 \ee

The internal energy scales directly with $T$ ($\sigma^2$). As the temperature (fluctuation intensity) goes up, the internal energy of the log-normal system rises.

\item 4. Entropy ($S$)

The Shannon (differential) entropy of a log-normal distribution has a well-known analytical formula:
 
\be
S = \ln(\sigma \sqrt{2\pi e}) + \mu
\ee

 As the temperature $T$ ($\sigma^2$) increases, the entropy increases. This matches core thermodynamics: hotter systems have more disorder, broader probability distributions, and higher information entropy.

 The parameter $\mu$ merely shifts the center of the distribution, adding a constant baseline to the entropy.

\subsection{The Logistic PDF}\label{logistic_pdf}

Let’s run the Logistic Probability Distribution Function through our thermodynamic dictionary.\footnote{In \cite{Bianchi:2025kna} it was found out that  the two dimensional PDF of the pinball scattering was fitted best to a logistic PDF}

The standard logistic PDF for a variable $x$ (where $-\infty < x < \infty$) is:
\be
P(x) = \frac{e^{-\frac{x-\mu}{s}}}{s \left( 1 + e^{-\frac{x-\mu}{s}} \right)^2}
\ee
To keep things clean, let's set the location parameter $\mu = 0$  so that  the simplified PDF reads
\be
P(x) = \frac{e^{-x/s}}{s \left( 1 + e^{-x/s} \right)^2}
\ee

\item 1. The Energy Landscape $E(x)$
Using our mapping $\beta E(x) = -\ln P(x) - \ln Z$:
$$\beta E(x) = -\ln \left( \frac{e^{-x/s}}{s \left( 1 + e^{-x/s} \right)^2} \right) - \ln Z$$

Breaking it down:
$$\beta E(x) = \frac{x}{s} + \ln(s) + 2 \ln\left( 1 + e^{-x/s} \right) - \ln Z$$

If we define our inverse temperature as $\beta = \frac{1}{s}$ (meaning $s = T = \frac{1}{\beta}$), the energy function simplifies to:
$$E(x) = x + 2T \ln\left( 1 + e^{-x/T} \right)$$
(plus constants).

Physical Meaning of this Energy Landscape:
* For large positive $x$ ($x \gg T$): $\ln(1 + e^{-x/T}) \approx e^{-x/T} \approx 0$, so the energy simplifies to $E(x) \approx x$. This is a linear potential (like a constant gravitational pull), causing the probability to drop off exponentially in the positive tail.
 For large negative $x$ ($x \ll -T$): $e^{-x/T}$ becomes huge, so $\ln(1 + e^{-x/T}) \approx \ln(e^{-x/T}) = -x/T$. Plugging this into the energy: $E(x) \approx x + 2T(-x/T) = x - 2x = -x$. This means the energy in the negative tail is $-x$, which creates a symmetric exponential drop-off on the other side.
 Overall shape: Unlike the log-normal or Gaussian distributions (which have parabolic/quadratic $(\ln x)^2$ or $x^2$ wells), the logistic distribution corresponds to a asymmetric logarithmic well that grows linearly at the extremes. This is why the logistic distribution has "fat tails" (heavy tails compared to a Gaussian)—the energy penalty for extreme values is linear, not quadratic.

\item 2. Temperature ($T$) 

As established, $s = T$ is the scale parameter of the logistic distribution.
 As $T \to 0$, the logistic distribution collapses into a Dirac delta function (a spike at $\mu$). The system freezes.
 As $T$ increases, the logistic distribution spreads out widely, flattening the peak and raising the probability of extreme outliers.

 \item 3. Internal energy

The internal energy $U = \langle E(x) \rangle$ is the expectation value of our energy landscape. For a standard logistic distribution, the mean is $\mu = 0$, and the expected value of the energy evaluates cleanly to depend directly on the temperature 
$U \propto T$.
Just like a classical ideal gas or a harmonic system, the internal energy scales linearly with temperature ($U = C T$), showing no anomalous thermodynamic divergences.

\item 4. Entropy ($S$)
The differential Shannon entropy of the logistic distribution has a beautiful, exact analytical solution:
\be
S = \int_{-\infty}^{\infty} -P(x) \ln P(x) \, dx = \ln(s) + 2=\ln(T) + 2\ee

 Just like the log-normal distribution, the entropy scales logarithmically with temperature ($S \sim \ln T$).
 This is the classic signature of systems where available phase space expands exponentially with the scale parameter. Higher temperature $T$ means a wider distribution, more uncertainty, and higher entropy.


\section{Thermodynamics of ratios of adjacent spacings}\label{Toros}

The PDFs associated with ratios was discussed in \cite{Atas2013}
\subsection{Poisson ratios }

Let the (independent) PDF of the Poisson  level spacings  
\be
p(s)=e^{-s}, \qquad s>0 .
\ee
Define the ratio of adjacent spacings
\be
r=\frac{s_{n+1}}{s_n}, \qquad r\in(0,\infty).
\ee
For the Poisson ensemble, the probability density of \(r\) is
\be\label{Poissonrati}
p(r)=\frac{1}{(1+r)^2}, \qquad r>0.
\ee
\item Differential entropy-

The differential entropy is
\be
S
= -\int_{0}^{\infty} \frac{1}{(1+r)^2}\,\ln\!\left(\frac{1}{(1+r)^2}\right)\,dr
= 2 
\ee

\item  The  Energy ($E(r)$)-

For  the  energy function $\beta E(r)$ using  
\be
\beta E(r) = -\ln P(r) = -\ln\left( \frac{1}{(1+r)^2} \right)-\ln Z = 2 \ln(1 + r) -\ln Z
\ee

\item Internal energy-

The internal energy (the ensemble average of this energy) is:
\be
\beta \langle E \rangle = \int_0^\infty E(r) P(r) \, dr = \int_0^\infty \frac{2 \ln(1 + r)}{(1 + r)^2} \, dr= S -\ln Z=2-\ln Z
\ee

\item 

 Effective Temperature ($T$)-
 
The partition function $Z$ is calculated by integrating the unnormalized Boltzmann factor over the domain:
\be
Z = \int_0^\infty e^{-\beta E(r)} dr = \int_0^\infty e^{-2\beta  \ln(1+r)} dr = \int_0^\infty (1 + r)^{-2\beta} dr =
\left[ \frac{(1 + r)^{1 - \frac{2}{T}}}{1 - \frac{2}{T}} \right]_0^\infty
\ee

For the integral to converge (preventing the upper bound from blowing up to infinity), we require the exponent to be strictly negative:
\be
1 - 2\beta < 0 \implies T < 2
\ee

If one assign $T=1$ then $Z=1$ and $E(r) = 2\ln(1+r)$)

\subsection{ PDF of ratios of spacings for Gaussian $\beta$–ensemble}
In addition to the spacings also the ratio of adjacent spacing is a measure of chaotic behavior.
\be
r_n\equiv \frac{\delta_{n+1}}{\delta_n} \qquad \tilde r_n\equiv Min\{r_n,\frac{1}{r_n}\}
\ee

For the Gaussian $\beta$–ensemble, the probability density of the ratio of
adjacent level spacings
\be\label{Atas}
r = \frac{s_{n+1}}{s_n}, \qquad r \ge 0,
\ee
is well approximated by the universal form (Atas et al.)
\be
p_\beta(r)
=
\frac{1}{Z_\beta}\,
\frac{(r+r^2)^{\beta}}{(1+r+r^2)^{1+\frac{3}{2}\beta}},
\ee
where $Z_\beta$ is a normalization constant given by
\be
Z_\beta
=
\int_0^\infty
\frac{(r+r^2)^{\beta}}{(1+r+r^2)^{1+\frac{3}{2}\beta}}\,dr.
\ee

The differential entropy of the ratio distribution is given by

\be
S_\beta
=
\ln Z_\beta
-
\beta
\int_0^\infty p_\beta(r)\,\ln(r+r^2)\,dr
+
\left(1+\frac{3}{2}\beta\right)
\int_0^\infty p_\beta(r)\,\ln(1+r+r^2)\,dr.
\ee

There is no closed-form expression in elementary functions for general
$\beta$; the entropy must be evaluated numerically.

For the classical ensembles, one finds (in nats):
\be
\begin{aligned}
\beta = 1 \; (\text{GOE}): & \qquad S_1 \approx 1.27,\\
\beta = 2 \; (\text{GUE}): & \qquad S_2 \approx 1.20,\\
\beta = 4 \; (\text{GSE}): & \qquad S_4 \approx 1.11.
\end{aligned}
\ee

The monotonic decrease of $H_\beta$ with increasing $\beta$ reflects the
strengthening of level repulsion and the growing rigidity of the spectrum
in the Gaussian $\beta$–ensemble.

\begin{figure}[h!]
  \centering
  \includegraphics[width=0.8\linewidth]{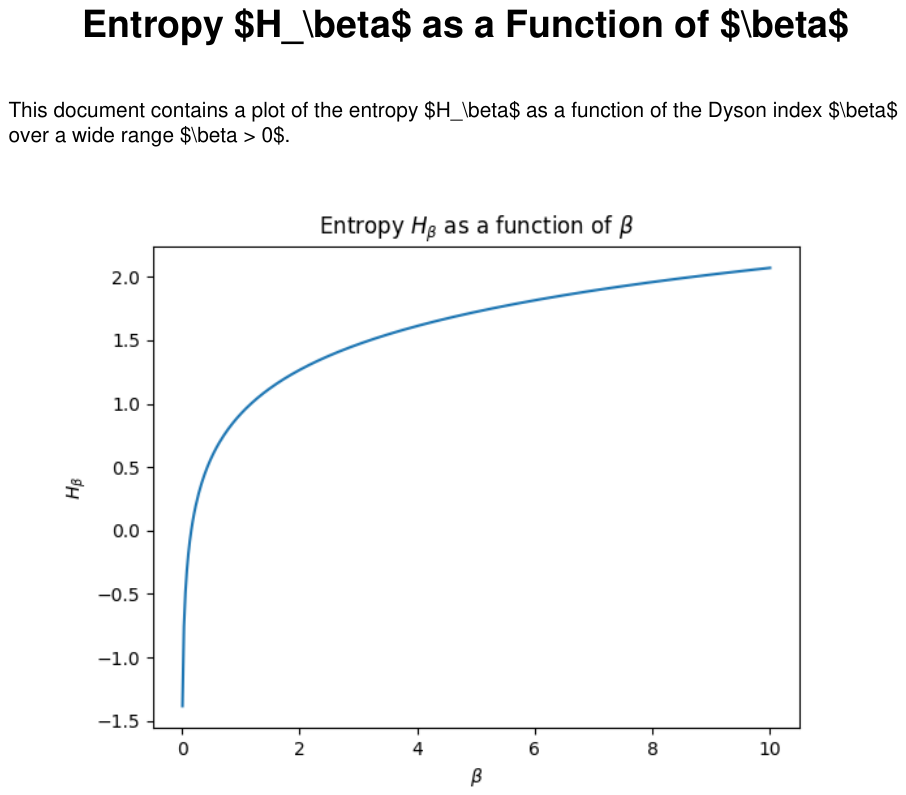}
  \caption{Entropy $H_\beta$ of the ratio distribution in the Gaussian
  $\beta$–ensemble as a function of $\beta$.}\label{Sbeta}
\end{figure}

\subsubsection{GOE ratios}

 The Atas et al.\cite{Atas:2013dis} PDF for GOE ($\beta = 1$)
  is given by:
\be
P_1(r) = \frac{27}{8} \frac{r + r^2}{(1 + r+r^2)^{5/2}}, \qquad r \in [0, \infty)
\ee
\item
 Energy ($E(r)$)-
 
We define the information-theoretic energy (surprisal) by taking the negative logarithm of the PDF:

\be
\beta E(r) = -\ln P_1(r) -\ln Z=
 -\ln\left(\frac{27}{4}\right) - \ln(r + r^2) + 4\ln(1 + r)-\ln Z
 \ee

 For $r \to 0$ (Level repulsion):
\be
E(r) \approx -\ln(r) \to +\infty
\ee
 For $r \to \infty$ (Suppression of extreme ratio fluctuations):
\be
E(r) \approx 2\ln(r) \to +\infty
\ee

\item  Effective Temperature ($T$)-
To extract the temperature, we rewrite the PDF into the standard Gibbs-Boltzmann form:

$$P(r) = \frac{1}{Z} \exp\left(-\frac{E_{\text{phys}}(r)}{T}\right)$$

Expressing the PDF via its exponential argument:
$$P_1(r) = \frac{27}{4} \exp\left( \ln(r + r^2) - 4\ln(1 + r) \right)$$

Using logarithmic properties:
$$\ln(r + r^2) - 4\ln(1 + r) = \ln(r) + \ln(1+r) - 4\ln(1+r) = \ln(r) - 3\ln(1+r)$$

Rearranging into a potential energy format:
$$P_1(r) = \frac{27}{4} \exp\left( -\left[ 3\ln(1+r) - \ln(r) \right] \right)$$

Defining the physical energy as $E_{\text{phys}}(r) = 3\ln(1+r) - \ln(r)$, comparison to the Boltzmann factor yields the canonical scaling temperature:
$$T_{\text{ratio}} = 1$$


\section{Thermodynamics of PDFs of "extrema hights"}\label{ToPoe}   

\subsection{GOE}
In the context of Random Matrix Theory (RMT), the Gaussian Orthogonal Ensemble (GOE) describes systems that are invariant under orthogonal transformations (typically systems with time-reversal symmetry).

The "wavefunctions" in this context are the eigenvectors of the random matrix. Because the GOE is defined by a probability measure that is invariant under orthogonal transformations, the distribution of the eigenvectors is governed by the Haar measure on the Orthogonal Group $O(N)$.

Here is the breakdown of the distribution of these wavefunctions:

 1. The Distribution of Eigenvectors
If $H$ is an $N \times N$ matrix drawn from the GOE, its eigenvectors $\psi_1, \psi_2, \dots, \psi_N$ form an orthonormal basis of $\mathbb{R}^N$.

Because the GOE is invariant under any orthogonal transformation $O$ (i.e., $H \to O^T H O$ has the same probability distribution), the set of eigenvectors is uniformly distributed over the orthogonal group $O(N)$.

This implies that any individual eigenvector $\psi$ is a random vector on the unit sphere $S^{N-1}$ in $\mathbb{R}^N$.

 2. Distribution of Components (Porter-Thomas Distribution)
If we look at a single component $\psi_i$ of a normalized eigenvector $\psi = (\psi_1, \psi_2, \dots, \psi_N)^T$ (where $\sum \psi_i^2 = 1$), the distribution of that component as $N \to \infty$ is Gaussian.

Specifically, for a large matrix, the components $\psi_i$ are independent and identically distributed (i.i.d.) Gaussian variables with:
 Mean: $\langle \psi_i \rangle = 0$
 Variance: $\langle \psi_i^2 \rangle = 1/N$

Consequently, the distribution of the intensity (or probability density) of the wavefunction at a specific site, $P = |\psi_i|^2$, follows the Porter-Thomas distribution:

$$P(x) = \frac{1}{\sqrt{2\pi \langle x \rangle x}} \exp\left( -\frac{x}{2\langle x \rangle} \right)$$

where $x = |\psi_i|^2$ and $\langle x \rangle = 1/N$. This is a Chi-squared distribution with one degree of freedom ($\chi^2_1$).

 3. Physical Significance
 Ergodicity/Delocalization: The fact that the components are Gaussian-distributed with variance $1/N$ means that the wavefunction is "spread out" over all $N$ basis states. This is the hallmark of quantum chaos or delocalized states.
 Comparison to Localization: In contrast, if a system were localized (e.g., Anderson localization), the wavefunction would be concentrated on a small number of sites, and the distribution of $|\psi_i|^2$ would deviate significantly from the Porter-Thomas distribution (often becoming exponential or having a power-law tail).
 Basis Independence: Because the GOE is invariant under orthogonal transformations, this distribution is independent of the basis chosen, provided the basis is fixed and the matrix ensemble is defined relative to it.

 Summary
 Global distribution: The eigenvectors are uniformly distributed according to the Haar measure on $O(N)$.
 Local distribution: The components of the eigenvectors are Gaussian distributed.
 Intensity distribution: The squared components follow the Porter-Thomas distribution ($P(x) \propto x^{-1/2} e^{-Nx/2}$), which is the signature of fully chaotic, delocalized wavefunctions.

\subsubsection{The thermodynamics}

If we switch from the Gaussian distribution of the eigenvector components $x$ to the Chi-squared distribution (specifically, the Porter-Thomas distribution for the intensity $I = x^2$), the resulting thermodynamics will look mathematically different in its exact formulas, but physically equivalent in its underlying reality.

Let's walk through the exact mapping for the intensity $I$:

 1. The Intensity PDF
The Porter-Thomas distribution for the intensity $I$ (where $\langle I \rangle = 1$) is:
$$P(I) = \frac{1}{\sqrt{2\pi I}} e^{-I / 2} \quad \text{for } I \ge 0$$

 2. The Energy Landscape $E(I)$
Using our thermodynamic mapping $\beta E(I) = -\ln P(I) - \ln Z$:
$$\beta E(I) = -\ln \left( \frac{1}{\sqrt{2\pi I}} e^{-I / 2} \right) - \ln Z$$
$$\beta E(I) = \frac{I}{2} + \frac{1}{2}\ln(I) + \frac{1}{2}\ln(2\pi) - \ln Z$$

If we define the effective energy function for intensity $I$:
$$E(I) = \frac{1}{2\beta} I - \frac{1}{2\beta} \ln(I) + \text{constants}$$

Physical Meaning of this Energy:
Unlike the pure harmonic oscillator we got for $x$ ($E \propto I$), the energy for the intensity $I$ includes a logarithmic potential term ($-\frac{1}{2\beta} \ln(I)$).
 The linear term $\frac{I}{2}$ acts like a restoring force pushing $I$ back toward zero.
 The logarithmic term $-\ln(I)$ acts as a repulsive barrier near $I = 0$, preventing the intensity from dropping below zero (since probabilities for $I < 0$ are physically impossible).
 This is precisely the effective potential of a 1D Brownian particle or a 2D radial Gaussian projected onto a single degree of freedom (often seen in random matrix level-spacing or intensity distributions).

 3. The Partition Function ($Z$) and Internal Energy ($U$)
Let's calculate the partition function to ensure normalization:
$$Z = \int_{0}^{\infty} e^{-\beta E(I)} dI = \int_{0}^{\infty} \frac{1}{\sqrt{2\pi I}} e^{-I / 2} dI = 1$$
(Because the Porter-Thomas PDF for intensity is already normalized to 1, the partition function $Z = 1$, which means $\ln Z = 0$!)

Now, let's calculate the internal energy $U = \langle I \rangle$:
$$U = \int_{0}^{\infty} I \cdot P(I) \, dI = \int_{0}^{\infty} I \left( \frac{1}{\sqrt{2\pi I}} e^{-I / 2} \right) dI$$
Using standard gamma function integrals for the chi-squared distribution, the mean intensity is:
$$U = \langle I \rangle = 1$$
(In physical units, $U = 1$ corresponds to $\frac{1}{2} k_B T$, fixing our temperature scale).

 4. The Entropy ($S$)
Let's compute the Shannon entropy $S = -k_B \int_{0}^{\infty} P(I) \ln P(I) \, dI$:
Substitute $\ln P(I) = -\frac{1}{2}\ln(2\pi) - \frac{1}{2}\ln(I) - \frac{I}{2}$:
$$S = -k_B \int_{0}^{\infty} P(I) \left( -\frac{1}{2}\ln(2\pi) - \frac{1}{2}\ln(I) - \frac{I}{2} \right) dI$$

We can break this into three expectation values:
1. $\int P(I) \left( -\frac{1}{2}\ln(2\pi) \right) dI = -\frac{1}{2}\ln(2\pi)$
2. $-\frac{1}{2} \int P(I) \ln(I) \, dI$: For a $\chi^2$ distribution with 1 degree of freedom, the expectation of $\ln(I)$ is known (it involves the Euler-Mascheroni constant $\gamma_E$ and $\ln 2$): $\langle \ln I \rangle = -\gamma_E - \ln 2 \approx -1.27$.
3. $-\frac{1}{2} \int P(I) I \, dI = -\frac{1}{2} \langle I \rangle = -\frac{1}{2}$ (since $\langle I \rangle = 1$).

Summing these up gives a specific, finite numerical value for the entropy of the intensity distribution, which differs from the entropy of the raw component $x$ because we squashed the domain from $(-\infty, \infty)$ to $[0, \infty)$ and warped the coordinate via $I = x^2$.

 Summary
 Does it change the thermodynamics? Mathematically, yes. The energy function gains a $\ln(I)$ term to act as a hard wall at $I=0$, and the numerical value of the entropy changes because we changed variables.
 Does it change the physics? No. Whether you look at the raw eigenvector components $x$ (Gaussian) or their intensities $I = x^2$ ($\chi^2$), you are describing the exact same underlying quantum chaotic system. The laws of thermodynamics (like internal energy $U = 1$ and the First Law) remain entirely consistent; you are merely choosing a different coordinate system (or "phase space representation") to write down the effective Hamiltonian.

\subsection{G$\beta$ ensemble}
   
Yes, there is a direct and very important analog of the Porter-Thomas distribution for the general $\beta$-ensemble.

In the context of Random Matrix Theory, the Porter-Thomas distribution describes the fluctuations of the squared components of the eigenvectors. Specifically, if $v = (v_1, v_2, \dots, v_n)^T$ is a normalized eigenvector of a G$\beta$E matrix, we are interested in the distribution of the intensity $x = |v_i|^2$.

 1. The General Porter-Thomas Distribution
For a G$\beta$E, the distribution of a single component $|v_i|^2$ follows a Gamma distribution (or a $\chi^2$ distribution, depending on the normalization).

The probability density function for $x = |v_i|^2$ is given by:
$$P_\beta(x) = \frac{(\frac{\beta n}{2})^{\frac{\beta}{2}}}{\Gamma(\frac{\beta}{2})} x^{\frac{\beta}{2} - 1} e^{-\frac{\beta n}{2} x}$$

 2. Recovering the Classical Cases
This formula perfectly recovers the standard Porter-Thomas results when we plug in the classical values of $\beta$:

 $\beta = 1$ (GOE):
$$P_1(x) = \sqrt{\frac{n}{2\pi x}} e^{-\frac{nx}{2}}$$
This is the classic Porter-Thomas distribution used in nuclear physics. It is a Chi-squared distribution with 1 degree of freedom (scaled by $1/n$). It has a "heavy tail" near zero, meaning there is a significant probability of finding very small components.

 $\beta = 2$ (GUE):
$$P_2(x) = n e^{-nx}$$
This is an Exponential distribution. The components are much more "regular" than in the GOE case.

 $\beta = 4$ (GSE):
$$P_4(x) = 2n^2 x e^{-2nx}$$
This is a Chi-squared distribution with 4 degrees of freedom. The density is zero at $x=0$, meaning the components are "pushed away" from zero more strongly than in the $\beta=1$ or $\beta=2$ cases.

 3. Physical Interpretation
The Porter-Thomas distribution is a measure of eigenvector ergodicity.

 The $\beta$ dependence: As $\beta$ increases, the distribution $P_\beta(x)$ becomes more peaked around the mean value $1/n$. This reflects the fact that for higher $\beta$, the eigenvectors are "more uniform" and less likely to have extreme fluctuations in their components.
 The "Fluctuation" scale: The fact that the distribution is independent of the index $i$ (due to the rotational invariance of the Haar measure) confirms that the eigenvectors are delocalized. If the system were localized (like in an Anderson insulator), the distribution would not follow this Gamma form; instead, it would be dominated by a delta function at zero and a few large values near 1.

 4. Why is this called "Porter-Thomas"?
The name originates from the 1956 work of Charles Porter and Robert Thomas, who studied the fluctuations of neutron resonance widths in atomic nuclei. They observed that the reduced widths followed a distribution that matched the statistics of the components of GOE eigenvectors.

In modern terms, they were observing that the "coupling" of a quantum state to a decay channel is proportional to the intensity of the eigenvector component at that channel, and that these intensities fluctuate according to the Porter-Thomas law.

 Summary for your research:
If you are working with a general $\beta$ system, you can treat the squared components of your eigenvectors as Gamma-distributed random variables with shape parameter $\alpha = \beta/2$ and rate parameter $\theta = n\beta/2$.

This is a powerful tool because it allows you to test if a system is "G$\beta$E-like" simply by looking at the histogram of its eigenvector components: if the histogram fits the Gamma distribution $P_\beta(x)$, your system is likely in the chaotic/delocalized regime corresponding to that specific $\beta$.
\subsection{The entropy of the Porter Thomas distribution}

To calculate the Shannon entropy of the Porter-Thomas distribution for the G$\beta$E, we look at the entropy of the distribution of the squared components $x = |v_i|^2$.

As established, the components follow a Gamma distribution (specifically, a scaled $\chi^2$ distribution):
$$P_\beta(x) = \frac{(\frac{\beta n}{2})^{\frac{\beta}{2}}}{\Gamma(\frac{\beta}{2})} x^{\frac{\beta}{2} - 1} e^{-\frac{\beta n}{2} x}$$

 1. The Shannon Entropy Formula
The Shannon entropy $S$ is defined as:
$$S = -\int_0^\infty P_\beta(x) \ln(P_\beta(x)) \, dx$$

For a Gamma distribution with shape parameter $k = \beta/2$ and rate parameter $\theta = \beta n / 2$, the entropy is given by the standard formula:
$$S = \ln \Gamma(k) - (k-1)\psi(k) - \ln(\theta) + k$$
where $\psi(k) = \frac{\Gamma'(k)}{\Gamma(k)}$ is the digamma function.

 2. Substituting the G$\beta$E Parameters
Plugging in $k = \beta/2$ and $\theta = \beta n / 2$:

$$S(\beta, n) = \ln \Gamma\left(\frac{\beta}{2}\right) - \left(\frac{\beta}{2} - 1\right)\psi\left(\frac{\beta}{2}\right) - \ln\left(\frac{\beta n}{2}\right) + \frac{\beta}{2}$$

We can simplify this by grouping the terms involving $n$:
$$S(\beta, n) = \ln\left(\frac{2}{\beta n}\right) + \ln \Gamma\left(\frac{\beta}{2}\right) - \left(\frac{\beta}{2} - 1\right)\psi\left(\frac{\beta}{2}\right) + \frac{\beta}{2}$$

 3. Interpretation of the Result
This entropy represents the "uncertainty" or "spread" of the eigenvector components.

 Dependence on $n$: The term $\ln(1/n)$ is the dominant part. As the dimension $n$ increases, the entropy decreases. This makes sense: in a higher-dimensional space, the "available" space for the eigenvector components is more constrained by the normalization condition $\sum |v_i|^2 = 1$, making the distribution more "concentrated" around the mean $1/n$.
 Dependence on $\beta$:
$\beta=1$ (GOE): $S \approx \ln(n) + \text{constant}$. The distribution is broad (heavy-tailed), leading to higher entropy.
 $\beta=2$ (GUE): The distribution is exponential. The entropy is lower than the GOE case.
 $\beta \to \infty$: As $\beta$ increases, the distribution becomes a delta function at $1/n$ (the "perfectly delocalized" state). The entropy approaches $-\infty$ (or more accurately, the distribution becomes deterministic), reflecting the extreme rigidity of the system.

 4. Connection to "Participation Entropy"
In physics, this is often related to the Participation Entropy $S_q$. The Shannon entropy of the components is specifically the limit of the Rényi entropy as $q \to 1$:
$$S_{Shannon} = -\sum_{i=1}^n \langle |v_i|^2 \ln |v_i|^2 \rangle$$

For a G$\beta$E, the expectation value $\langle |v_i|^2 \ln |v_i|^2 \rangle$ can be calculated using the properties of the Gamma distribution:
$$\langle |v_i|^2 \ln |v_i|^2 \rangle = \frac{1}{n} \left( \psi\left(\frac{\beta}{2} + 1\right) - \ln\left(\frac{\beta n}{2}\right) \right)$$

Summing over all $n$ components:
$$S_{Shannon} = \ln\left(\frac{\beta n}{2}\right) - \psi\left(\frac{\beta}{2} + 1\right)$$

(Note: This is a cleaner, more common form used in literature for the entropy of the eigenvector components of a G$\beta$E.)

 Summary
The Shannon entropy of the Porter-Thomas distribution for the G$\beta$E is:
$$S = \ln\left(\frac{\beta n}{2}\right) - \psi\left(\frac{\beta}{2} + 1\right)$$
This formula quantifies how "spread out" the eigenvector is. A higher entropy indicates a more "random" or "chaotic" eigenvector, while a lower entropy indicates a more "ordered" or "localized" structure.

\section{Thermodynamics for higher-dimensional PDFs}\label{TfhdP}

Next we would like to move  from 1D probability density functions (like the Wigner surmise/Poisson spacings) to higher dimensions PDFs.  

In higher dimensions, the "PDF" becomes a {\bf joint probability distribution}  $p(x_1, x_2, \dots, x_N)$, and the geometry of the space completely changes. 
We need to distinguish between the case where the $JPDF$ is a {\bf  product of probabilities} for instance in two dimensions
\be
p_{\vec X}(\vec X)= p_{XY}(x,y) = p_X(x)p_Y(y)
\ee

In this case each PDF can be treated separately as a one dimensional one. The total Shanon entropy is obviously
\be
S(X,Y)= S(X)+ S(Y)
\ee
We will consider from here only the cases where the JPDF is not a product of PDFs.  
\item 1. Entropy

For the JPDF $p(\vec x)$\footnote{ In this section we simply the notation and write $p(\vec x)$ instead of $p_{\vec X}(\vec X) $} Shannon's entropy is given by 
\be
S(\vec X)=-\int p(\vec x) ln p(\vec x)
\ee
\item 2. Energy function

Identifying the JPDF with a Gibbs probability, namely 
\be
p(\vec x)= \frac{e^{\beta E(\vec x)}}{Z}
\ee
We find that the energy function reads
\be
\beta E(\vec x)=- ln p(\vec x) -ln Z 
\ee
So that up to a constant $\beta E(\vec x)=- ln p(\vec x)$.
\item Temperature

Here are the most important and useful higher-dimensional PDFs, mapped by analogy to what you've been working with:

To find the temperature associated with the Gaussian Orthogonal Ensemble (GOE) and the broader Gaussian $\beta$-Ensemble (GBE), we first need to look at their exact joint probability density functions (PDFs) for the eigenvalues.

Unlike the isolated Wigner semicircle distribution (which is just the thermodynamic limit of the GOE), the full GOE/GBE PDF is naturally written in the exact Gibbs Boltzmann form, making the identification of energy and temperature mathematically rigorous.

Next we consider too special cases of the thermodynamics of higher dimensional PDFs:
(i) The Gaussian $\beta$-Ensemble (GBE) PDF.
(ii) Multivariate Gaussian distribtuion.

\subsection{Multivariate Gaussian distribtuion}

For a Multivariate Gaussian (Normal) Distribution, we can solve for the entropy, energy, and temperature completely analytically and explicitly. In fact, it is one of the few high-dimensional PDFs where every thermodynamic quantity has a clean, closed-form expression.


\item  1. The Multivariate Gaussian PDF

Let $\vec{x}$ be an $N$-dimensional vector. The PDF is:
\be  
P(\vec{x}) = \frac{1}{(2\pi)^{N/2} |\Sigma|^{1/2}} \exp\left( -\frac{1}{2} (\vec{x} - \vec{\mu})^T \Sigma^{-1} (\vec{x} - \vec{\mu}) \right)
\ee
where:
 $\vec{\mu}$ is the mean vector.
$\Sigma$ is the $N \times N$ covariance matrix (which governs the spread/shape).
$|\Sigma|$ is the determinant of $\Sigma$.

\item  2. Identifying the Energy Function and Temperature ($\beta$)

Let's take the negative log of the PDF to find $\beta E(\vec{x})$:
\be
-\ln P(\vec{x}) = \frac{N}{2}\ln(2\pi) + \frac{1}{2}\ln|\Sigma| + \frac{1}{2} (\vec{x} - \vec{\mu})^T \Sigma^{-1} (\vec{x} - \vec{\mu})
\ee

If we match this to the thermodynamic form $\beta E(\vec{x}) - (-\ln Z)$, we can naturally split it into a position-dependent energy and a normalization/partition term:

The Inverse Temperature ($\beta$):
Let's define the thermodynamic inverse temperature as $\beta = 1$ (or scale it by a thermal energy parameter $k_B T$). For pure probability, setting $\beta = 1$ is standard, meaning the covariance matrix $\Sigma$ directly dictates the "temperature" or stiffness of the potential well.

The Energy Function $E(\vec{x})$:
\be
E(\vec{x}) = \frac{1}{2} (\vec{x} - \vec{\mu})^T \Sigma^{-1} (\vec{x} - \vec{\mu})\ee
This is the exact potential energy of a multi-dimensional particle in an anisotropic harmonic trap, where $\Sigma^{-1}$ plays the role of the spring-constant tensor.
 The Partition Function ($Z$):
From the constant terms on the right-hand side:
$$\ln Z = -\frac{N}{2}\ln(2\pi) - \frac{1}{2}\ln|\Sigma| \implies Z = (2\pi)^{N/2} |\Sigma|^{1/2}$$
(This matches the known normalization constant of a multivariate Gaussian).

\item 3. Internal Energy ($U$)
The internal energy is the expected value of the energy function:
$$U = \langle E(\vec{x}) \rangle = \int P(\vec{x}) \left[ \frac{1}{2} (\vec{x} - \vec{\mu})^T \Sigma^{-1} (\vec{x} - \vec{\mu}) \right] d^N\vec{x}$$

Using the cyclic property of the trace and the definition of the covariance matrix ($\langle (\vec{x} - \vec{\mu})(\vec{x} - \vec{\mu})^T \rangle = \Sigma$):
\be
U = \frac{1}{2} \text{Tr} \left( \Sigma^{-1} \langle (\vec{x} - \vec{\mu})(\vec{x} - \vec{\mu})^T \rangle \right) = \frac{1}{2} \text{Tr}(\Sigma^{-1} \Sigma) = \frac{1}{2} \text{Tr}(I_N)=\frac{N}{2} 
\ee

This is the Equipartition Theorem in action!
Each of the $N$ independent dimensions contributes $\frac{1}{2}$ to the internal energy (just like $\frac{1}{2}k_B T$ per degree of freedom in classical statistical mechanics).

---

\item 4.  Shannon / Gibbs Entropy ($S$) 

Now we can evaluate the entropy $S = -\int P(\vec{x}) \ln P(\vec{x}) d^N\vec{x}$.

Substitute our log-PDF into the integral:
\be 
S = \int P(\vec{x}) \left[ \frac{N}{2}\ln(2\pi) + \frac{1}{2}\ln|\Sigma| + E(\vec{x}) \right] d^N\vec{x}=\frac{1}{2} \ln \left( (2\pi e)^N |\Sigma| \right)
\ee

 What does this analytical entropy tell us?

1. Volume of Uncertainty ($|\Sigma|$): 

The entropy depends only on the determinant of the covariance matrix, $|\Sigma|$, which represents the generalized "volume" of the uncertainty ellipsoid. The mean $\vec{\mu}$ doesn't affect the entropy at all—shifting the center of a Gaussian doesn't change how spread out it is.

2. Dimension Scaling ($N$):

Entropy scales linearly with dimension $N$. Each added dimension adds $\frac{1}{2}\ln(2\pi e) + \frac{1}{2}\ln(\sigma_i^2)$ of entropy.

3. Connection to Thermodynamic Free Energy:

Recall the thermodynamic relation $F=-\ln Z = -\beta U + S$.
Let's check it:
 $\ln Z = -\frac{N}{2}\ln(2\pi) - \frac{1}{2}\ln|\Sigma|$ and 
 $\beta U - S = \frac{N}{2} - \left[ \frac{N}{2}\ln(2\pi) + \frac{1}{2}\ln|\Sigma| + \frac{N}{2} \right] = -\frac{N}{2}\ln(2\pi) - \frac{1}{2}\ln|\Sigma| = \ln Z =-F$.


 \subsection{The Gaussian $\beta$-Ensemble (GBE) PDF}. 

The joint probability density function of the $n$ eigenvalues $\vec{x} = (x_1, x_2, \dots, x_N)$ for the Gaussian $\beta$-Ensemble is given by:

$$p(\vec{x}) = \frac{1}{Z} \prod_{1 \le i < j \le N} |x_i - x_j|^\beta \exp\left( -\frac{\beta}{2} \sum_{i=1}^N x_i^2 \right)$$

 \item 1. Entropy

 \begin{align}
h_{N,\beta}
={}&
\frac{N}{2}\ln\left(\frac{\pi}{a}\right)
+\frac{N}{2}
\nonumber\\
&+\sum_{j=1}^N
\ln\Gamma\left(1+\frac{j\beta}{2}\right)
-N\ln\Gamma\left(1+\frac{\beta}{2}\right)
\nonumber\\
&-\frac{\beta}{2}
\left[
\sum_{j=1}^N
j\psi\left(1+\frac{j\beta}{2}\right)
-N\psi\left(1+\frac{\beta}{2}\right)
\right].
\label{eq:entropy_final}
\end{align}

 \item 2. Extracting the Energy $E(\vec{x})$

To find the temperature, we map this PDF directly to the standard canonical Boltzmann distribution:
$$p(\vec{x}) = \frac{1}{Z} \exp\left( -\frac{E(\vec{x})}{T} \right)$$

By taking the negative logarithm of the GBE PDF, we can read off the total energy $E(\vec{x})$ of the eigenvalue configuration:

$$-\ln p(\vec{x}) = -\ln\left(\frac{1}{Z}\right) - \sum_{i < j} \beta \ln |x_i - x_j| + \frac{\beta}{2} \sum_{i=1}^N x_i^2$$

Matching this to $\frac{E(\vec{x})}{T}$, we define the physical Hamiltonian (energy) of the Coulomb gas analogy:

$$E(\vec{x}) = \sum_{i=1}^N \frac{1}{2} x_i^2 - \frac{1}{\beta} \sum_{i < j} \ln |x_i - x_j|$$

 3. The Temperature Associated with GBE and GOE

Now, we rewrite the exponent of the GBE distribution to explicitly expose the temperature $T$:

$$p(\vec{x}) = \frac{1}{Z} \exp\left( -\frac{1}{\left(\frac{2}{\beta}\right)} \left[ \sum_{i=1}^N \frac{1}{2} x_i^2 - \frac{1}{\beta} \sum_{i < j} \ln |x_i - x_j| \right] \right)$$

If we compare it to  the Boltzmann factor 
\be 
P \propto \exp\left( -\frac{E}{T_{\text{effective}}} \right)
\ee
we get that 
\be
T_{\text{effective}}=\frac{1}{\beta}
\ee
This follows Dyson \cite{Dyson:1962b} analogy of the JPD with the Coulomb gas in 1d confined in a harmonic oscillator potential ($x_i^2$)   where like-charges repel each other logarithmically ($-\ln|x_i - x_j|$).
 As $\beta$ increases (from GOE $\rightarrow$ GUE $\rightarrow$ GSE), the "temperature" of the eigenvalue electrostatic repulsion effectively drops, meaning the eigenvalues repel each other more strongly and exhibit tighter level repulsion with less thermal fluctuation.

\item 1. The Direct Higher-Dim Analogy: Joint Matrix Ensembles (RMT)

Instead of looking at the spacing between adjacent locations of the extrema points (1D), we would like to  look at the joint probability of all $N$ locations simultaneously.

 The PDF: The Joint Probability Density Function (JPDF) for the Gaussian Orthogonal/Unitary Ensembles (GOE/GUE):
$$P(\lambda_1, \lambda_2, \dots, \lambda_N) = \frac{1}{Z_N} \prod_{i < j} |\lambda_i - \lambda_j|^\beta e^{-\frac{1}{2} \sum \lambda_i^2}$$
 Why it’s useful: That product term $\prod |\lambda_i - \lambda_j|^\beta$ is a Repulsion Term (analogous to level repulsion in 1D). It acts like a 1D Coulomb gas (Dyson's gas model), where eigenvalues push away from each other. It describes complex quantum systems, heavy nuclei, chaotic systems, and even the zeros of the Riemann zeta function.

 2. The Multi-Dimensional "Flat" PDF: The Uniform Distribution on Manifolds
Just like a flat 1D PDF represents maximal entropy over an interval $[a,b]$, higher-dimensional flat PDFs represent maximal ignorance over geometric shapes:
 The Hypercube: $P(\vec{x}) = \frac{1}{(b-a)^N}$ (Independent uniform variables).
 The Hypersphere: Uniform distribution restricted to the surface of a high-dimensional sphere ($\sum x_i^2 = R^2$).
 Why it’s useful: Crucial for Monte Carlo integration, statistical mechanics (the microcanonical ensemble, where phase space is uniformly filled on a constant-energy shell), and high-dimensional geometry (where you encounter the "curse of dimensionality," e.g., most of the volume of a hypersphere lives near its thin surface).

 3. The Multivariate Gaussian (The King of Higher Dimensions)
Just as the 1D Gaussian is ubiquitous, the Multivariate Normal Distribution is the foundational tool for continuous multi-variable data.
 The PDF:
$$P(\vec{x}) = \frac{1}{(2\pi)^{k/2} |\Sigma|^{1/2}} \exp\left( -\frac{1}{2} (\vec{x} - \vec{\mu})^T \Sigma^{-1} (\vec{x} - \vec{\mu}) \right)$$
(where $\vec{\mu}$ is the mean vector and $\Sigma$ is the covariance matrix).
 Why it’s useful: It completely captures linear correlations between variables. It is the steady-state of the Central Limit Theorem in vector spaces, heavily used in signal processing, Kalman filtering, Gaussian Process regression, and portfolio optimization.

 4. The Dirichlet Distribution (The PDF of Probabilities)
If you have a set of variables that must sum to 1 (like probabilities in a categorical distribution or fractions of a mixture), the 1D Beta distribution generalizes to the Dirichlet distribution in higher dimensions.
 The PDF:
$$P(x_1, \dots, x_K) = \frac{1}{B(\vec{\alpha})} \prod_{i=1}^K x_i^{\alpha_i - 1} \quad \text{where } \sum x_i = 1$$
Why it’s useful: It is the "prior of priors" in Bayesian statistics for categorical data (Bayesian multinomial modeling) and is widely used in Machine Learning for topic modeling (like Latent Dirichlet Allocation to categorize documents).

 5. The Wishart Distribution (The PDF of Covariance Matrices)
If a 1D $\chi^2$ (Chi-squared) distribution describes the sum of squared Gaussian variables, the Wishart distribution is its matrix-valued generalization.
 The PDF: It describes the distribution of the $p \times p$ sample covariance matrix derived from multivariate Gaussian samples.
 Why it’s useful: Essential in multivariate statistics, statistical physics, and financial covariance matrix estimation (e.g., cleaning up noisy financial correlation matrices).

 6. Gibbs / Boltzmann Distributions on Graphs (Markov Random Fields)
In statistical mechanics and machine learning, when variables interact locally (like spins on a grid), we use higher-dimensional joint PDFs of the form:
 The PDF:
$$P(\vec{x}) = \frac{1}{Z} \exp\left( -\sum_{(i,j)} V(x_i, x_j) \right)$$
 Why it’s useful: This is the bedrock of Statistical Physics (Ising model, Potts model) and Artificial Intelligence (Boltzmann machines, Hopfield networks, image segmentation). It models how local interactions give rise to global macroscopic states (like phase transitions or image features).

It is actually a profound feature of probability theory that every single one of these higher-dimensional PDFs naturally maps to thermodynamics.

Because of the identity we established earlier—$\beta E(\vec{x}) = -\ln P(\vec{x}) - \ln Z$—any joint probability distribution $P(\vec{x})$ can be instantly reinterpreted as a Boltzmann distribution for a physical system.

However, for some of these PDFs, "thermodynamics" is just a mathematical trick; for others, it is an exact, literal physical reality. Here is how they rank in terms of how naturally thermodynamics applies to them:

---

 Tier 1: Literal Physical Thermodynamics (Exact Analogy)
For these distributions, energy and temperature are not just analogies—they are actual physical quantities governing particles or fields in space.

 1. Gibbs / Boltzmann Distributions on Graphs (e.g., Ising Model)
 The Energy: $E(\vec{x}) = \sum V(x_i, x_j)$ (explicitly built into the exponent).
 Why it’s natural: This *is* statistical mechanics. The math of AI (Boltzmann machines) and physics (magnetism, lattice gases) are identical here. Temperature ($T$) directly controls the amount of thermal fluctuations versus local alignment.
 2. Random Matrix Theory (RMT) Joint Matrix Ensembles
 The Energy: $E(\vec{\lambda}) = \frac{1}{2}\sum \lambda_i^2 - \beta \sum_{i < j} \ln|\lambda_i - \lambda_j|$
 Why it’s natural: As Freeman Dyson famously proved, the joint probability of eigenvalues maps exactly to a 1D electrostatic "Coulomb gas" confined in a harmonic potential at a specific temperature. The repulsion term is literally electrostatic potential energy, and the Gaussian term is a confining harmonic trap.

---

 Tier 2: Information-Theoretic Thermodynamics (Very Natural)
For these distributions, thermodynamic quantities map directly to Information Theory (Shannon entropy, free energy). This is the realm of modern machine learning, Bayesian inference, and statistical data analysis.

 3. The Multivariate Gaussian Distribution
 The Energy:
$$E(\vec{x}) = \frac{1}{2} (\vec{x} - \vec{\mu})^T \Sigma^{-1} (\vec{x} - \vec{\mu})$$
(This is a multidimensional quadratic potential energy well—like a multi-spring system).*
 Why it’s natural: A multivariate Gaussian is the maximum entropy distribution for a system with fixed mean and covariance (variance/energy). The covariance matrix $\Sigma$ acts inversely like a temperature/stiffness tensor. In physics, this describes a multi-dimensional harmonic oscillator in a thermal bath (the Ornstein-Uhlenbeck process).
 4. The Multidimensional Uniform Distribution (Hypersphere / Hypercube)
 The Energy: $E(\vec{x}) = \text{constant}$ (as we just worked out!).
 Why it’s natural: This corresponds precisely to the Microcanonical Ensemble in statistical mechanics—a system isolated at a fixed total energy where all accessible microstates are equally likely. Temperature is formally infinite (or undefined), and all macroscopic properties are driven purely by entropy (the volume of the phase space, e.g., the surface area of the hypersphere).

 Tier 3: Abstract / Curvilinear Thermodynamics (Requires Transformation)
For these distributions, the PDF isn't written in an exponential form natively, but you can always force it by defining an "effective energy" $E(\vec{x}) = -\ln P(\vec{x})$.

 5. The Dirichlet Distribution
 The Energy: $E(\vec{x}) = -\sum (\alpha_i - 1) \ln x_i$
 Why it’s natural: Because the variables must sum to 1 ($\sum x_i = 1$), this represents a system on a simplex (a triangular phase space boundary). The "energy" pushes probabilities away from the edges or corners depending on $\alpha$. It maps to physical systems with particle number conservation or chemical equilibrium (like the Dirichlet process in Bayesian nonparametrics).
 6. The Wishart Distribution
 The Energy: $E(X) = -\frac{\nu - p - 1}{2}\ln|X| + \frac{1}{2}\text{Tr}(\Psi^{-1} X)$
 Why it’s natural: Since Wishart describes matrices, its "energy" involves matrix determinants and traces. This maps to the thermodynamics of multi-variable covariance structures, often appearing in quantum chaos and complex multi-antenna communication systems where matrices play the role of phase space.

---

 Summary
If you want to study thermodynamics where Energy, Temperature, and Entropy behave intuitively:
1. Start with the Multivariate Gaussian if you want quadratic potentials (harmonic traps).
2. Start with RMT/Coulomb Gases if you want interacting particles with forces.
3. Start with Gibbs/Markov Random Fields if you want lattice-based local interactions.
 




\end{itemize}
\end{document}